\documentclass[aps,prxquantum,twocolumn,notitlepage]{revtex4-1}
\usepackage{amsmath,amssymb,mathtools,braket}		
\usepackage{color,graphicx}
\usepackage{natbib} 
\usepackage{hyperref}
\usepackage{bm}
\usepackage[caption=false]{subfig}
\usepackage{epstopdf}
\usepackage{makecell}
\usepackage{booktabs}
\DeclareMathOperator{\Tr}{Tr}
\renewcommand{\Im}{\text{Im}}

\newcommand*{\rom}[1]{\expandafter\@slowromancap\romannumeral #1@}
\newcommand{\HO}{\hat{\mathcal{H}}}
	
\begin{document}
\title{Optimization of the resonator-induced phase gate for superconducting qubits}
\author{Moein Malekakhlagh}\email{Electronic address: moein.malekakhlagh@ibm.com}
\author{William Shanks}\email{Electronic address: willshanks@us.ibm.com}
\author{Hanhee Paik}\email{Electronic address: hanhee.paik@us.ibm.com}
\affiliation{IBM Quantum, IBM Thomas J. Watson Research Center, 1101 Kitchawan Rd, Yorktown Heights, NY 10598}
\date{\today}
\begin{abstract}
The resonator-induced phase gate is a two-qubit operation in which driving a bus resonator induces a state-dependent phase shift on the qubits equivalent to an effective $ZZ$ interaction. In principle, the dispersive nature of the gate offers flexibility for qubit parameters. However, the drive can cause resonator and qubit leakage, the physics of which cannot be fully captured using either the existing Jaynes-Cummings or Kerr models. In this paper, we adopt an ab-initio model based on Josephson nonlinearity for transmon qubits. The ab-initio analysis agrees well with the Kerr model in terms of capturing the effective $ZZ$ interaction in the weak-drive dispersive regime. In addition, however, it reveals numerous leakage transitions involving high-excitation qubit states. We analyze the physics behind such novel leakage channels, demonstrate the connection with specific qubits-resonator frequency collisions, and lay out a plan towards device parameter optimization. We show this type of leakage can be substantially suppressed using very weakly anharmonic transmons. In particular, weaker qubit anharmonicity mitigates both collision density and leakage amplitude, while larger qubit frequency moves the collisions to occur only at large anharmonicity not relevant to experiment. Our work is broadly applicable to the physics of weakly anharmonic transmon qubits coupled to linear resonators. In particular, our analysis confirms and generalizes the measurement-induced state transitions noted in Sank et al. \cite{Sank_Measurement-Induced_2016} and lays the groundwork for both strong-drive resonator-induced phase gate implementation and strong-drive dispersive qubit measurement.      
\end{abstract}
\maketitle
%%%%%%%%%%%%%%%%%%%%%%%%%%%%%%%%%%% Sec:Introduction %%%%%%%%%%%%%%%%%%%%%%%%%%%%%%%%%%%%%%%%
\section{Introduction}
\label{Sec:Intro}

Superconducting circuits \cite{Bouchiat_Quantum_1998, Nakamura_Coherent_1999, Friedman_Quantum_2000, Makhlin_Quantum-State_2001, Van_Quantum_2000, Martinis_Rabi_2002, Blais_Cavity_2004, Wallraff_Strong_2004, Koch_Charge_2007, Majer_Coupling_2007} provide a promising platform for quantum computation \cite{Shor_Algorithms_1994, Divincenzo_Quantum_1995, Barenco_Elementary_1995, Steane_Quantum_1998, Shor_Quantum_1998, Nielsen_Quantum_2002}. To ensure fault-tolerant quantum computation \cite{Shor_Fault_1996, Gottesman_Theory_1998, Kitaev_Fault_2003, Raussendorf_Fault_2007}, the underlying single- and two-qubit gates must satisfy certain error thresholds \cite{Gambetta_Building_2017}. Proposals for two-qubit gates fall into two major categories depending on the requirement for dynamic flux tunability. All-microwave architectures \cite{Paraoanu_Microwave_2006, Rigetti_Fully_2010, Chow_Microwave_2013, Cross_Optimized_2015} employ microwave pulses to entangle qubits, especially useful for fixed-frequency qubits such as the transmon \cite{Koch_Charge_2007}. 

The Resonator-Induced Phase (RIP) gate \cite{Cross_Optimized_2015, Paik_Experimental_2016, Puri_High-Fidelity_2016} exploits an effective dynamic $ZZ$ (controlled-phase) interaction activated by a microwave pulse applied to a mediating bus resonator coupled to the qubits. The effective $ZZ$ rate is proportional to the resonator photon number and qubit-resonator dispersive couplings and inversely proportional to the resonator-drive detuning \cite{Cross_Optimized_2015, Puri_High-Fidelity_2016}. A theoretical proposal for the RIP gate was introduced in Ref.~\cite{Cross_Optimized_2015} as a promising multi-qubit controlled-phase gate. 

The most notable advantage of the RIP gate is the forgiving qubit frequency requirements. The RIP gate was experimentally demonstrated to entangle two fixed-frequency transmon qubits from as close as 380 MHz apart to as far as 1.8 GHz \cite{Paik_Experimental_2016} with fidelity per Clifford \cite{Knill_RB_2008, Magesan_Scalable_2011} ranging $93\%$ to $97\%$ (estimated gate fidelity $96\%$ to $98\%$). One unexplained experimental observation was that the measured fidelity showed improvement at smaller resonator-drive detuning. This did not agree with the detuning dependence expected for measurement-induced dephasing which should cause more error at small detuning. Moreover, this was not consistent with the leakage predictions of the dispersive Kerr model for the RIP gate \cite{Cross_Optimized_2015}.

%%%%%%%%%%%%%%%%%%%%%%%%% Fig: RIPDifferentModels %%%%%%%%%%%%%%%%%%%%%%%%%%%%%%%
\begin{figure*}[t!]
\centering
\subfloat[\label{subfig:Model-AbInitio}]{%
\includegraphics[scale=0.275]{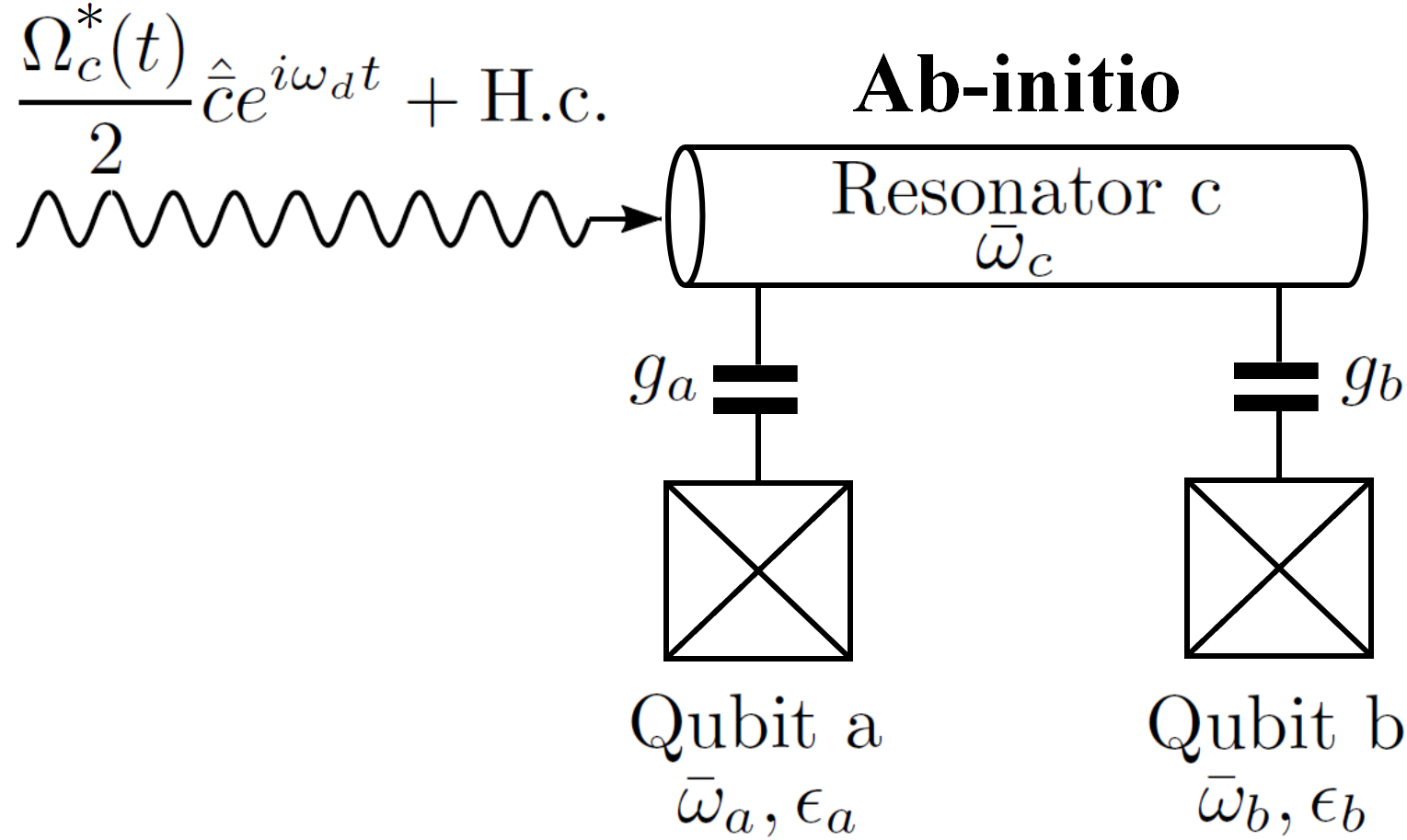}%
}
\subfloat[\label{subfig:Model-KM}]{%
\includegraphics[scale=0.280]{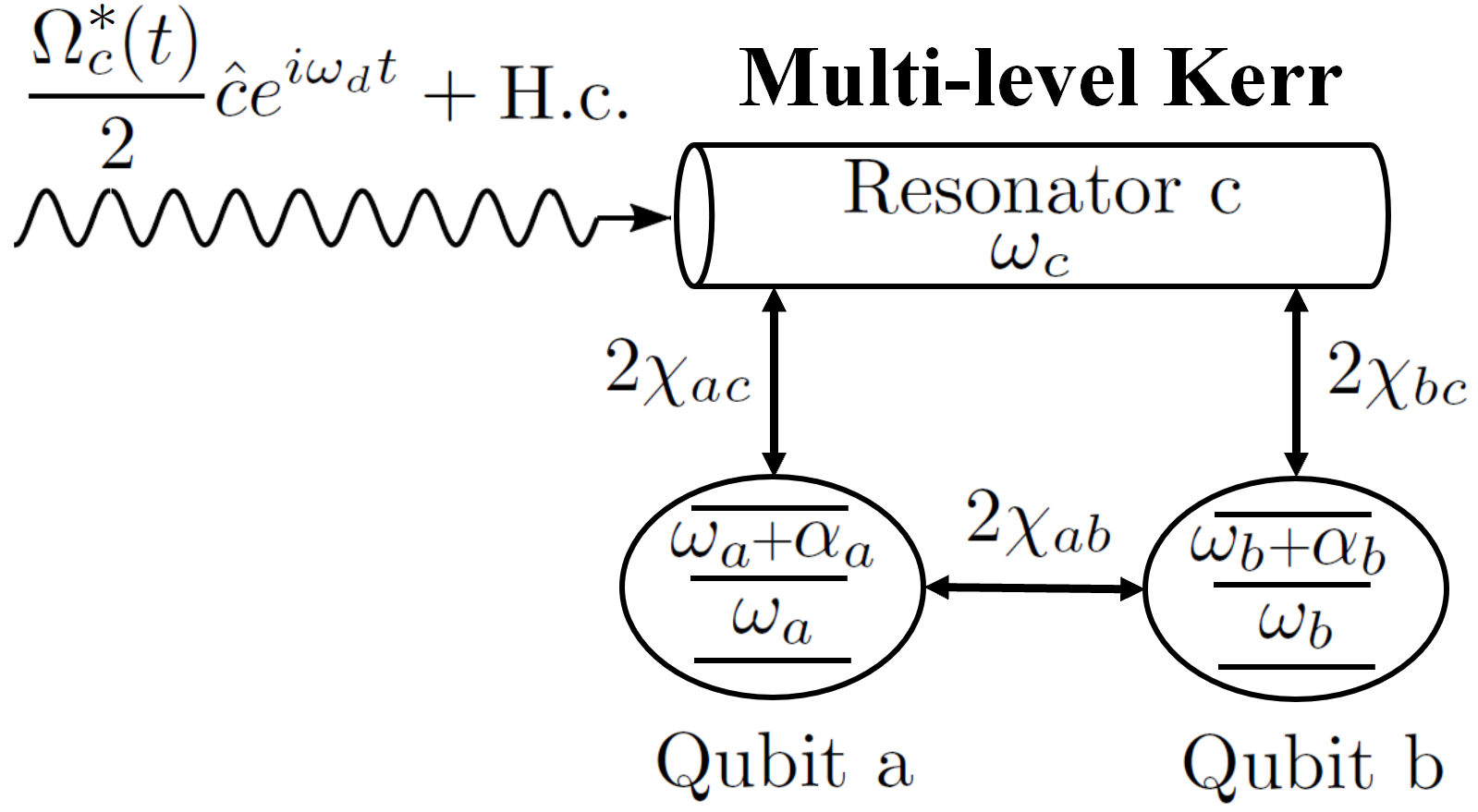}%
}
\subfloat[\label{subfig:Model-TLM}]{%
\includegraphics[scale=0.280]{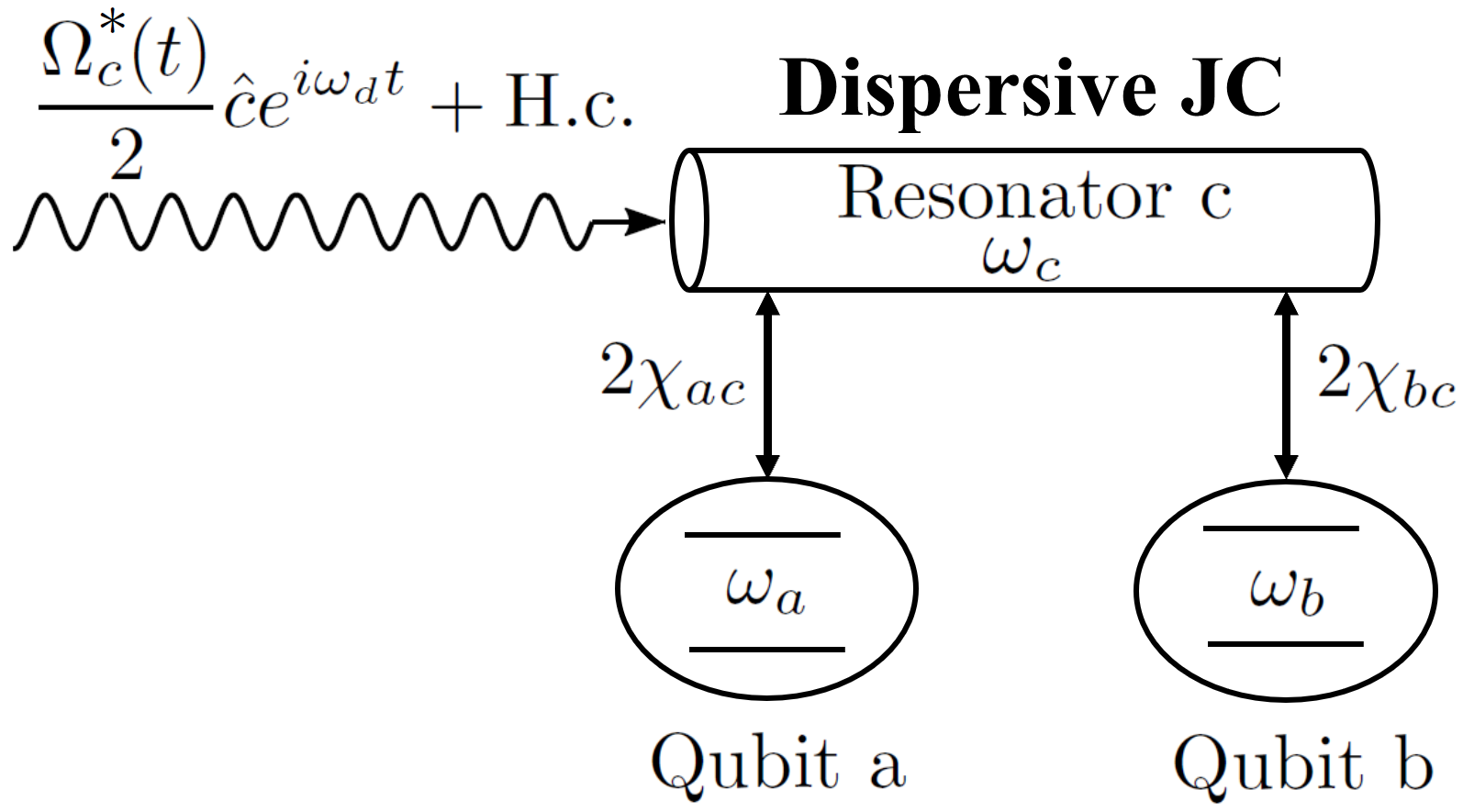}%
}
\caption{RIP gate under three levels of abstraction: (a) an ab-initio model based on the Josephson nonlinearity, (b) a multilevel Kerr model similar to Ref.~\cite{Cross_Optimized_2015} (see Appendix~\ref{App:KM}), and (c) a dispersive JC model similar to Ref.~\cite{Puri_High-Fidelity_2016} (see Appendix~\ref{App:TLM}). When referring to the phenomenological models, we adopt the same level of precision as introduced by the original studies~\cite{Cross_Optimized_2015, Puri_High-Fidelity_2016} as a point of comparison. For instance, \text{only} in the ab-initio model do we consider a RIP drive that acts on the \textit{bare} (shown with a bar) resonator mode although such a modification and more could be made to improve the former models.} 
\label{fig:Model-RIPDifferentModels}
\end{figure*}
%%%%%%%%%%%%%%%%%%%%%%%%%%%%%%%%%%%%%%%%%%%%%%%%%%%%%%%%%%%%%%%%%%%%%%%%%%%%%%%%%

In this work, we characterize the RIP gate operation over a broad span of system parameters for two transmon qubits coupled to a linear resonator and propose experimentally relevant choices for optimal design. To this aim, we introduce an ab-initio model, based on the Josephson nonlinearity, that accounts for high-excitation states of the qubits. We show that the ab-initio analysis agrees with as well as extends the findings of previous studies based on multilevel Kerr \cite{Cross_Optimized_2015, Paik_Experimental_2016} and dispersive Jaynes-Cummings (JC) \cite{Puri_High-Fidelity_2016} models. Our study is focused on the coherent dynamics of the RIP gate, with an emphasis on the effective gate interactions, and on leakage as the main source of error. To derive an effective RIP Hamiltonian \cite{Maricq_Application_1982, Grozdanov_Quantum_1988, Rahav_Effective_2003, Mirrahimi_Modeling_2010, Goldman_Periodically_2014, Wang_Photon_2020, Magesan_Effective_2020, Malekakhlagh_First-Principles_2020}, we employ time-dependent Schrieffer-Wolff Perturbation Theory (SWPT) \cite{Schrieffer_Relation_1966, Bravyi_Schrieffer_2011, Bukov_Schrieffer_2016, Malekakhlagh_Lifetime_2020, Petrescu_Lifetime_2020, Magesan_Effective_2020, Malekakhlagh_First-Principles_2020}. We make comparisons between the effective gate parameters based on the JC, Kerr and ab-initio models and analyze the validity and breakdown of each model. 

Our analysis of leakage reveals various distinct mechanisms that we categorize in terms of only the resonator (residual photons) \cite{Cross_Optimized_2015, Paik_Experimental_2016}, two qubits (qubit-qubit) similar to the cross-resonance (CR) gate \cite{Tripathi_Operation_2019, Malekakhlagh_First-Principles_2020, Hertzberg_Laser_2021, Zhang_High_2020}, one qubit and the resonator (qubit-resonator) similar to the standard readout scheme \cite{Sank_Measurement-Induced_2016, Lescanne_Escape_2019, Verney_Structural_2019}, and both qubits and the resonator (three-body). Revisiting the residual photons, we demonstrate order-of-magnitude improvement via DRAG \cite{Motzoi_Simple_2009, Gambetta_Analytic_2011, Malekakhlagh_Mitigating_2021} on the resonator. Qubit leakage, however, cannot be correctly captured by the phenomenological dispersive models \cite{Cross_Optimized_2015, Puri_High-Fidelity_2016}. This is due to the diagonal form of interactions with respect to the qubits subspace and the two-level assumption. Therefore, using the ab-initio model becomes essential. 

We present extensive ab-initio simulation results for qubit leakage and identify the leakage clusters in terms of collisions between high-excitation qubit states and computational states with high photon number, generalizing Ref.~\cite{Sank_Measurement-Induced_2016}. In particular, we show that qubit-resonator leakage involves very high excitation qubit states (5th-9th) compared to three-body leakage that involves also the medium-excitation states (2nd and above). These findings impose stringent trade-offs on parameter allocation for the RIP gate. As a general remedy for achieving both weaker leakage amplitude and less collision density, without compromising the effective RIP interaction, we propose to use (i) very weakly anharmonic transmons, with anharmonicity reduced down to approximately -200 MHz, and (ii) large qubit frequency of the order of 6 GHz.       
 	     
The paper is organized as follows. In Sec.~\ref{Sec:Model}, starting from the Josephson nonlinearity, we introduce an ab-initio model for the RIP gate and compare it with previous studies \cite{Cross_Optimized_2015, Puri_High-Fidelity_2016}. In Sec.~\ref{Sec:EffHam}, we demonstrate our analytical method, based on time-dependent SWPT, for deriving an effective Hamiltonian for the RIP gate. We then make a comparison of the effective gate parameters ($IZ$, $ZI$ and $ZZ$) between the considered models. Section~\ref{Sec:Leak} provides a characterization of leakage related to frequency collisions based on numerical simulation of the full dynamics. Lastly, in Sec.~\ref{Sec:Tkwys}, we summarize our findings towards system parameter optimization. 

There are ten appendices that will be referred to throughout the paper. In Appendices~\ref{App:TLM} and~\ref{App:KM}, using time-dependent SWPT, we provide the derivation of an effective gate Hamiltonian based on the dispersive JC \cite{Puri_High-Fidelity_2016} and the multilevel Kerr \cite{Cross_Optimized_2015} models, respectively. Appendix~\ref{App:NormModeHam} discusses the normal mode representation of the ab-initio Hamiltonian using a canonical (Bogoliubov) transformation. Appendix~\ref{App:DispTrans} provides the details of a displacement transformation and the effective dynamics for the resonator mode. In Appendix~\ref{App:ResRes}, we analyze the resonator response to basic RIP pulse shapes and provide a \textit{classical} characterization of resonator leakage. Appendix~\ref{App:MLM} presents the derivation of an approximate ab-initio model via normal ordering of the nonlinear interactions. In Appendix~\ref{App:EffRIPHam}, we derive an effective RIP gate Hamiltonian based on the approximate ab-initio model of Appendix~\ref{App:MLM}. Appendix~\ref{App:LeakMech} discusses leakage mechanisms based on a three-level toy model. In Appendix~\ref{App:FidITOLeak}, we characterize the average gate error due to leakage. Lastly, Appendix~\ref{App:NumMet} gives a brief summary of the numerical methods for leakage simulation.      
%%%%%%%%%%%%%%%%%%%%%%%%%%%%%%%%%%%%%%%%%%%%%%%%%%%%%%%%%%%%%%%%%%%%%%%%%%%%%%%%%

%%%%%%%%%%%%%%%%%%%%%%%%%%%%%%%%%% Sec:Model %%%%%%%%%%%%%%%%%%%%%%%%%%%%%%%%%%%%
\section{Model}
\label{Sec:Model}
For comparison, we consider several starting Hamiltonian models for the RIP gate and discuss how the underlying assumptions may limit precise analysis of certain aspects of the gate operation. To this aim, we analyze the RIP gate under the following levels of abstraction: (i) an \textit{exact} ab-initio model for our numerical simulations, (ii) an \textit{approximate} ab-initio model obtained by normal mode expansion of the former, (iii) a multilevel Kerr model similar to the one introduced in Ref.~\cite{Cross_Optimized_2015} and (iv) a dispersive JC model similar to the one in Ref.~\cite{Puri_High-Fidelity_2016} (see Fig.~\ref{fig:Model-RIPDifferentModels}).		

In the exact ab-initio model, we account for the Josephson nonlinearity of each qubit. The system and the drive Hamiltonian in the lab frame can be expressed in a \textit{unitless} quadrature form as \cite{Koch_Charge_2007, Malekakhlagh_NonMarkovian_2016, Malekakhlagh_Origin_2016, Malekakhlagh_Cutoff-Free_2017, Didier_Analytical_2018, Malekakhlagh_Lifetime_2020, Petrescu_Lifetime_2020}
\begin{subequations}
\begin{align}
\begin{split}
\HO_s &=\sum\limits_{j=a,b}\frac{\bar{\omega}_j}{4}\left[(\hat{\bar{y}}_j-y_{gj})^2-\frac{2}{\epsilon_j}\cos(\sqrt{\epsilon_j}\hat{\bar{x}}_j)\right]\\
&+\frac{\bar{\omega}_c}{4}\left(\hat{\bar{x}}_c^2+\hat{\bar{y}}_c^2\right)+\sum\limits_{j=a,b}g_j\hat{\bar{y}}_j\hat{\bar{y}}_c \;,
\end{split}
\label{eqn:Model-Starting Hs}\\
\HO_d(t)&=-[\Omega_{cx}(t)\cos(\omega_d t)+\Omega_{cy}(t)\sin(\omega_d t)]\hat{\bar{y}}_c \;,
\label{eqn:Model-Starting Hd}
\end{align}
\end{subequations}
where the qubit modes, the resonator mode and the drive are labeled as $a$, $b$, $c$ and $d$, respectively. The bar notation is used for quantities in the \textit{bare} frame, to distinguish from the \textit{normal} mode frame quantities that will be denoted with no bar. Moreover, $\bar{\omega}_j\equiv \sqrt{8E_{Cj}E_{Jj}}$ is the harmonic frequency, $\epsilon_j \equiv \sqrt{2E_{Cj}/E_{Jj}}=\varphi_{\text{j,ZPF}}^2$ is a \textit{small} unitless anharmonicity measure and $y_{gj}\equiv n_{gj}/n_{\text{j,ZPF}}$ is the unitless gate charge for qubit $j=a, b$. Qubits are capacitively coupled to a common bus resonator with strength $g_a$ and $g_b$, respectively. The flux (phase) and charge (number) quadratures are written in a unitless form as $\hat{\bar{x}}_k \equiv \hat{\bar{k}}+\hat{\bar{k}}^{\dag}$ and $\hat{\bar{y}}_k \equiv -i(\hat{\bar{k}}-\hat{\bar{k}}^{\dag})$ for $k=a, b, c$. The drive acts capacitively on the resonator mode with time-dependent pulse amplitudes $\Omega_{cx}(t)$ and $\Omega_{cy}(t)$, and carrier frequency $\omega_d$. We take $\Omega_{cy}(t)$ as the main RIP pulse, while $\Omega_{cx}(t)$ provides additional degree of freedom used for DRAG \cite{Motzoi_Simple_2009, Gambetta_Analytic_2011, Malekakhlagh_Mitigating_2021}. Hamiltonian~(\ref{eqn:Model-Starting Hs})--(\ref{eqn:Model-Starting Hd}) serves as our point of reference and is used for the full numerical simulation.

For our analytical calculations, starting from Eqs.~(\ref{eqn:Model-Starting Hs})--(\ref{eqn:Model-Starting Hd}), we derive an approximate ab-initio model by first solving for the normal modes up to the harmonic theory, and then expanding the nonlinearity in the normal mode frame \cite{Malekakhlagh_Lifetime_2020,Petrescu_Lifetime_2020} similar to the black-box quantization \cite{Nigg_BlackBox_2012} and energy participation ratio \cite{Minev_EPR_2020} methods. Normal mode expansion converges faster when the transmon qubits are weakly anharmonic ($\epsilon \lessapprox 0.2 $) and also when the effective interactions are implemented \textit{dispersively}, i.e. large detuning between the qubits and the resonator as well as between the qubits themselves compared to qubits anharmonicity. This is an alternative to \textit{few-quantum-level} models used commonly for systems operating in the straddling regime like the CR gate \cite{Magesan_Effective_2020, Tripathi_Operation_2019, Malekakhlagh_First-Principles_2020}. The normal mode representation of the RIP gate Hamiltonian reads 
\begin{subequations}
\begin{align} 
\begin{split}
\HO_s &=\sum\limits_{k=a,b,c}\tilde{\omega}_k\hat{k}^{\dag}\hat{k}+\sum\limits_{j=a,b}\sum\limits_{n=2}^{\infty} \frac{\bar{\omega}_j}{2}(-\epsilon_j)^{n-1}\\
&\times \frac{\left[\left(\sum\limits_{k=a,b,c}u_{jk}\hat{k}\right)+\text{H.c.}\right]^{2n}}{(2n)!} \;,
\end{split}
\label{eqn:Model-NormMode Hs}\\
\begin{split}
\HO_d(t) &= [\Omega_{cx}(t)\cos(\omega_d t)+\Omega_{cy}(t)\sin(\omega_d t)]\\
&\times i[\sum\limits_{k=a,b,c}v_{ck}\hat{k}-\text{H.c.}] \;,	
\end{split}
\label{eqn:Model-NormMode Hd}
\end{align}
\end{subequations}	
where $u_{jk}$ and $v_{jk}$, $j,k=a,b,c$, are the flux and charge hybridization coefficients that relate the corresponding bare and normal mode quadratures and are derived via a canonical (Bogoliubov) transformation \cite{Jellal_Two_2005, Merdaci_Entanglement_2020, Malekakhlagh_Lifetime_2020} (see Appendix~\ref{App:NormModeHam}). Moreover, the normal mode \textit{harmonic} frequencies are shown as $\tilde{\omega}_k$ in order to distinguish from the renormalized normal mode frequencies (no bar) that contain the static (Lamb) corrections.
  
Equations~(\ref{eqn:Model-NormMode Hs})--(\ref{eqn:Model-NormMode Hd}) demonstrate a rich variety of possible mixing both at the linear level through charge hybridization and the nonlinear level through flux hybridization. Equation~(\ref{eqn:Model-NormMode Hd}) shows that the RIP drive that is \textit{ideally} supposed to populate the resonator mode will also act on the normal qubit modes through charge hybridization. For typical RIP parameters, the cross-drive on the qubits can be $10$\% of the intended resonator drive. Furthermore, flux hybridization in Eq.~(\ref{eqn:Model-NormMode Hs}) leads to numerous nonlinear processes even up to the quartic expansion. To handle such complexity, we employ SNEG \cite{Zitko_Sneg_2011} in Mathematica, which allows sybmolic manipulation of bosonic operators, and derive an approximate ab-initio model for the RIP gate (see Appendix~\ref{App:MLM}). 

To compare to previous studies, we also consider starting models similar to the ones introduced in Refs.~\cite{Puri_High-Fidelity_2016,Cross_Optimized_2015} (see Fig.~\ref{fig:Model-RIPDifferentModels}). Reference~\cite{Cross_Optimized_2015} models the transmon qubits as weakly anharmonic Kerr oscillators as
\begin{align}
\begin{split}
\HO_{s,\text{Kerr}} &=\sum\limits_{j=a,b}\left[\omega_j \hat{n}_j+\frac{\alpha_j}{2}\hat{n}_j(\hat{n}_j-1)\right]\\
&+\omega_c\hat{n}_c+\sum\limits_{j,k=a,b,c\atop j> k}2\chi_{jc}\hat{n}_j\hat{n}_c \;,
\end{split}
\label{eqn:Model-Def of H_s,Kerr}  
\end{align}
with the anharmonicity and dispersive couplings denoted by $\alpha_j$ and $2\chi_{jk}$ for $j,k=a,b,c$. The dispersive JC model \cite{Puri_High-Fidelity_2016} is similar to the Kerr model, but with two-level qubits as 
\begin{align}
\HO_{s,\text{JC}}=\sum\limits_{j=a,b}\frac{\omega_{j}}{2}\hat{\sigma}_j^{z}+\omega_c \hat{n}_c+\sum\limits_{j=a,b} \chi_{jc}\hat{n}_c\hat{\sigma}_j^z \;,
\label{eqn:Model-Def of H_s,JC} 
\end{align}
where $\hat{\sigma}_j^z\equiv \ket{1_j}\bra{1_j}-\ket{0_j}\bra{0_j}$. Moreover, in both studies, it is assumed that the drive acts only on the \textit{normal} resonator mode as 
\begin{align}
\HO_d(t)=\frac{1}{2}\left[\Omega_c^*(t)\hat{c}e^{i\omega_d t}+\Omega_c(t)\hat{c}^{\dag}e^{-i\omega_d t}\right] \;,	
\label{eqn:Model-Def of H_d}
\end{align}
where $\Omega_c(t)\equiv \Omega_{cy}(t)-i \Omega_{cx}(t)$. Equations~(\ref{eqn:Model-Def of H_s,Kerr})--(\ref{eqn:Model-Def of H_d}) have been modified with respect to the original studies to be consistent with our convention of denoting the full dispersive shift by $2\chi$ and the drive amplitude by $\Omega_c(t)$.

In terms of capturing the dynamic $ZZ$ (RIP) interaction, we find in the following that the dipsersive JC model is valid in the parameter regime $2\chi_{ac},2\chi_{bc} \ll |\Delta_{cd}| \ll |\Delta_{ad}|, |\Delta_{bd}|$, the multilevel Kerr model in $2\chi_{ac},2\chi_{bc}<|\Delta_{cd}| \ll |\Delta_{ad}|, |\Delta_{bd}|$ and the approximate ab-initio model in a broader resonator-drive detuning range of $2\chi_{ac},2\chi_{bc} < |\Delta_{cd}| < |\Delta_{ad}|, |\Delta_{bd}|$ (see Sec.~\ref{Sec:EffHam}). In terms of capturing qubit leakage, however, the dispersive JC model cannot be used due to its two-level construction. The multilevel Kerr model is also unable to correctly predict qubit leakage for two reasons. Firstly, the starting Hamiltonian is \textit{diagonal} with respect to the qubits preventing transitions out of the computational subspace. This can in principle be improved by adding phenomenological direct drive terms on the qubit modes. However, more importantly, Kerr estimates for transmon eigenenergies become less valid for high-excitation states, for which we observe considerable leakage due to frequency collisions. Therefore, using the exact ab-initio model is necessary in the characterization of qubit leakage discussed in Sec.~\ref{Sec:Leak}.

%%%%%%%%%%%%%%%%%%%%%%%%%%%%%%%%%%%%%%%%%%%%%%%%%%%%%%%%%%%%%%%%%%%%%%%%%%%%%%%%%

%%%%%%%%%%%%%%%%%%%%%%%%%%%%%%%%% Sec: EffHam %%%%%%%%%%%%%%%%%%%%%%%%%%%%%%%%%%%%%%%%%%%%%%%%%%%%%%
\section{Effective RIP gate Hamiltonian}
\label{Sec:EffHam}

In Cross et al.~\cite{Cross_Optimized_2015}, the gate operation was described within a Lindblad formalism by modeling the qubits as multilevel Kerr oscillators, and analytical estimates for the effective $ZZ$ rate were derived using the generalized P-representation \cite{Drummond_Generalised_1980}. Here, we analyze effective RIP interactions via SWPT and make a comparison between the aforementioned starting models. The effective RIP Hamiltonian takes the form
\begin{align}
\HO_{\text{RIP,eff}}(t) \equiv \omega_{iz}(t)\frac{\hat{I}\hat{Z}}{2}+\omega_{zi}(t)\frac{\hat{Z}\hat{I}}{2}+\omega_{zz}(t)\frac{\hat{Z}\hat{Z}}{2} \;,
\label{eqn:EffHam-Def of H_RIP,eff(t)}
\end{align}
with a dynamic frequency shift for each qubit along with a two-qubit $ZZ$ interaction which consists of \textit{static} and \textit{dynamic} contributions. Our analytical method implements a series of unitary transformations from the lab frame to the \textit{effective} diagonal frame of Eq.~(\ref{eqn:EffHam-Def of H_RIP,eff(t)}).

In Sec.~\ref{Subsec:SWPT}, we discuss a generic derivation of the effective Hamiltonian using time-dependent SWPT. In Sec.~\ref{Subsec:Kerr}, we apply it to the multilevel Kerr model, as a simpler case that captures the essential mechanism for the effective $ZZ$ interaction. Section~\ref{Subsec:Compare} compares the effective Hamiltonians found by applying the approach to all three models. Furthermore, detailed derivations of the effective RIP Hamiltonians based on the dispersive JC, the multilevel Kerr and the ab-initio models can be found in Appendices~\ref{App:TLM}, \ref{App:KM} and~\ref{App:EffRIPHam}, respectively.   	 
%%%%%%%%%%%%%%%%%%%%%%%%% SubSec: SWPT %%%%%%%%%%%%%%%%%%%%%%%%%%%%%%%%%%%%%%%%%	
\subsection{Effective Hamiltonian via SWPT}
\label{Subsec:SWPT}
To arrive at the effective Hamiltonian introduced in Eq.~(\ref{eqn:EffHam-Def of H_RIP,eff(t)}), we devise a unitary transformation $\hat{U}_{\text{diag}}(t)$ that maps the lab frame to the diagonal frame as
\begin{align}
\HO_{I,\text{eff}}(t)\equiv \hat{U}_{\text{diag}}^{\dag}(t)\left[\HO_{s}+\HO_{d}(t)-i\partial_t\right]\hat{U}_{\text{diag}}(t) \;.
\label{eqn:EffHam-Def of H_I,eff(t)}
\end{align}
It can be decomposed into intermediate unitary transformations as
\begin{align}
\hat{U}_{\text{diag}}(t) \equiv \hat{D}[d_c(t)] \hat{U}_0(t) \hat{U}_{\text{SW}}(t) \;,
\label{eqn:EffHam-U_diag decomp}
\end{align}
where $\hat{D}[d_c(t)]$, $\hat{U}_0(t)$ and $\hat{U}_{\text{SW}}(t)$ denote a coherent displacement transformation of the resonator mode, transformation to the interaction frame, and finally a SW transformation, respectively. In the following, we discuss each transformation in more detail. 

A typical RIP drive can populate the resonator mode with several photons. The resonator response can therefore be effectively described in terms of classical (coherent) and quantum fluctuation parts. Formally, this is achieved by the displacement transformation
\begin{align}
\hat{D}[d_c(t)]\equiv e^{d_c(t)\hat{c}^{\dag}-d_c^*(t)\hat{c}} \;,
\label{eqn:EffHam-Def of D[d_c]}
\end{align}
where $\hat{D}^{\dag}[d_c(t)]\hat{c}\hat{D}[d_c(t)]=\hat{c}+d_c(t)$. We then set $d_c(t)$ to cancel out terms linear in $\hat{c}$ and $\hat{c}^{\dag}$ in the displaced frame Hamiltonian. Up to the quartic expansion of the Josephson nonlinearity, this is equivalent to the response of a classical Kerr oscillator to the input RIP pulse. Under the rotating-wave approximation, one finds
\begin{align}
\dot{\eta}_c(t)+i\Delta_{cd}\eta_c(t)+i\alpha_c|\eta_c(t)|^2\eta_c(t)=-\frac{i}{2}\Omega_{c}(t)\;,
\label{eqn:EffHam-Cond for eta_c(t)}
\end{align}
where $\eta_c(t)\equiv d_c(t)e^{i\omega_d t}$ is the slowly-varying response amplitude, $\Delta_{cd}\equiv \omega_c -\omega_d$ is the resonator-drive detuning and $\alpha_c$ is the effective anharmonicity for the resonator [see Appendices~\ref{App:TLM} and~\ref{App:DispTrans} for the derivations based on the JC ($\alpha_c=0$) and the ab-initio models, respectively]. In Appendix~\ref{App:ResRes}, based on Eq.~(\ref{eqn:EffHam-Cond for eta_c(t)}), we have analyzed the resonator response and leakage in detail. Importantly, we show that using DRAG \cite{Motzoi_Simple_2009, Gambetta_Analytic_2011, Malekakhlagh_Mitigating_2021} can be helpful in suppressing the residual photons \cite{Cross_Optimized_2015} (see Sec.~\ref{Sec:Leak} and Appendix~\ref{App:ResRes}). 

Next, in the displaced frame, we split the Hamiltonian into \textit{bare} and \textit{interaction} contributions as $\HO_0+ \HO_{\text{int}}(t)$. The interaction frame Hamiltonian is then found by the unitary transformation $\hat{U}_0\equiv \exp(-i\HO_{0} t)$ as $\HO_{I}(t) \equiv \hat{U}_0^{\dag}(t) \HO_{\text{int}}(t)\hat{U}_0(t)$. We note that there is flexibility in defining what bare and interaction contributions are. In particular, in Ref.~\cite{Puri_High-Fidelity_2016}, cross-Kerr terms were \textit{not} kept as bare contribution, which is justified for the explored parameter regime $|\Delta_{cd}|\gg |2\chi_{ac}|,|2\chi_{bc}|$. When referring to the dispersive JC model, we follow the same approximation as a point of comparison (see Appendix~\ref{App:TLM}). In Ref.~\cite{Cross_Optimized_2015} and the ab-initio model, cross-Kerr interaction terms are kept in the bare Hamiltonian allowing for more precise modeling of the effective gate parameters for resonator-drive detunings comparable to the dispersive shift, i.e. $|\Delta_{cd}| \sim |2\chi_{ac}|, |2\chi_{bc}|$.

We then apply time-dependent SWPT to diagonalize the interaction frame Hamiltonian. The SW transformation is in principle a unitary transformation of the form,
\begin{align}
\hat{U}_{\text{SW}}(t)=e^{-i\hat{G}(t)} \;,
\label{Eq:ResRes-Def Of U_SW}
\end{align}
where we expand the generator $\hat{G}(t)$ and the resulting effective Hamiltonian up to an arbitrary order in the interaction \cite{Gambetta_Analytic_2011, Magesan_Effective_2020, Malekakhlagh_Lifetime_2020, Petrescu_Lifetime_2020, Malekakhlagh_First-Principles_2020, Petrescu_Accurate_2021, Malekakhlagh_Mitigating_2021}. Here, we implement the perturbation up to the second order. The first order perturbation can be summarized as 
\begin{subequations}
\begin{align}
&\HO_{I,\text{eff}}^{(1)}(t)=\mathcal{S}\left(\HO_{I}(t)\right) \;,
\label{eqn:EffHam-H_I,eff^(1) Cond}\\
&\dot{\hat{G}}_1(t)=\mathcal{N}\left(\HO_{I}(t)\right) \;,
\label{eqn:EffHam-G1 Cond}
\end{align}
while the second order reads (see Appendix~C of Ref.~\cite{Malekakhlagh_First-Principles_2020})
\begin{align}
&\HO_{I,\text{eff}}^{(2)}(t)=\mathcal{S}\Big(i[\hat{G}_1(t),\HO_I(t)]-\frac{i}{2}[\hat{G}_1(t),\dot{\hat{G}}_1(t)]\Big) \;,
\label{eqn:EffHam-H_I,eff^(2) Cond}\\
&\dot{\hat{G}}_2(t)= \mathcal{N}\Big(i[\hat{G}_1(t),\HO_I(t)]-\frac{i}{2}[\hat{G}_1(t),\dot{\hat{G}}_1(t)]\Big) \;.
\label{eqn:EffHam-G2 Cond}
\end{align}
\end{subequations}
In Eqs.~(\ref{eqn:EffHam-H_I,eff^(1) Cond})--(\ref{eqn:EffHam-G2 Cond}), $\mathcal{S}(\bullet)$ and $\mathcal{N}(\bullet)$ denote the diagonal and off-diagonal parts of an operator. In summary, at each order in perturbation, we keep the diagonal contributions in the effective Hamiltonian and solve for a non-trivial SW generator that removes the off-diagonal part. For instance, first order off-diagonal terms in Eq.~(\ref{eqn:EffHam-G1 Cond}) can produce diagonal contributions through 2nd order mixings that appear in terms of commutators in Eq.~(\ref{eqn:EffHam-H_I,eff^(2) Cond}).  

Given the effective Hamiltonian in an extended Hilbert space, we read off the relevant gate parameters as
\begin{subequations}
\begin{align}
&\omega_{iz}(t)\equiv \frac{1}{2}\text{Tr}\left\{\left(\hat{I}_a \otimes \hat{Z}_b \otimes \ket{0_c} \bra{0_c} \right)\cdot \HO_{I,\text{eff}}(t)\right\} \;,
\label{eqn:EffHam-Def of w_iz}\\
&\omega_{zi}(t)\equiv \frac{1}{2}\text{Tr}\left\{\left(\hat{Z}_a \otimes \hat{I}_b \otimes \ket{0_c} \bra{0_c} \right)\cdot \HO_{I,\text{eff}}(t)\right\} \;, 
\label{eqn:EffHam-Def of w_zi}\\
&\omega_{zz}(t)\equiv \frac{1}{2}\text{Tr}\left\{\left(\hat{Z}_a \otimes \hat{Z}_b \otimes \ket{0_c} \bra{0_c} \right)\cdot \HO_{I,\text{eff}}(t)\right\},
\label{eqn:EffHam-Def of w_zz}
\end{align}
\end{subequations}
where $\hat{Z}\equiv \ket{0}\bra{0}-\ket{1}\bra{1}$. We have distinguished between the \textit{physical} and the \textit{logical} Pauli operators, used in Eqs.~(\ref{eqn:Model-Def of H_s,JC}) and~(\ref{eqn:EffHam-Def of w_iz})--(\ref{eqn:EffHam-Def of w_zz}) such that $\hat{Z}=-\hat{\sigma}^z$. The effective Hamiltonian is defined in the displaced frame, and hence the zero-photon subspace in Eqs.~(\ref{eqn:EffHam-Def of w_iz})--(\ref{eqn:EffHam-Def of w_zz}) corresponds to no excitations beyond the coherent photon number $|\eta_c(t)|^2$. In experimental results, it is more common to report the \textit{full} $ZZ$ rate, which is twice the value quoted in this paper.

In order to calibrate a controlled-phase gate of rotation angle $\theta_{zz}$ we need to set $\int_{0}^{\tau}dt'\omega_{zz}(t') = \theta_{zz}$, where $\omega_{zz}(t)=\omega_{zz}^{(0)}+\omega_{zz}^{(2)}(t)+O(\HO_{I}^4)$. In particular, $\theta_{zz}=\pi/2$ is equivalent to CNOT up to single-qubit Hadamard and $Z$ rotations as
\begin{subequations}
\begin{align}
&\quad \quad \hat{U}_{\text{CZ}}=\exp \left[-i\frac{\pi}{4} \left(\hat{I}\hat{I}-\hat{I}\hat{Z}-\hat{Z}\hat{I}+\hat{Z}\hat{Z} \right)\right] \;,
\label{eqn:EffHam-U_cz ITO U_zz} \\
& \quad \quad \hat{U}_{\text{CX}}=\hat{I}\hat{H} \cdot \hat{U}_{\text{CZ}} \cdot \hat{I}\hat{H} \;,
\label{eqn:EffHam-U_cx ITO U_cz}
\end{align}
\end{subequations}
with $\hat{U}_{\text{CZ}}$, $\hat{U}_{\text{CX}}$ and $\hat{H}$ denoting the controlled-$Z$, controlled-$X$ and Hadamard operations. 

The procedure outlined in this section can be extended to a Lindblad master equation to arrive at an effective $ZZ$ rate that accounts for the resonator decay rate $\kappa_c$ \cite{Cross_Optimized_2015}. The result can be obtained by replacing $\omega_c\rightarrow \omega_c-i\kappa_c/2$ and reading off the real part of Eq.~(\ref{eqn:EffHam-Def of w_zz}) (see Cross et al.~\cite{Cross_Optimized_2015} for more detail).
%%%%%%%%%%%%%%%%%%%%%%%%%%%%%%%%%%%%%%%%%%%%%%%%%%%%%%%%%%%%%%%%%%%%%%%%%%%%%%%%

%%%%%%%%%%%%%%%%%%%%%%%%% SubSec: multilevel Kerr model %%%%%%%%%%%%%%%%%%%%%%%%
\subsection{Kerr model}
\label{Subsec:Kerr}
We next analyze the effective Hamiltonian and gate parameters based on the multilevel Kerr model (see Appendix~\ref{App:KM} for derivation). The perturbative description of the effective Hamiltonian is valid when $|\chi_{jc} \eta_c(t)| < |\Delta_{cd}|$ for $j=a,b$.

\begin{figure}
\centering
\includegraphics[scale=0.40]{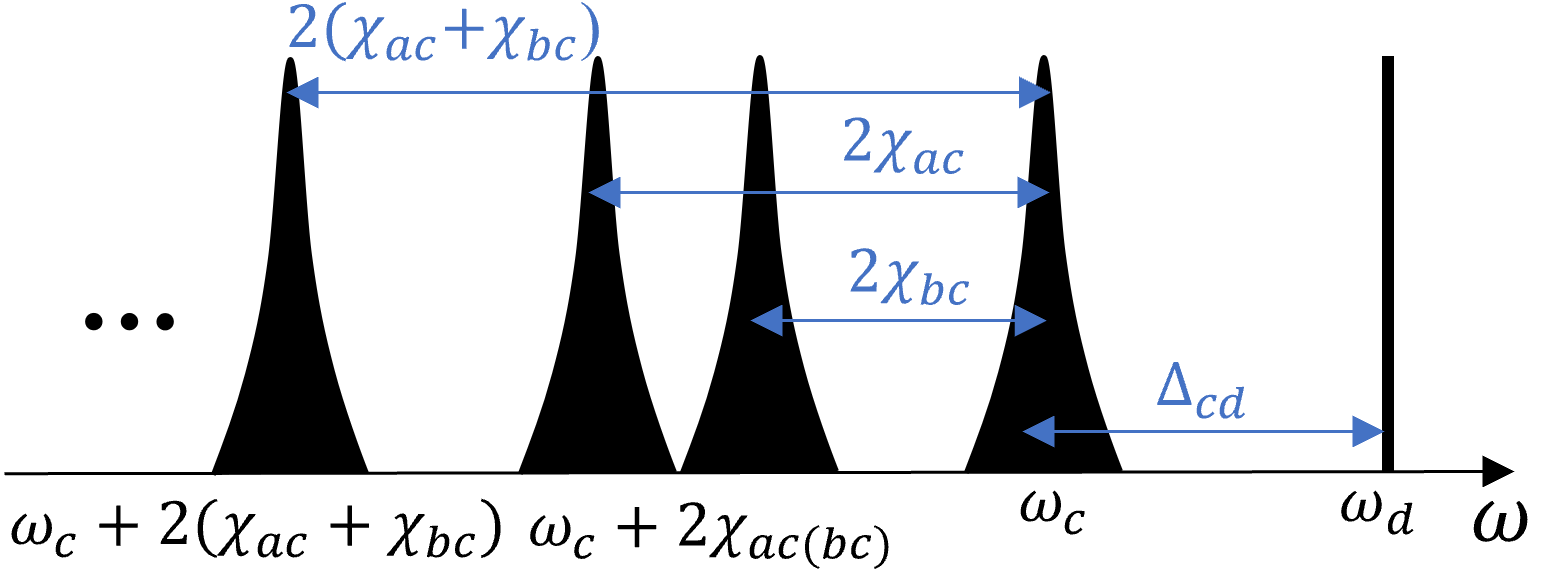}
\caption{Schematic of the four (computational) qubit-state-dependent frequencies for the resonator based on the Kerr model.}
\label{fig:EffHam-KerrModelFreqSchematic}
\end{figure}

The lowest order correction to the effective Hamiltonian reads
\begin{align}
\HO_{I,\text{eff}}^{(1)}(t)=2\chi_{ac}|\eta_c(t)|^2\hat{n}_a+2\chi_{bc}|\eta_c(t)|^2\hat{n}_b \;,
\label{eqn:EffHam-H_I,eff^(1) Sol}	
\end{align}
consisting of a dynamic frequency shift equal to $2\chi$ per resonator photon number for each qubit. Up to the second order, we find
\begin{subequations}
\begin{align}
\begin{split}
\HO_{I,\text{eff}}^{(2)}(t)=&-8\chi_{ac}\chi_{bc}\hat{\mathcal{A}}_{\eta}(t)\hat{n}_a\hat{n}_b\\
&-4\chi_{ac}^2\hat{\mathcal{A}}_{\eta}(t)\hat{n}_a^2\\ 
&-4\chi_{bc}^2\hat{\mathcal{A}}_{\eta}(t)\hat{n}_b^2 \;,\\ 
\end{split}
\label{eqn:EffHam-H_I,eff^(2) Sol}
\end{align}
which contains an effective number-number interaction as well as an anharmonic shift of the qubit eigenfrequencies. The drive dependence of the effective interaction can be compactly written in terms of the response function
\begin{align}
&\hat{\mathcal{A}}_{\eta}(t) \equiv \Im\left\{\int^{t}dt'\eta_c(t)\eta_c^*(t')e^{i\hat{\Delta}_{cd}(t-t')}\right\} \;, 
\label{eqn:EffHam-Def of A_eta(na,nb)}\\
&\hat{\Delta}_{cd} \equiv \Delta_{cd}+2\chi_{ac}\hat{n}_a+2\chi_{bc}\hat{n}_b \;,
\label{eqn:EffHam-Def of hat(Delta)_cd}
\end{align}
\end{subequations}
where the operator-valued detuning $\hat{\Delta}_{cd}$ encodes the qubit-state-dependent phase for the resonator (see Fig.~\ref{fig:EffHam-KerrModelFreqSchematic}). In the adiabatic limit, for which the spectral content of the drive has negligible overlap with the resonator-drive detuning $\Delta_{cd}$, we can approximate $\hat{\mathcal{A}}_{\eta}(t)$ as
\begin{align}
\begin{split}
\hat{\mathcal{A}}_{\eta} (t)=\frac{|\eta_c(t)|^2}{\hat{\Delta}_{cd}}+\frac{\Im\left\{\eta_c(t)\dot{\eta}_c^*(t)\right\}}{\hat{\Delta}_{cd}^2}+O\left(\frac{\eta_c(t)\ddot{\eta}_c^*(t)}{\hat{\Delta}_{cd}^3}\right), 
\end{split}
\label{eqn:EffHam-Adiabatic A_eta}
\end{align}
where the first and the second terms are known as the \textit{dynamic} and \textit{geometric} \cite{Pechal_geometric_2012, Bohm_Geometric_2013, Cross_Optimized_2015} contributions. The geometric correction becomes pertinent for relatively fast pulses with spectral widths comparable to the resonator-drive detuning. Because such pulses lead to significant resonator leakage (see Appendix~E), we consider only the leading dynamic contribution which is proportional to the photon number $|\eta_c(t)|^2$. 

%%%%%%%%%%%%%%%%%% Figure: RIP Frequency Schematic%%%%%%%%%%%%%%%%%%%%%%%
%%%%%%%%%%%%%%%%%%%%%%%%%%%%%%%%%%%%%%%%%%%%%%%%%%%%%%%%%%%%%%%%%%%%%%%%%%

Using Eqs.~(\ref{eqn:EffHam-Def of w_iz})--(\ref{eqn:EffHam-Def of w_zz}), we read off the gate parameters by projecting the effective Hamiltonian in Eqs.~(\ref{eqn:EffHam-H_I,eff^(1) Sol}) and~(\ref{eqn:EffHam-H_I,eff^(2) Sol}) onto the computational subspace. The lowest order Hamiltonian contains dynamic frequency shifts for the qubits as 
\begin{subequations}
\begin{align}
&\omega_{iz}^{(1)}(t) = -2\chi_{bc}|\eta_c(t)|^2 \;,
\label{eqn:EffHam-w_iz^(1)}\\
&\omega_{zi}^{(1)}(t) = -2\chi_{ac}|\eta_c(t)|^2  \;.
\label{eqn:EffHam-w_zi^(1)}
\end{align}
\end{subequations}
The second order dynamic frequency shifts read  
\begin{subequations}
\begin{align}
\begin{split}
\omega_{iz}^{(2)}(t)=\frac{1}{2}\left[\frac{2\chi_{bc}(\Delta_{cd}+4\chi_{bc})}{\Delta_{cd}+2\chi_{bc}}-\frac{\Delta_{cd}^2}{\Delta_{cd}+2\chi_{ac}} \right.\\
\left.+\frac{\Delta_{cd}^2}{\Delta_{cd}+2(\chi_{bc}+\chi_{ac})}\right]|\eta_c(t)|^2 \;,
\end{split}
\label{eqn:EffHam-w_iz^(2)}
\end{align}
\begin{align}
\begin{split}
\omega_{zi}^{(2)}(t)=\frac{1}{2}\left[\frac{2\chi_{ac}(\Delta_{cd}+4\chi_{ac})}{\Delta_{cd}+2\chi_{ac}}-\frac{\Delta_{cd}^2}{\Delta_{cd}+2\chi_{bc}} \right.\\
\left.+\frac{\Delta_{cd}^2}{\Delta_{cd}+2(\chi_{ac}+\chi_{bc})}\right]|\eta_c(t)|^2 \;.
\end{split}
\label{eqn:EffHam-w_zi^(2)}
\end{align}

%%%%%%%% Fig: Comparison between theories %%%%%%%%%%%%%%%%%%%%%%%%%%%%%%%%%%%%%%%
\begin{figure}[t!]
\centering
\includegraphics[scale=0.220]{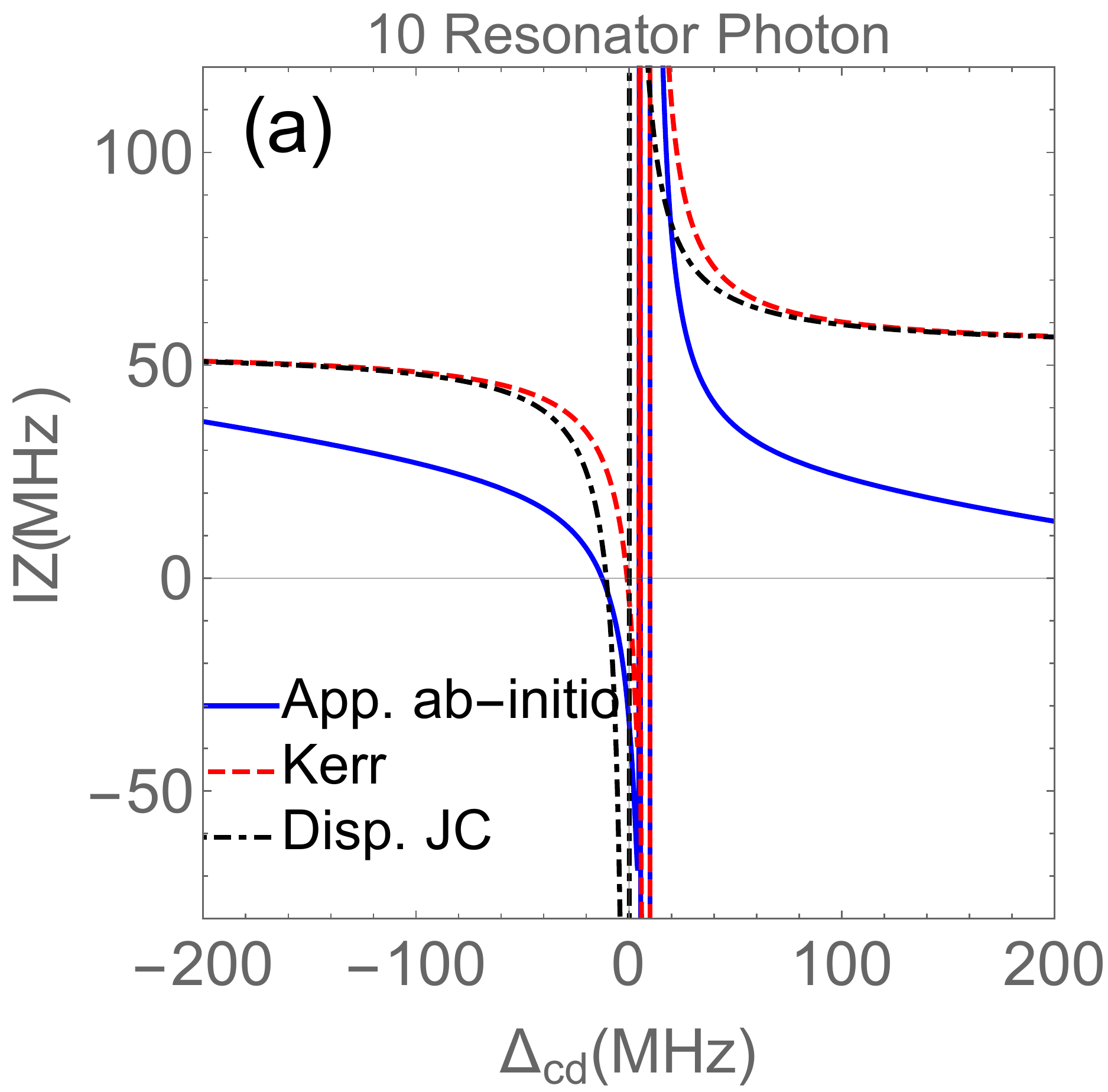}
\includegraphics[scale=0.220]{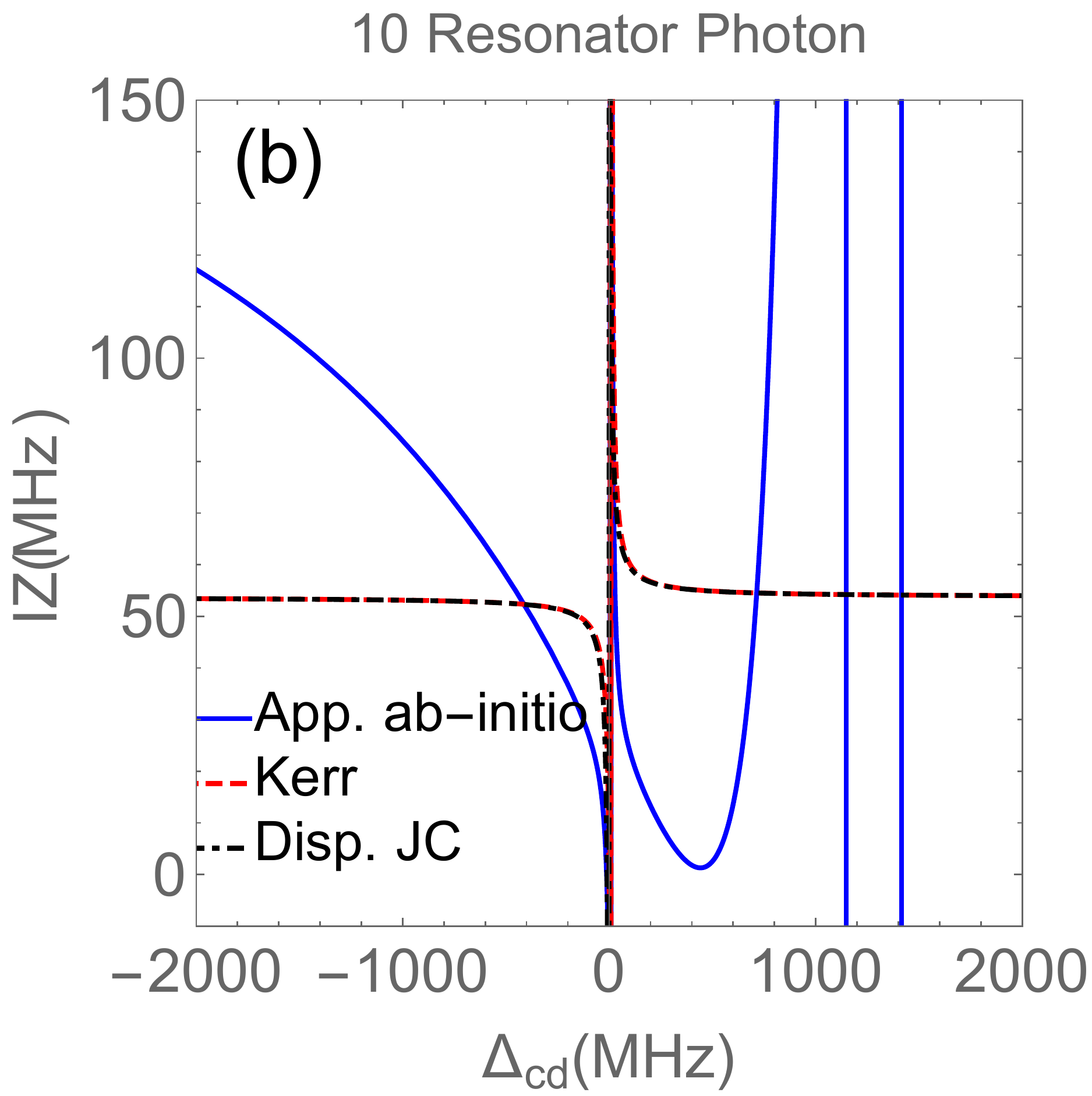}\\
\includegraphics[scale=0.220]{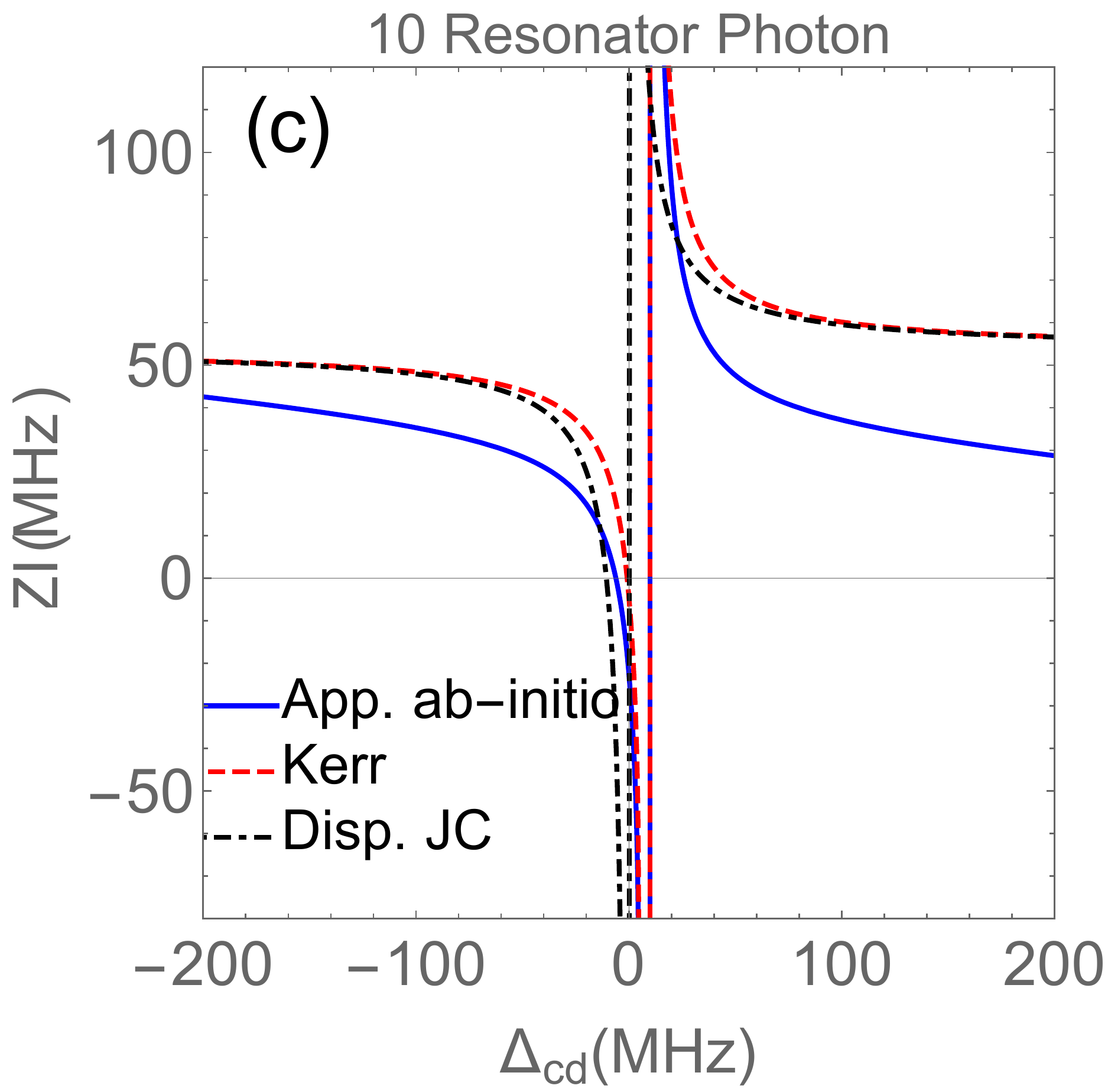}
\includegraphics[scale=0.220]{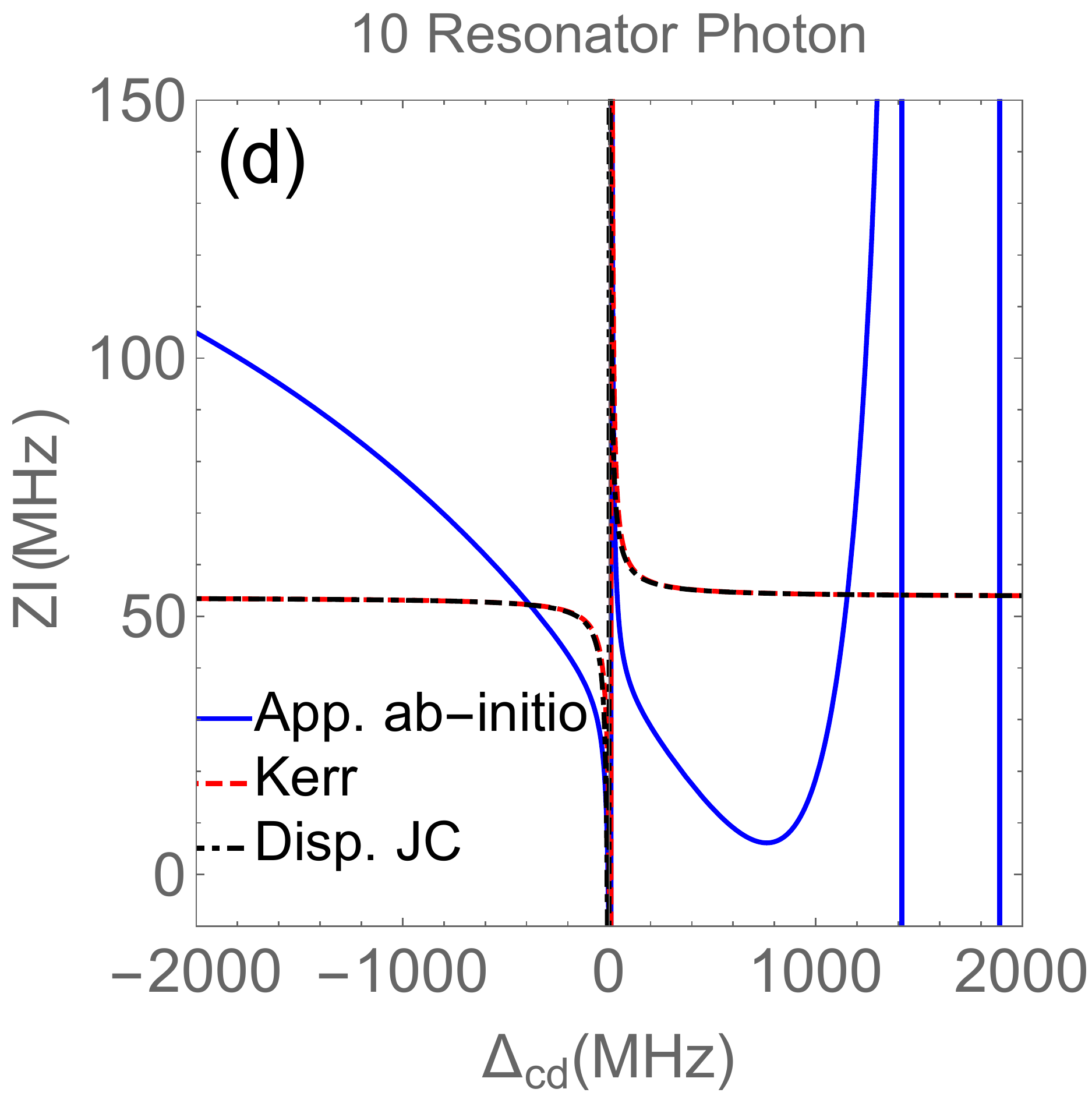}\\
\includegraphics[scale=0.225]{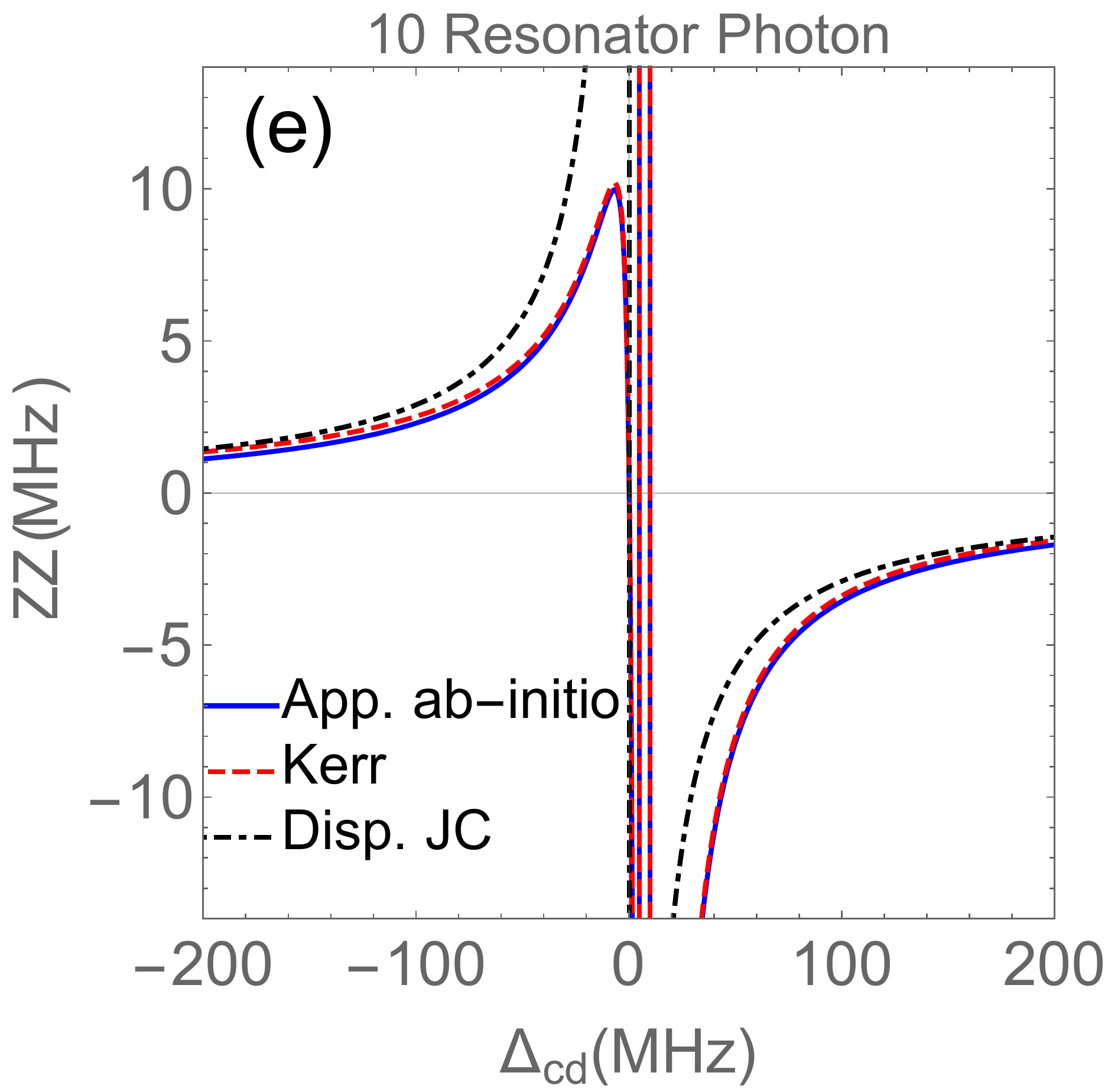}
\includegraphics[scale=0.220]{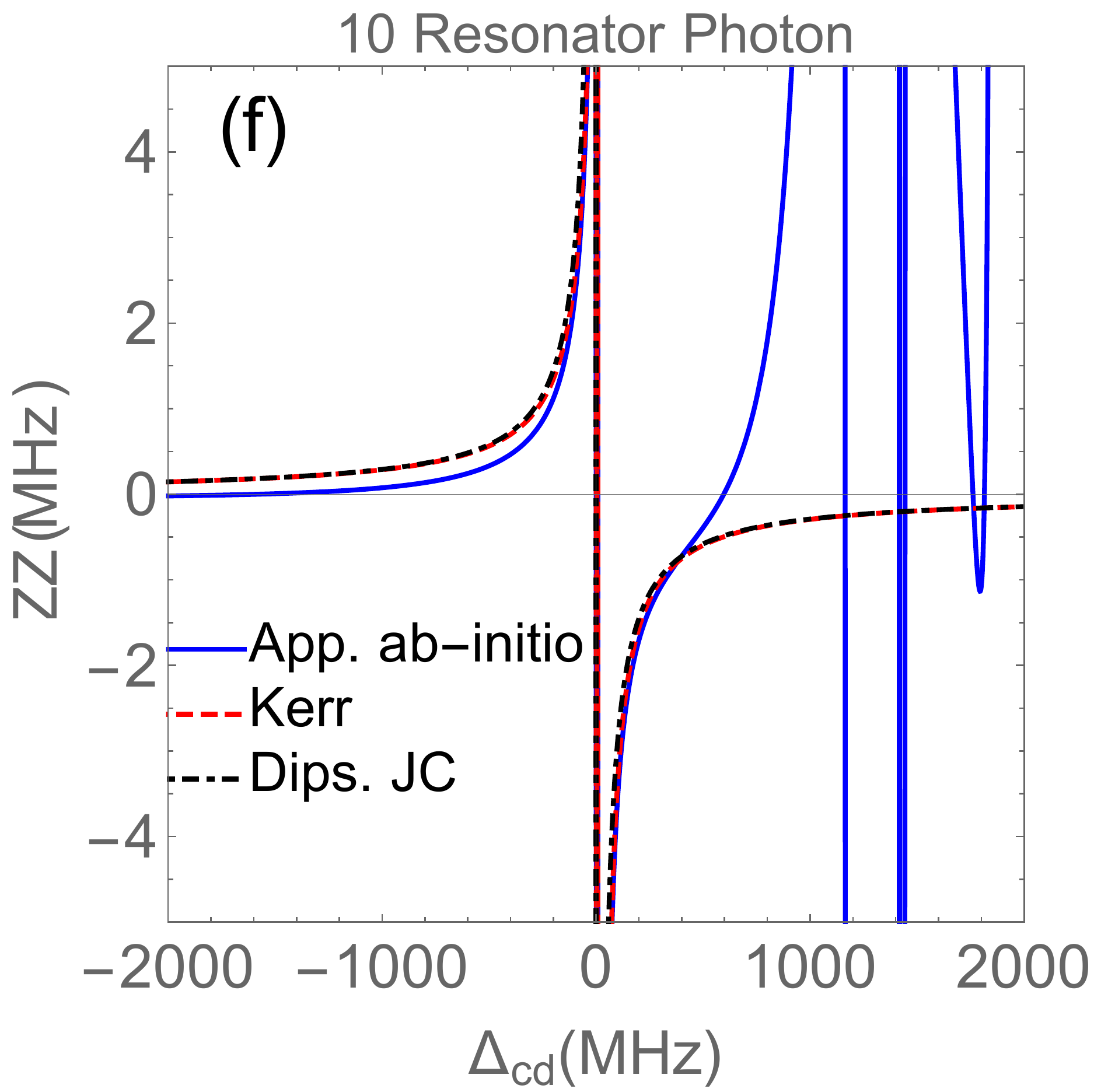}
\caption{Comparison between the phenomenological and the approximate ab-initio models under adiabatic approximation as a function of $\Delta_{cd}$ and maximum photon number $|\eta_c(t)|^2\rightarrow |\eta_{c,\text{ss}}|^2\approx(\Omega_c/2\Delta_{cd})^2=10$. For the ab-initio model, system parameters are $E_{Ja}/2\pi=14250$, $E_{Ca}/2\pi=255$ ($\bar{\omega}_a/2\pi \approx 5391.660$), $E_{Jb}/2\pi=17000$, $E_{Cb}/2\pi=275$ ($\bar{\omega}_b/2\pi \approx 6115.550$), $\bar{\omega}_c/2\pi =7000$, with couplings $g_a/2\pi=150$ and $g_b/2\pi=85$, which up to the sextic expansion yields $2\chi_{ac}/2\pi \approx -4.863$ and $2\chi_{bc}/2\pi\approx -4.910$ (all in MHz). Gate charge is set to zero for simplicity. Phenomenological models adopt the same dispersive shifts and normal mode frequencies. The left column shows small--medium resonator-drive detuning where the RIP gate is typically implemented. The right column shows the same traces plotted over a wider detuning range to demonstrate that the ab-initio theory captures the renormalization of the gate parameters due to additional interaction forms involving qubit resonances (see Appendix~\ref{App:EffRIPHam}).}
\label{fig:EffHam-GateParamsComparison}
\end{figure}
%%%%%%%%%%%%%%%%%%%%%%%%%%%%%%%%%%%%%%%%%%%%%%%%%%%%%%%%%%%%%%%%%%%%%%%%%%%%%%%%%

Furthermore, there is a \textit{dynamic} $ZZ$ term corresponding to the \textit{intended} RIP interaction,
\begin{align}
\omega_{zz}^{(2)}(t) =\frac{-4\chi_{ac}\chi_{bc}(\Delta_{cd}+\chi_{ac}+\chi_{bc})\Delta_{cd}|\eta_c(t)|^2}{(\Delta_{cd}+2\chi_{ac})(\Delta_{cd}+2\chi_{bc})[\Delta_{cd}+2(\chi_{ac}+\chi_{bc})]},
\label{eqn:EffHam-w_zz^(2)}
\end{align}
\end{subequations}
which is proportional to the resonator photon number and the dispersive shift for each qubit. There are four distinct poles at $\Delta_{cd}\approx 0$ [hidden in $\eta_c(t)$ in the adiabatic limit], $\Delta_{cd}=-2\chi_{ac}$, $\Delta_{cd}=-2\chi_{bc}$ and $\Delta_{cd}=-2(\chi_{ac}+2\chi_{bc})$ corresponding to resonance between the drive frequency and the resonator frequency for the computational states $\ket{0_a0_b}$, $\ket{1_a 0_b}$, $\ket{0_a 1_b}$ and $\ket{1_a 1_b}$, respectively (see Fig.~\ref{fig:EffHam-KerrModelFreqSchematic}). In the limit where $\chi_{ac}=\chi_{bc}=\chi$, we recover the expression quoted in Refs.~\cite{Cross_Optimized_2015, Paik_Experimental_2016} as $-4\chi^2 \Delta_{cd}|\eta_c(t)|^2/[(\Delta_{cd}+2\chi)(\Delta_{cd}+4\chi)]$. Moreover, if the detuning is much larger than the dispersive shifts, we reach a simpler expression as $-4\chi^2 |\eta_c(t)|^2/\Delta_{cd}$ in agreement with the dispersive JC model \cite{Puri_High-Fidelity_2016}. 

Note that the RIP interaction sits on top of a static $ZZ$ rate that is captured only by the \textit{multilevel} models (Kerr and ab-initio). In terms of an effective qubit-qubit exchange interaction
\begin{subequations}
\begin{align}
J \approx \left[\frac{(\bar{\Sigma}_{ab}-2\bar{\omega}_c)}{2\bar{\Delta}_{ac}\bar{\Delta}_{bc}}-\frac{(\bar{\Sigma}_{ab}+2\bar{\omega}_c)}{2\bar{\Sigma}_{ac}\bar{\Sigma}_{bc}}\right]g_ag_b\;,
\label{eqn:EffHam-Def of J}
\end{align}
with $\bar{\Sigma}_{jk}\equiv \bar{\omega}_j+\bar{\omega}_k$ and $\bar{\Delta}_{jk}\equiv \bar{\omega}_j-\bar{\omega}_k$, the multilevel Kerr model \cite{Cross_Optimized_2015, Magesan_Effective_2020} predicts
\begin{align}
\omega_{zz}^{(0)} \approx \frac{J^2}{\Delta_{ab}-\alpha_b}-\frac{J^2}{\Delta_{ab}+\alpha_a}\;.
\label{eqn:EffHam-w_zz^(0)}
\end{align}
\end{subequations}
Static $ZZ$ interaction reduces the on/off contrast of the $ZZ$ gate and can be detrimental to most gate operations. However, there are various methods for its suppression: (i) large qubit-qubit detuning compared to qubit anharmonicity [see Eq.~(\ref{eqn:EffHam-w_zz^(0)})], (ii) fast tunable couplers \cite{Chen_Qubit_2014}, (iii) destructive interference between multiple couplers \cite{Mundada_Suppression_2019, Kandala_Demonstration_2020}, (iii) combining qubits with opposite anharmonicity \cite{Ku_Suppression_2020}, and (iv) dynamics Stark tones (siZZle) \cite{Wei_Quantum_2021, Mitchell_Hardware_2021}. In particular, method (iii) has shown significant improvement in reducing the static $ZZ$ down to 0.1 KHz for a test RIP device consisting of six qubits \cite{Kumph_Novel_APS2021}. 

%%%%%%%%%%%%%%%%%%%%%%%%%%%%%%%%%%%%%%%%%%%%%%%%%%%%%%%%%%%%%%%%%%%%%%%%%%%%%%%%  

\subsection{Comparison and discussion}
\label{Subsec:Compare}

We next study the RIP gate parameters in more detail and make a comparison between the various starting models introduced in Sec.~\ref{Sec:Model}. Figure~\ref{fig:EffHam-GateParamsComparison} presents the effective RIP gate parameters as a function of resonator-drive detuning for fixed photon number.

In particular, in terms of agreement for the \textit{dynamic} $ZZ$ rate, we recognize three regions of operation depending on the relation between the resonator-drive detuning $\Delta_{cd}$, qubit-drive detunings $\Delta_{ad},\Delta_{bd}$ and the dispersive shifts $2\chi_{ac}$ and $2\chi_{bc}$:
\begin{itemize}
\item [(i)]  Small detuning ($|\Delta_{cd}|\sim~2|\chi_{ac}|, \ 2|\chi_{bc}|$) ---  The ab-initio and the Kerr \cite{Cross_Optimized_2015} models agree well and both predict a local maximum for the $ZZ$ rate for $\Delta_{cd}<0$ [Fig.~\ref{fig:EffHam-GateParamsComparison}(e)]. The local maximum is a result of having multiple relatively close poles at $\Delta_{cd}=-2\chi_{ac},-2\chi_{bc}$ and $\Delta_{cd}=-2\chi_{ac}-2\chi_{bc}$.

\item [(ii)] Medium detuning ($2|\chi_{ac}|, \ 2|\chi_{bc}|\ll |\Delta_{cd}| \ll |\Delta_{ad}|,|\Delta_{bd}|$) --- RIP drive frequency is closer to the resonator than to the qubit frequencies. Moreover, resonator-drive detuning is sufficiently larger than the dispersive shifts such that the qubit-state dependence of the poles is less noticeable. Consequently, all models agree. 

\item [(iii)]  Large detuning ($ 2|\chi_{ac}|, \ 2|\chi_{bc}|\ll |\Delta_{cd}|\sim |\Delta_{ad}|,|\Delta_{bd}|$) --- Resonator-drive and qubit-drive detunings are comparable. Therefore, there are additional processes, beyond dispersive JC and Kerr interactions, that renormalize the gate parameters (see Appendices~\ref{App:MLM} and~\ref{App:EffRIPHam}). This region is not necessarily relevant to RIP gate implementation but is a natural extension of our comparison which quantifies the validity of the phenomenological models [Fig.~\ref{fig:EffHam-GateParamsComparison}(f)].\\
\end{itemize}

In terms of the RIP interaction, the ab-initio model agrees well with the former phenomenological models in their \textit{intended} resonator-drive detuning regimes. However, according to Fig.~\ref{fig:EffHam-GateParamsComparison}, we observe a large deviation between the dynamic frequency shifts predicted by the ab-initio theory compared to the phenomenological models in all regions of operation. We find the source of this deviation to be contributions of the form $\HO_{\text{qd}}(t) \equiv \sum_{j=a,b} [\lambda_j(t)\hat{j}e^{i\omega_d t}+\text{H.c.}]$ that act as \textit{direct} drive terms on the qubits. Here, $\lambda_j(t)$ denotes the \textit{effective} drive amplitude on qubit $j=a,b$ and contains contributions from \textit{linear} charge hybridization and nonlinear mode mixing through flux hybridization (see Appendices~\ref{App:MLM} and \ref{App:EffRIPHam}).

In summary, larger dispersive shifts, stronger RIP drive amplitude and smaller resonator-drive detuning all contribute to a stronger $ZZ$ interaction. However, we have to account \textit{also} for the trade-off between a strong dynamic $ZZ$ rate and the corresponding unwanted increase in both resonator and qubit leakage. Our analysis of leakage reveals further restrictions on the choice for qubit frequency, anharmonicity and RIP drive parameters, which we discuss in Sec.~\ref{Sec:Leak} and Appendix~\ref{App:ResRes}.       

%%%%%%%%%%%%%%%%%%%%%%%%%%%%%% Sec: Quantum Leakage %%%%%%%%%%%%%%%%%%%%%%%%%%%%%
\section{Leakage}
\label{Sec:Leak}

%%%%%%%%%%%%%%%%%%%%%%% Fig: Universe of Collisions %%%%%%%%%%%%%%%%%%%%%%%%%%%%%
\begin{figure}[t!]
\centering
\includegraphics[scale=0.35]{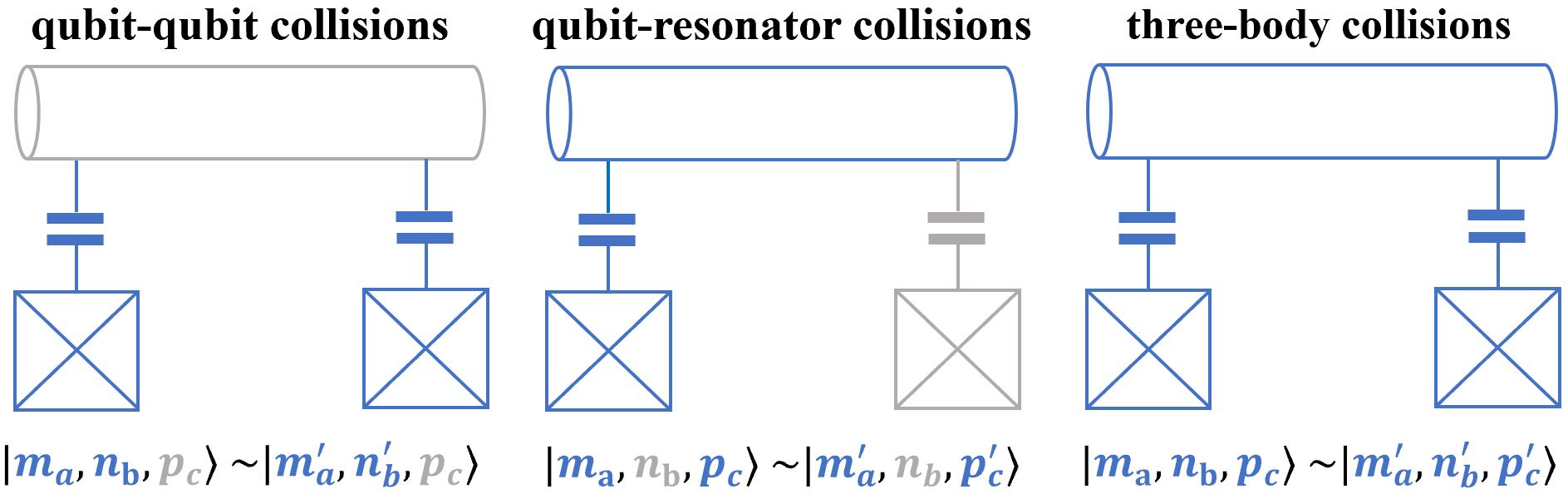}
\caption{Universe of collisions involving two transmon qubits that are connected via a common bus resonator. The possibilities can be summarized as (i) qubit-qubit collisions \cite{Magesan_Effective_2020, Malekakhlagh_First-Principles_2020, Hertzberg_Effects_2020}, (ii) qubit-resonator collisions \cite{Sank_Measurement-Induced_2016} and (iii) three-body collisions. Symbol ``$\sim$'' denotes collision (degeneracy) between system states.}
\label{fig:Leak-CollisionUniverse}
\end{figure}
%%%%%%%%%%%%%%%%%%%%%%%%%%%%%%%%%%%%%%%%%%%%%%%%%%%%%%%%%%%%%%%%%%%%%%%%%%%%%%%%

%%%%%%%%%%%%%%%%%%%%%%% Fig: 2D Sweep Leakage %%%%%%%%%%%%%%%%%%%%%%%%%%%%%%%%%%
\begin{figure}[t!]
\centering
\includegraphics[scale=0.144]{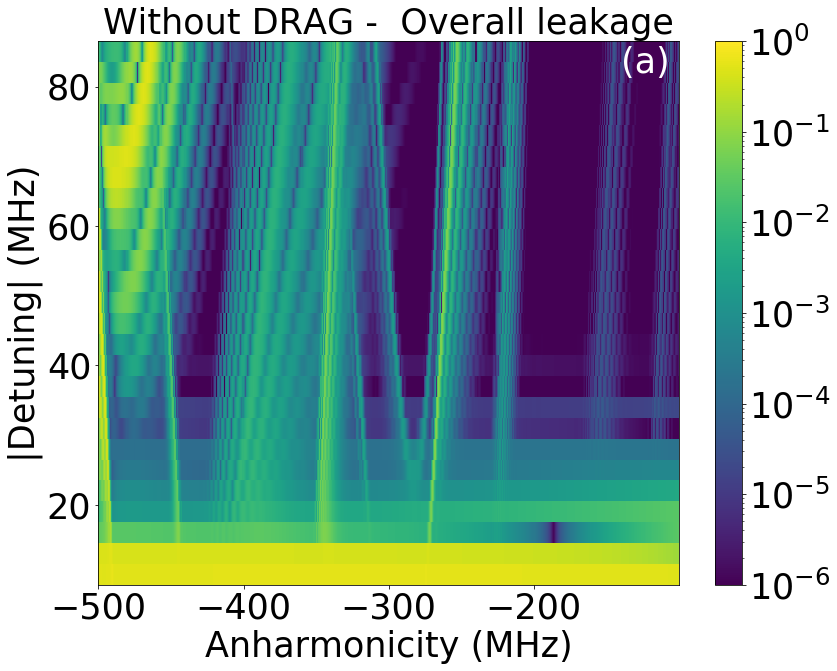}
\includegraphics[scale=0.144]{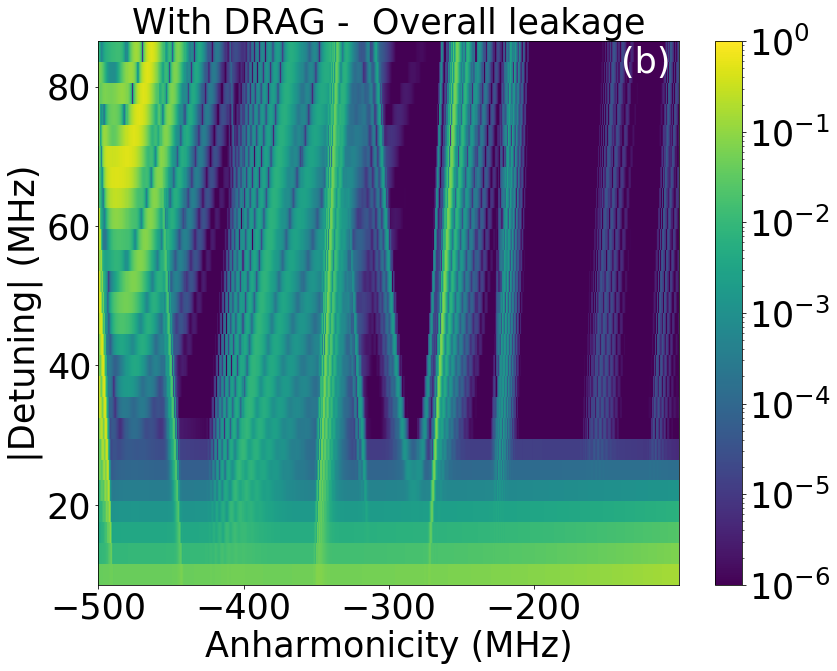}\\
\includegraphics[scale=0.144]{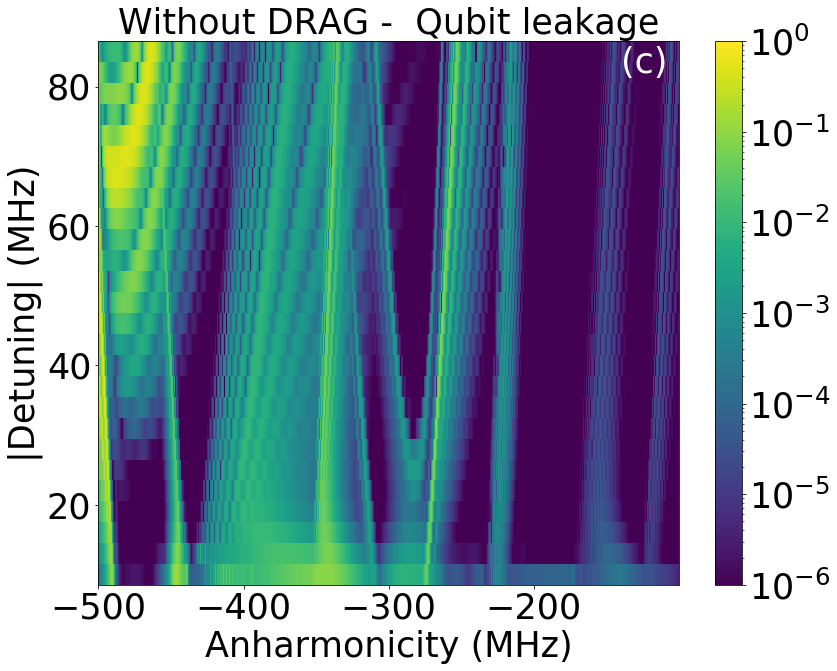}
\includegraphics[scale=0.144]{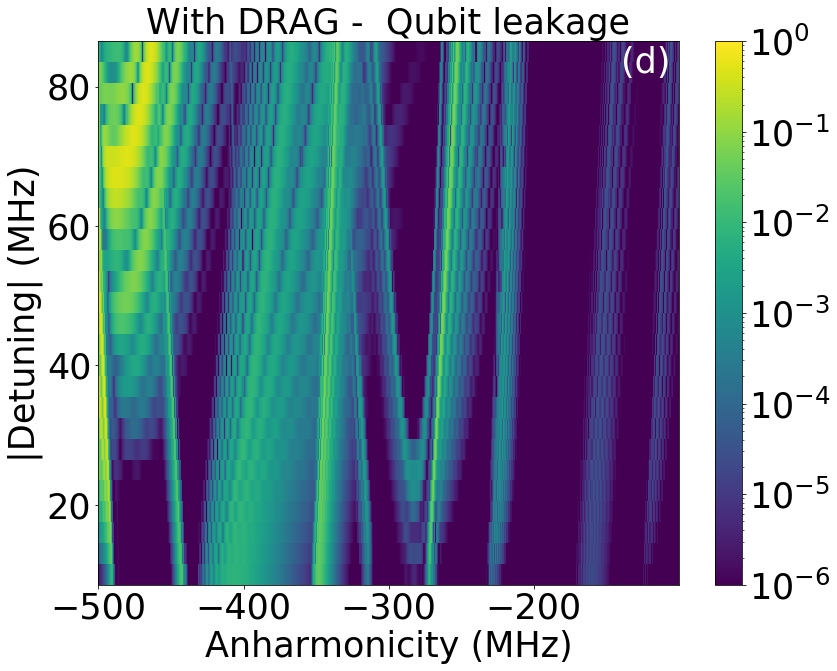}\\
\includegraphics[scale=0.144]{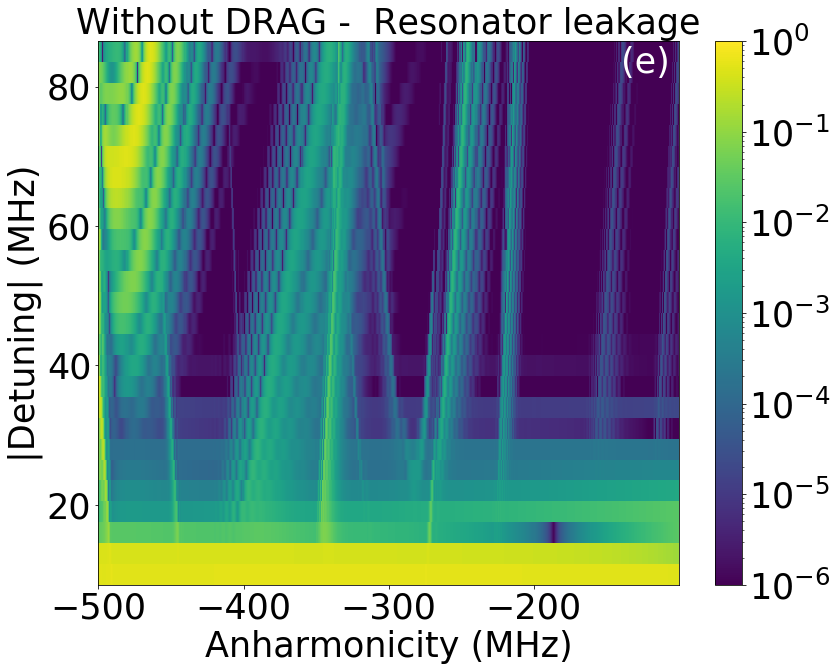}
\includegraphics[scale=0.144]{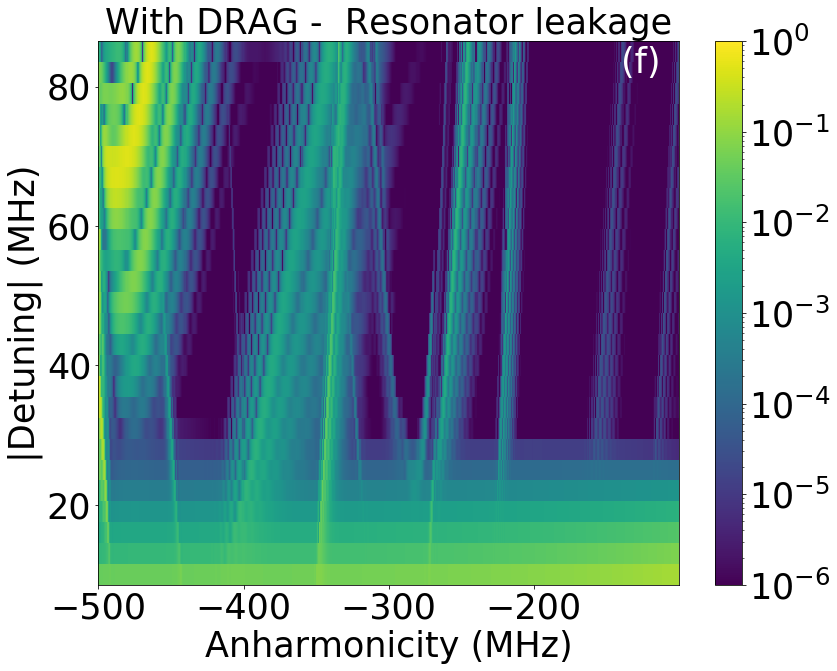}\\
\caption{Qubit-resonator case --- Overall, qubit and resonator leakages for $\omega_a/2\pi \approx 5140$ MHz, $\omega_c/2\pi \approx 6971$ MHz, $2\chi_{ac}/2\pi \approx -5.57$ MHz, fixed maximum photon number $[\Omega_c/(2\Delta_{cd})]^2\approx 16$ and fixed gate charge set to $0.37$ as a function of qubit anharmonicity and resonator-drive detuning. The initial state is set as $\ket{\Psi(0)}=(1/\sqrt{2})(\ket{0_a}+\ket{1_a})\otimes \ket{0_c}$ to allow leakage from both qubit states in a single run. The results here are for $\omega_d>\omega_c$. We confirmed numerically that $\omega_d<\omega_c$ leads to more resonator leakage consistent with additional $2\chi$-shifted poles in Eq.~(\ref{eqn:EffHam-Def of hat(Delta)_cd}) and Fig.~\ref{fig:EffHam-KerrModelFreqSchematic}. The resonator is driven with a nested cosine pulse [Eq.~\ref{eqn:Leak-Def of NCPulse}] of $\tau=200$ ns. The left (right) column present results without (with) DRAG on the resonator given by Eqs.~(\ref{eqn:Leak-Def of Omcx})--(\ref{eqn:Leak-Def of Omcy}).}
\label{fig:Leak-2DSweepWithDRAG}
\end{figure}
%%%%%%%%%%%%%%%%%%%%%%%%%%%%%%%%%%%%%%%%%%%%%%%%%%%%%%%%%%%%%%%%%%%%%%%%%%%%%%%%%

RIP gate leakage can be separated into a background leakage, also called residual photons, due to the non-adiabaticity of the drive pulse and the resonator, and leakage to specific system states due to frequency collisions. The background leakage is comparably easy to understand and control, as its ratio to the intended photon number depends primarily on the overlap of the pulse's spectrum with the resonator frequency via the collective quantity $\Delta_{cd} \tau$, with $\tau$ being the pulse (rise) time. Therefore, to keep the background leakage constant, while making the pulse twice as fast, the most basic solution is to double $\Delta_{cd}$ and $\Omega_c$. A more involved control scheme for mitigating the background leakage, however, is to filter the pulse spectrum at $\Delta_{cd}$. We show that adding a DRAG pulse to the resonator works as an effective filter (see also Appendix~\ref{App:ResRes}).                

Focusing on the leakage due to frequency collisions, depending on what circuit elements participate in the exchange of excitations, three categories arise (see Fig.~\ref{fig:Leak-CollisionUniverse}): (i) Qubit-qubit collisions \cite{Magesan_Effective_2020, Malekakhlagh_First-Principles_2020, Hertzberg_Effects_2020}, (ii) qubit-resonator collisions \cite{Sank_Measurement-Induced_2016}, and (iii) three-body collisions. Qubit-qubit collisions occur in the zero (fixed)-photon subspace of the resonator when the qubits' detuning is approximately an integer multiple of the qubits' anharmonicity. Having qubit-qubit detuning away from the straddling regime minimizes the leakage due to such inter-qubit collisions. Qubit-resonator collisions occur between high- and low-excitation states of one of the qubits and the resonator while the other qubit is in a fixed state. More involved three-body collisions are also possible where both qubits and the resonator participate.

%%%%%%%%%%%%%%%%%%%%%%% Fig: Qubit and Resonator States %%%%%%%%%%%%%%%%%%%%%%%%%
\begin{figure}[t!]
\centering
\includegraphics[scale=0.35]{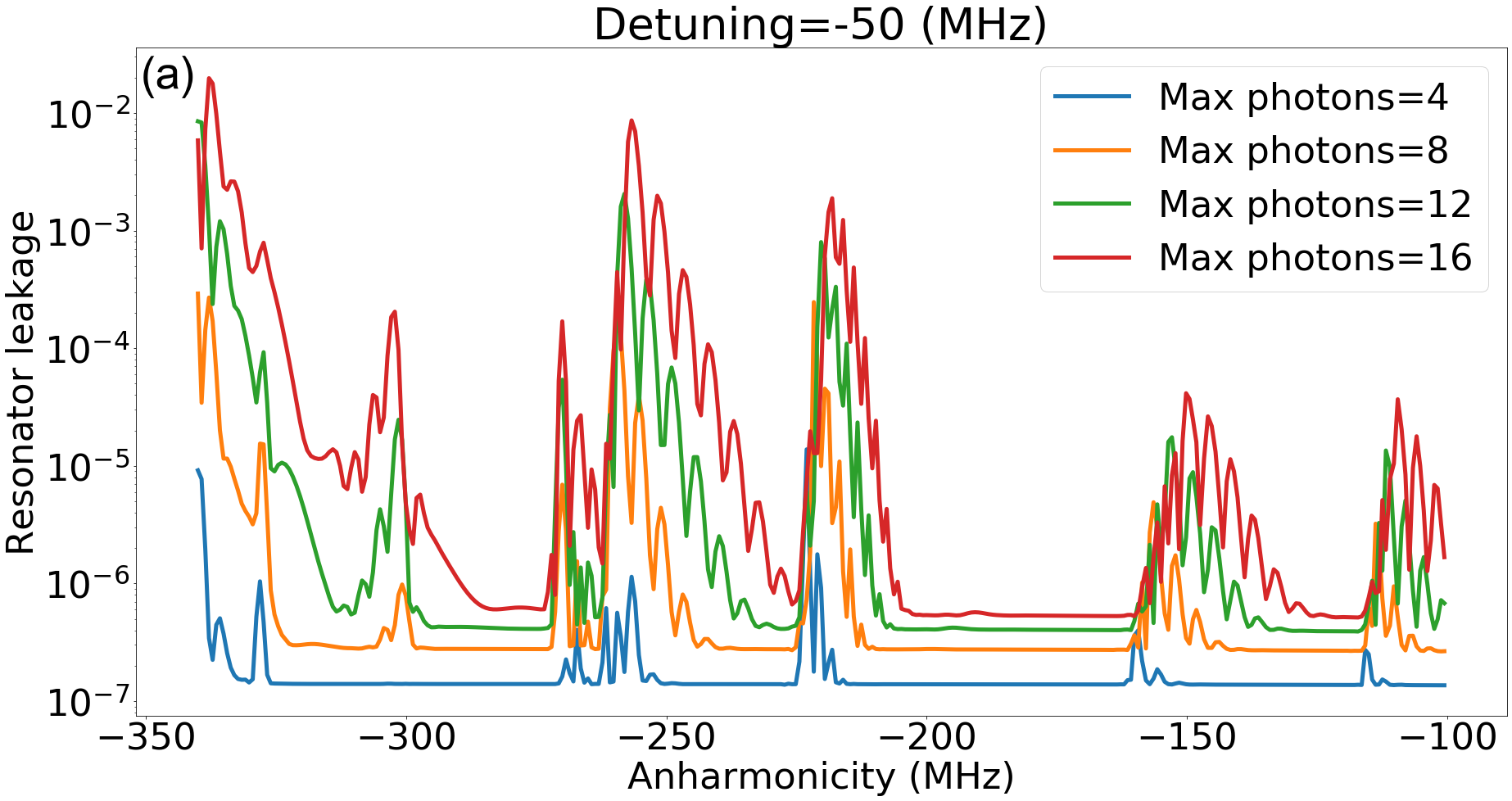}\\
\includegraphics[scale=0.35]{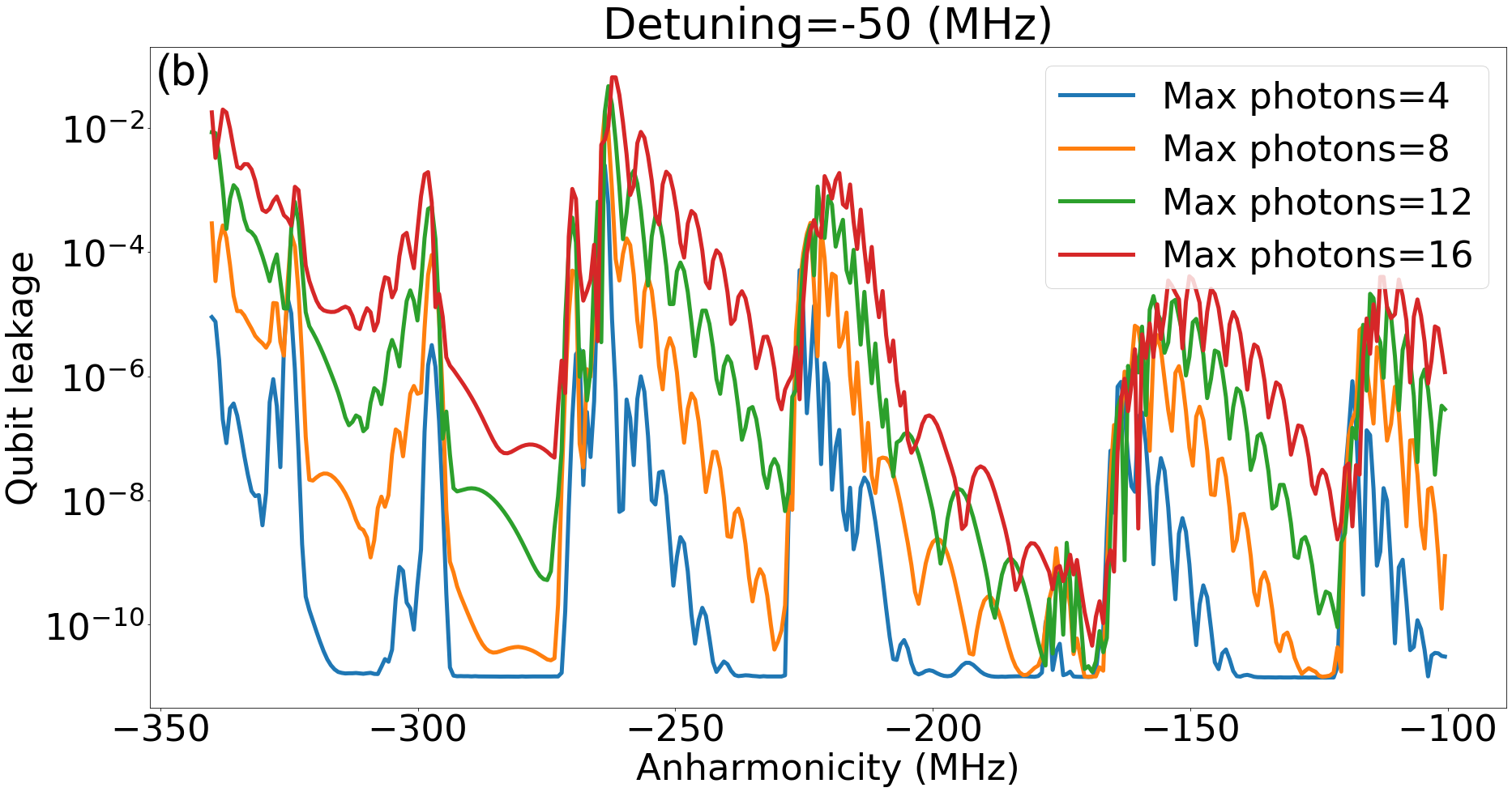}
\caption{Qubit-resonator case --- (a) Resonator leakage, (b) qubit leakage for the same parameters as in Fig.~\ref{fig:Leak-2DSweepWithDRAG}, constant detuning $\Delta_{cd}/2\pi=-50$ MHz and varying photon number.}
\label{fig:Leak-QuResLeakage}
\end{figure}
%%%%%%%%%%%%%%%%%%%%%%%%%%%%%%%%%%%%%%%%%%%%%%%%%%%%%%%%%%%%%%%%%%%%%%%%%%%%%%%%%
%%%%%%%%%%%%%%%%%%%%%%% Fig: Most Leaked States %%%%%%%%%%%%%%%%%%%%%%%%%%%%%%%%%
\begin{figure}[t!]
\centering
\includegraphics[scale=0.35]{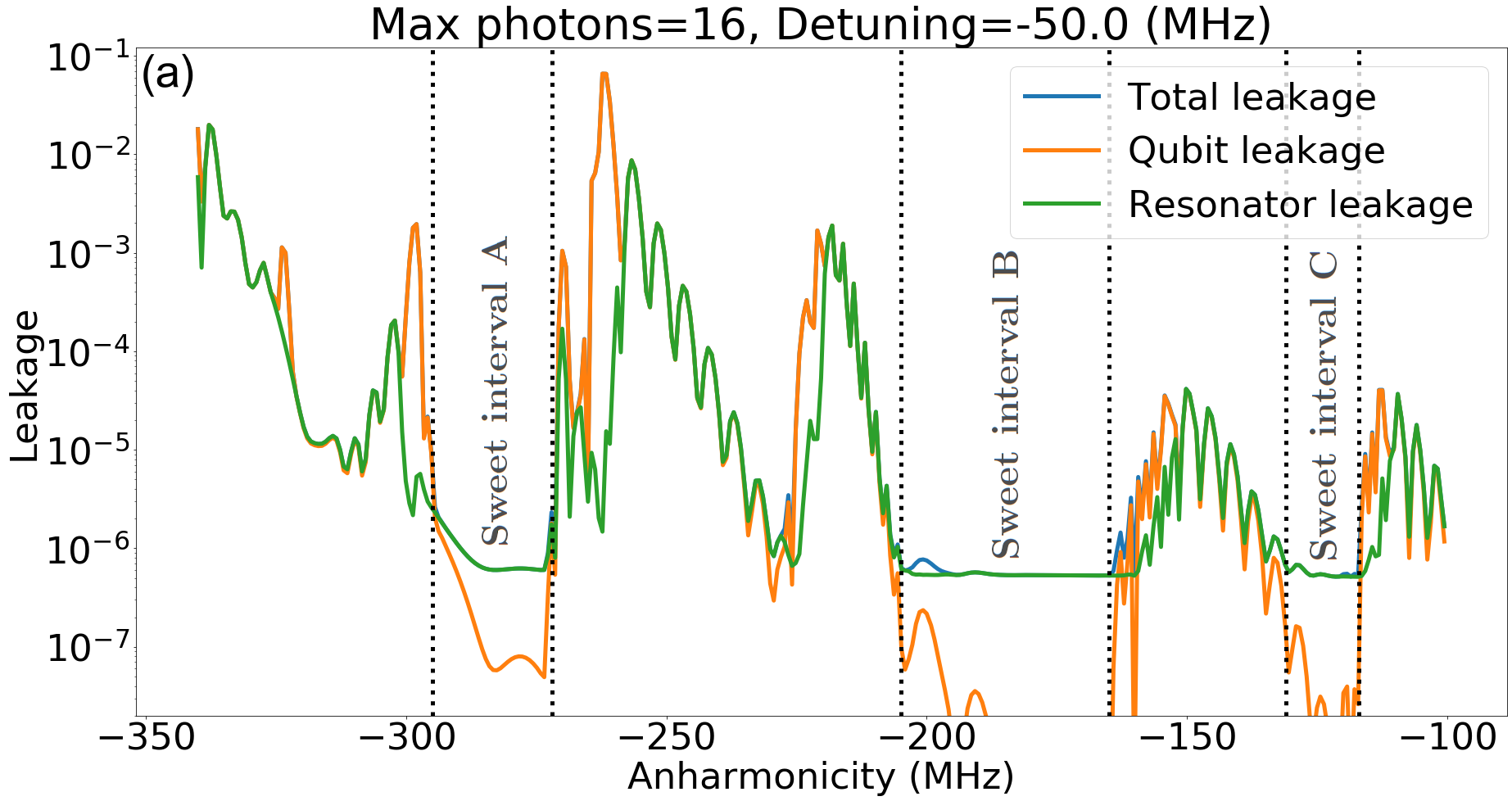}\\
\includegraphics[scale=0.35]{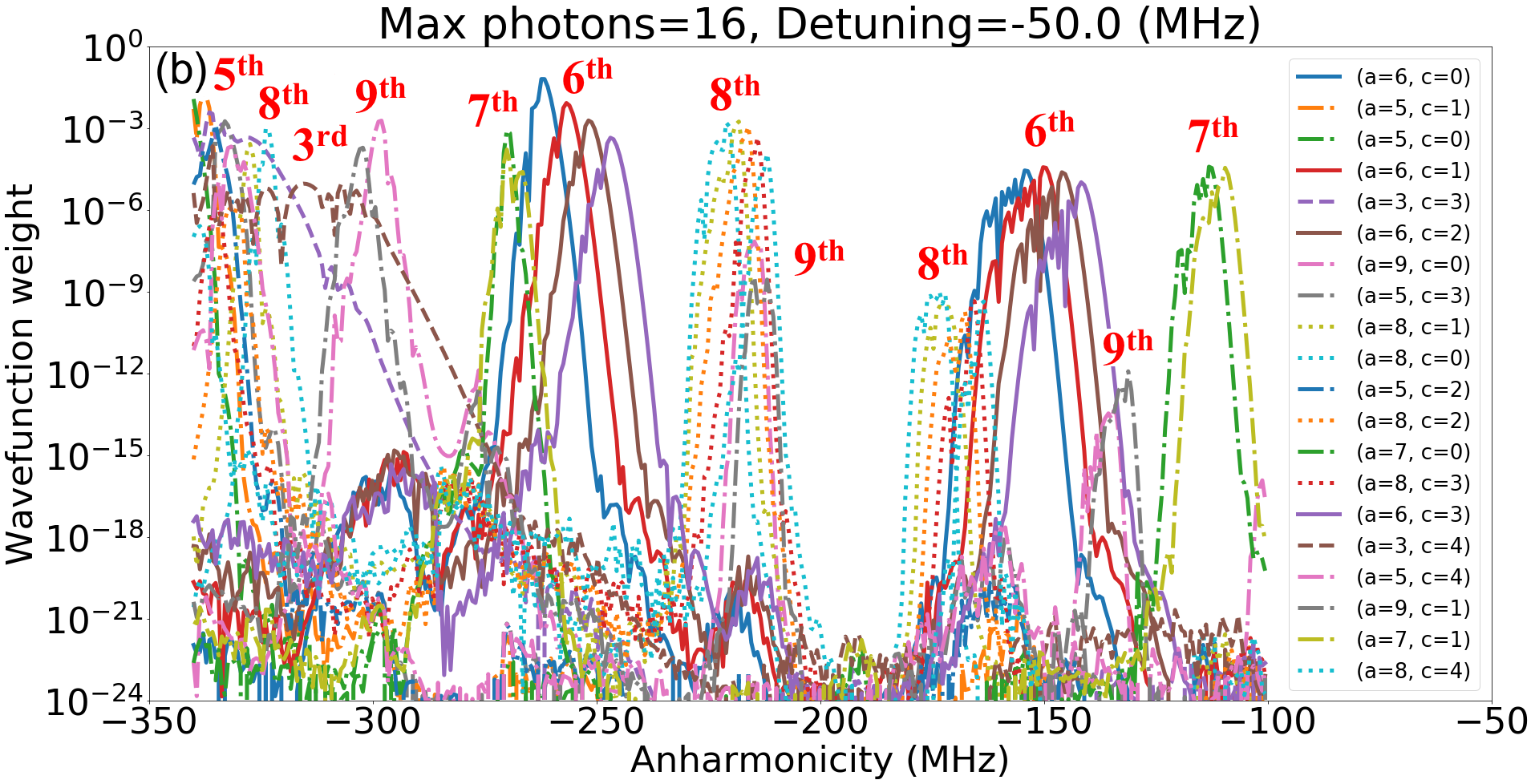}
\caption{Qubit-resonator case --- (a) Resonator, qubit and total leakage, (b) decomposition of leakage into qubit$\otimes$resonator eigenstates for the same parameters as in Fig.~\ref{fig:Leak-QuResLeakage} and 16 maximum photons. The \textit{dominant} leakage clusters in (b) correspond to specific high-excitation qubit states and variant photon number. They can be understood in terms of the frequency collisions in Fig.~\ref{fig:Leak-HighLowStateCollision} and Table~\ref{tab:Leak-ReadoutCollisions}.}
\label{fig:Leak-MostLeakedStates}
\end{figure}
%%%%%%%%%%%%%%%%%%%%%%%%%%%%%%%%%%%%%%%%%%%%%%%%%%%%%%%%%%%%%%%%%%%%%%%%%%%%%%%

\subsection{Qubit-resonator leakage}
\label{SubSec:READLikeLeak}
We first consider a single transmon qubit coupled to a driven resonator which serves as a simpler version of the RIP gate that exhibits the essential leakage mechanisms for the qubit-resonator category. In particular, we observe leakage clusters to certain high-excitation qubit states and associate them with collisions between high-excitation and computational states of the qubit with different photon number. This analysis has implications beyond the RIP gate and specifically for optimal design of qubit readout. Similar drive-induced collisions have been studied in Ref.~\cite{Sank_Measurement-Induced_2016}. Here, by sweeping qubit parameters, we categorize a series of collisions that extends Ref.~\cite{Sank_Measurement-Induced_2016} (see Table~\ref{tab:Leak-ReadoutCollisions}). 

For numerical simulation of the qubit-resonator case, we adopt the exact ab-initio Hamiltonian~(\ref{eqn:Model-Starting Hs})--(\ref{eqn:Model-Starting Hd}) with just one transmon qubit. To correctly model leakage to high-excitation states, we keep 10 transmon and 48 resonator eigenstates. The technical details of the numerical simulations are described in Appendix~\ref{App:NumMet}. For experimentally relevant comparisons, we numerically search for the ab-initio system parameters in Eqs.~(\ref{eqn:Model-Starting Hs})--(\ref{eqn:Model-Starting Hd}) that keep certain relevant quantities fixed, while sweeping others. Given that the ab-initio and the pulse parameter space is large [8D (12D) for the single (two) qubit cases], our results are presented in terms of continuous sweeps only in select quantities such as qubit anharmonicity and resonator-drive detuning, with approximately constant cuts in other quantities such as the qubit frequency, resonator frequency, resonator photon number and dispersive shift. 

We drive the resonator with the pulse shape 
\begin{align}
P_{\text{nc}}(t;\tau) \equiv \frac{1}{2}\left\{\cos\left[\pi\cos\left(\pi\frac{t}{\tau}\right)\right]+1\right\} \;,
\label{eqn:Leak-Def of NCPulse}
\end{align}
which we call a nested cosine of duration $\tau$. Compared to a Gaussian pulse, the nested cosine pulse leads to improved resonator leakage due to smoother ramps \cite{Cross_Optimized_2015} (see Appendix~\ref{App:ResRes}). We then compute the overall leakage as the residual occupation of system states at the \textit{end} of the pulse. More specifically, resonator leakage is computed as the probability of finding the system with a non-zero photon number summed over all possible qubit states, and qubit leakage is found as the probability of finding the qubit outside the computational subspace summed over all possible resonator states.

%%%%%%%%%%%%%%%%%%%%%%% Fig: HighLowStateCollisions %%%%%%%%%%%%%%%%%%%%%%%%%%%%
\begin{figure}[t!]
\centering
\includegraphics[scale=0.290]{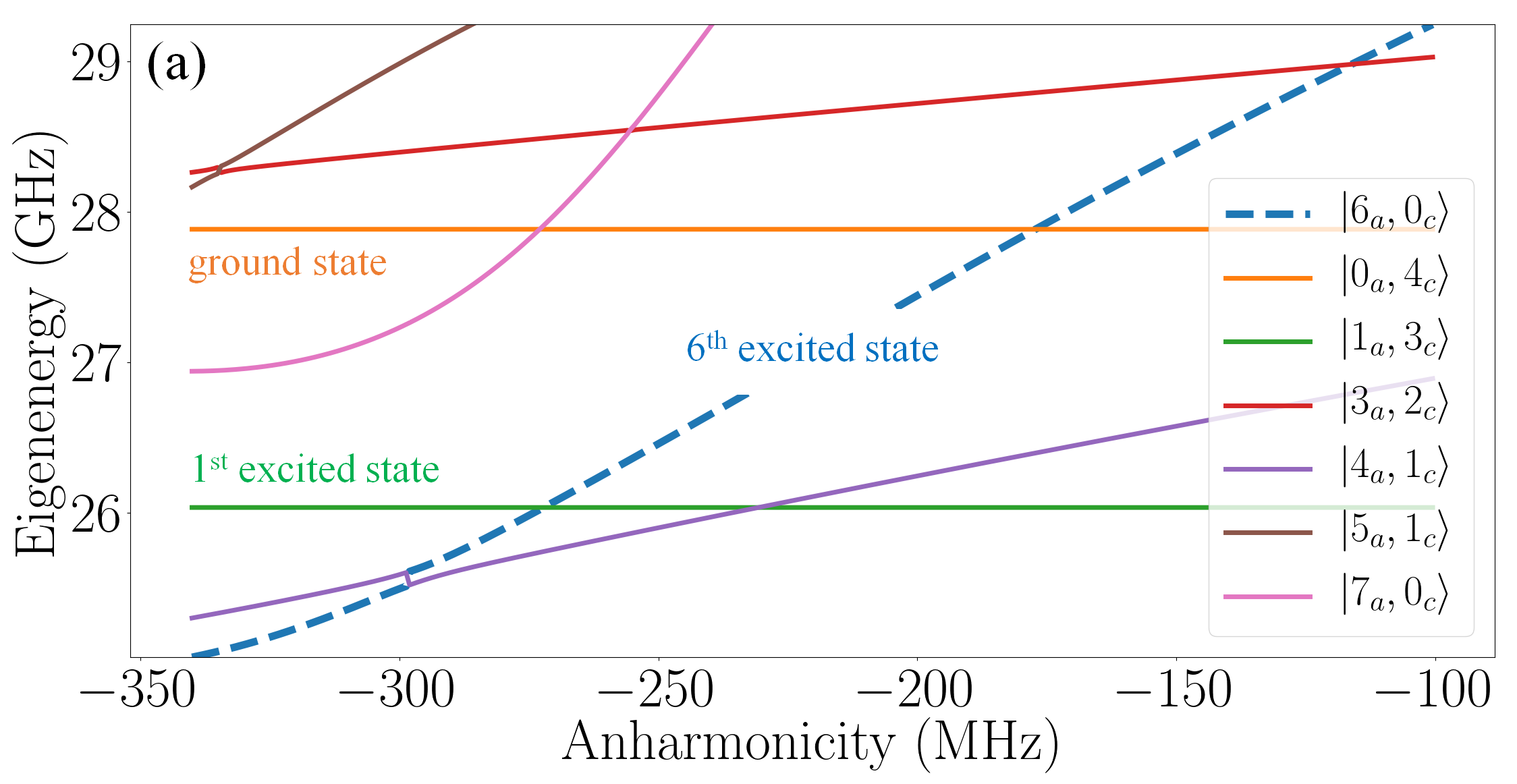}\\
\includegraphics[scale=0.290]{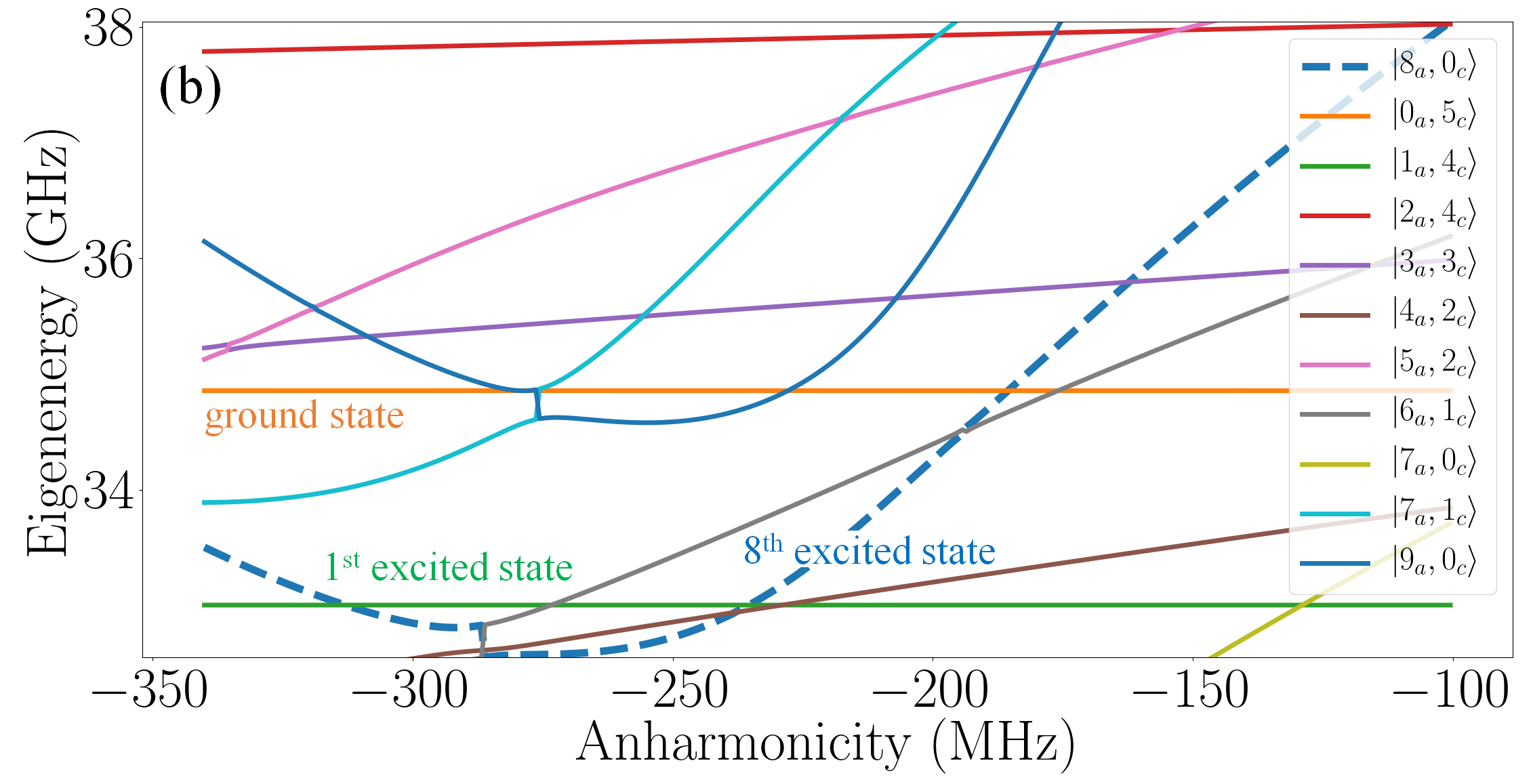}
\caption{Qubit-resonator case --- Examples of collisions between high-excitation and computational states for the same system parameters as in Figs.~\ref{fig:Leak-QuResLeakage}--\ref{fig:Leak-MostLeakedStates} \textit{without drive}. States are labeled by the maximum overlap with the corresponding uncoupled case, hence avoided crossings appear as \textit{fictitious} jumps. (a) Collisions between state $\ket{6_a,0_c}$ and computational states $\ket{0_a,4_c}$ (similar to Ref.~\cite{Sank_Measurement-Induced_2016}) and $\ket{1_a,3_c}$. (b) Collisions between state $\ket{8_a,0_c}$ and computational states $\ket{0_a,5_c}$ and $\ket{1_a,4_c}$. Note the possibility of \textit{multiple} collisions between the \textit{same} states due to e.g. bending of the 8th excited state at large anharmonicity [see Eq.~(\ref{eqn:Leak-WhyStatesBend})]. This is in agreement with observing \textit{two} leakage clusters to the 6th excited state and \textit{three} to the 8th excited state in Fig.~\ref{fig:Leak-MostLeakedStates}(b).}
\label{fig:Leak-HighLowStateCollision}
\end{figure}
%%%%%%%%%%%%%%%%%%%%%%%%%%%%%%%%%%%%%%%%%%%%%%%%%%%%%%%%%%%%%%%%%%%%%%%%%%%%%%%%%

We first analyze the case where the qubit and resonator frequencies are approximately fixed at $\omega_a/2\pi \approx 5140$~MHz and $\omega_c/2\pi \approx 6971$~MHz and the qubit-resonator dispersive coupling at $2\chi_{ac}/2\pi \approx -5.57$~MHz. Results for other qubit frequencies and dispersive shifts are also shown at the end of this section. Figure~\ref{fig:Leak-2DSweepWithDRAG} shows the overall, resonator and qubit leakage for 16 photons as a 2D sweep of detuning and qubit anharmonicity, while comparing between pulses with and without DRAG to suppress drive at the resonator frequency. The DRAG pulse has the generic form 
\begin{subequations}
\begin{align}
&\Omega_{cy}(t)=\Omega_c P_{\text{nc}}(t;\tau) \;, 
\label{eqn:Leak-Def of Omcx}\\
&\Omega_{cx}(t)=(1/ \Delta_D)\Omega_c\dot{P}_{\text{nc}}(t;\tau) \;. 
\label{eqn:Leak-Def of Omcy}
\end{align}
\end{subequations}
Setting the DRAG coefficient to $\Delta_D=\Delta_{cd}$ suppresses the pulse spectrum at the resonator-drive detuning (see Appendix~\ref{App:ResRes}). This is confirmed in Fig.~\ref{fig:Leak-2DSweepWithDRAG} where the background leakage is reduced by at least one order of magnitude at small $\Delta_{cd}$. Besides the resonator leakage at small $\Delta_{cd}$, there are leakage clusters as well as sweet intervals with suppressed leakage as a function of qubit anharmonicity. We observe that qubit-resonator leakage amplitude is generally suppressed at smaller qubit anharmonicity. This is understood as nonlinear interactions connecting the underlying states that increase in powers of $E_C$. For the chosen parameters, in particular, setting the detuning to be larger than 30 MHz and anharmonicity smaller than -200 MHz keeps leakage below the desired threshold of $10^{-5}$. 

Figure~\ref{fig:Leak-QuResLeakage} shows the resonator and the qubit leakage for constant detuning $\Delta_{cd}/2\pi \approx -50$~MHz and varying resonator photon number. Both types of leakage are \textit{universally} increased at stronger drive. A decomposition of the overall leakage in terms of individual system states in Fig.~\ref{fig:Leak-MostLeakedStates} reveals that the dominant clusters can be associated with specific high-excitation qubit states and variant photon number. In particular, we observe considerable leakage to the 5th--9th excited states of the transmon qubit. Furthermore, clusters to a particular high-excitation qubit state can appear multiple times (twice for the 6th and 7th and three times for the 8th and 9th).

%%%%%%%%%%%%%%%%%%%%%%% Table: ReadoutLike Collisions %%%%%%%%%%%%%%%%%%%%%%%%%%%%
\begin{table*}[t!]
\begin{tabular}{|c|c|c|c|}
\hline
States & Condition (Kerr) & Kerr estimate of $\alpha_a/2\pi$ (MHz) & Numerical estimate of $\alpha_a/2\pi$ (MHz)\\
\hline
$\ket{5_a,n_c} \sim \ket{0_a,n_c+3}$ & $5\omega_a +10\alpha_a \approx 3\omega_c-10n_c\chi_{ac}$ & $-478.700+2.785 n_c$ & $-355.213+0.609 n_c$ \\
\hline
$\ket{6_a,n_c} \sim \ket{0_a,n_c+4}$ & $6\omega_a+15\alpha_a \approx 4\omega_c-12n_c\chi_{ac}$ & $-197.067+2.228n_c$ & $-177.230+1.292 n_c$\\
\hline
$\ket{6_a,n_c} \sim \ket{1_a,n_c+3}$ & $5\omega_a+15\alpha_a \approx 3\omega_c-2(5n_c-3)\chi_{ac}$ & $-320.247+1.857 n_c$ & $-272.136-0.981 n_c$\\
\hline
$\ket{7_a,n_c} \sim \ket{0_a,n_c+4}$ & $7\omega_a+21\alpha_a\approx 4\omega_c-14n_c\chi_{ac}$ & $-385.524+1.857 n_c$ & $-272.892-0.423 n_c$\\
\hline
$\ket{7_a,n_c} \sim \ket{1_a,n_c+4}$ & $6\omega_a+21\alpha_a \approx 4\omega_c - 4(3n_c-2)\chi_{ac}$ & $-141.823+1.591 n_c $ & $-129.007+0.868 n_c$\\
\hline
$\ket{8_a,n_c} \sim \ket{0_a,n_c+5}$ & $8\omega_a+28\alpha_a \approx 5\omega_c-16 n_c \chi_{ac}$ & $-223.750+1.591 n_c$ & $-185.549+0.853 n_c$\\
\hline
$\ket{8_a,n_c} \sim \ket{1_a,n_c+4}$ & $7\omega_a +28\alpha_a \approx 4\omega_c -2(7n_c-4)\chi_{ac}$ & $-289.939+1.393n_c$ & \makecell{$ -235.336+1.096 n_c$ \\ $\bullet -313.738 +708 n_c$}\\
\hline
$\ket{9_a,n_c} \sim \ket{0_a,n_c+5}$ & $9\omega_a +36\alpha_a \approx 5\omega_c -18n_c\chi_{ac}$ & $-316.806+1.393n_c$ & \makecell{$-227.839+0.807 n_c$ \\ $\bullet  -280.415 +1.549 n_c$}\\
\hline
$\ket{9_a,n_c} \sim \ket{1_a,n_c+5}$ & $8 \omega_a +36\alpha_a \approx 5 \omega_c -2(8n_c-5)\chi_{ac}$ & $-174.801 + 1.238 n_c$ & $-147.268+1.549 n_c$\\
\hline
\end{tabular}
\caption{Examples of qubit-resonator frequency collisions between the high-excitation and computational subspaces observed in Figs.~\ref{fig:Leak-2DSweepWithDRAG}--\ref{fig:Leak-MostLeakedStates}. The leftmost column shows the colliding quantum states (qubit $\otimes$ resonator), the second provides a collision condition based on the \textit{undriven} Kerr spectrum, and the third and the fourth provide experimental estimates for qubit anharmonicity assuming the same parameters as in Figs.~\ref{fig:Leak-QuResLeakage} and~\ref{fig:Leak-MostLeakedStates} and based on Kerr and exact ab-initio models, respectively. The Kerr model approximately captures the order by which these clusters happen, but the estimate is less valid especially for collisions involving higher qubit states and occurring at larger anharmonicity. In particular, the ab-initio analysis reveals the possibility of \textit{additional} collisions between the same states [shown with a bullet, see e.g. Fig.~\ref{fig:Leak-HighLowStateCollision}(b)].}
\label{tab:Leak-ReadoutCollisions}
\end{table*}
%%%%%%%%%%%%%%%%%%%%%%%%%%%%%%%%%%%%%%%%%%%%%%%%%%%%%%%%%%%%%%%%%%%%%%%%%%%%%%%%

Analytical modeling of leakage requires advanced time-dependent methods such as SWPT or Magnus expansion \cite{Magnus_Exponential_1954, Blanes_Magnus_2009, Blanes_Pedagogical_2010}. It is in principle possible to use time-dependent SWPT to compute leakage rates. However, generally, SWPT is more suitable for computing effective (resonant) rates, while leakage rates are more easily found via Magnus. For the RIP gate, the two methods are connected via 
\begin{align}
\hat{U}_{I}(t,0) =\hat{U}_{\text{diag}}(t)\hat{U}_{I,\text{eff}}(t,0)\hat{U}^{\dag}_{\text{diag}}(0) \;,
\label{eqn:Leak-UI=Ud*UIeff*Ud'}
\end{align}
where $\hat{U}_{I}(t,0)$ and $\hat{U}_{I,\text{eff}}(t,0)$ are the overall and the effective time-evolution operators, and $\hat{U}_{\text{diag}}(t)$ is the mapping similar to Eq.~(\ref{eqn:EffHam-U_diag decomp}). For a physical process to cause leakage, it \textit{must} be off-diagonal with respect to the computational eigenstates. Hence, in SWPT, the information about leakage is encoded indirectly through $\hat{U}_{\text{diag}}(t)$. In Magnus, however, we perform an expansion \textit{directly} on $\hat{U}_{I}(t,0)$ (see Appendix~\ref{App:LeakMech}). 

Regardless of the method, the contribution from each physical process appears as the Fourier transform of the underlying time-dependent interaction rate evaluated at the corresponding system transition frequency. Therefore, leakage can be characterized given the following information: (i) system transition frequencies, (ii) interaction matrix elements (connectivity between the eigenstates), and (iii) spectral content of the interaction (pulse shape). Once the pulse spectrum has a non-negligible overlap with a particular system transition frequency, we expect an increase in the leakage. Even though capturing the precise leakage amplitude is a difficult task, estimating where leakage occurs in the parameter space is feasible based on the frequency collisions in the system spectrum.  

Consequently, regions in system parameter space where a computational state becomes \textit{degenerate} with a high-excitation qubit state provide a bridge for leakage given that such states are connected directly or indirectly by the underlying interaction. Figure~\ref{fig:Leak-HighLowStateCollision} provides two examples of such collisions as a function of qubit anharmonicity. Firstly, state $\ket{6_a,0_c}$ collides with states $\ket{0_a,4_c}$ and $\ket{1_a,3_c}$ at $\alpha_a /2\pi\approx -177.230$ MHz and $\alpha_a /2\pi \approx -272.136$ MHz, respectively, explaining the \textit{two} observed leakage clusters in Fig.~\ref{fig:Leak-MostLeakedStates}(b). The clustering is due to additional collisions with increasing photon number, e.g. between $\ket{6_a,1_c}$ and $\ket{0_a,5_c}$ etc (see Table~\ref{tab:Leak-ReadoutCollisions}). Secondly, state $\ket{8_a,0_c}$ collides \textit{once} with state $\ket{0_a,5_c}$ at $\alpha_a/2\pi\approx -185.549$ MHz and \textit{twice} with state $\ket{1_a,4_c}$ at $\alpha_a /2\pi\approx -235.336$ MHz and $\alpha_a/2\pi\approx -313.738$ MHz, in agreement with the \textit{three} leakage clusters in Fig.~\ref{fig:Leak-MostLeakedStates}(b). Such repeated collisions between the same states \textit{cannot} be predicted by the Kerr model which is \textit{linear} in qubit anharmonicity. It is a signature that the sextic, and possibly higher order, expansion of the Josephson potential is relevant at larger anharmonicity as (assuming $g_a=0$ and $n_{ga}=0$)
\begin{align}
\begin{split}
\HO_{qa}&=(8E_{Ca}E_{Ja})^{1/2}\hat{a}^{\dag}\hat{a}-\frac{E_{Ca}}{12} \left(\hat{a}+\hat{a}^{\dag}\right)^{4}\\
&+\frac{(2E_{Ca}^3/E_{Ja})^{1/2}}{360} \left(\hat{a}+\hat{a}^{\dag}\right)^{6}+O\left(\frac{E_{Ca}^2}{E_{Ja}}\right) \;.
\end{split}
\label{eqn:Leak-WhyStatesBend}
\end{align}
Moreover, the observation that the high- and low-excitation qubit states cross, as opposed to undergoing an anti-crossing, reveals that such states are not coupled via nonlinearity but rather through an \textit{unwanted} projection of the RIP drive over the corresponding transition. A summary of the observed qubit-resonator collisions as well as a comparison between ab-initio and Kerr predictions is given in Table~\ref{tab:Leak-ReadoutCollisions}.

%%%%%%%%%%%%%%%%%%%%%%% Fig: Dependence on Gate Charge %%%%%%%%%%%%%%%%%%%%%%%%
\begin{figure}[t!]
\centering
\includegraphics[scale=0.196]{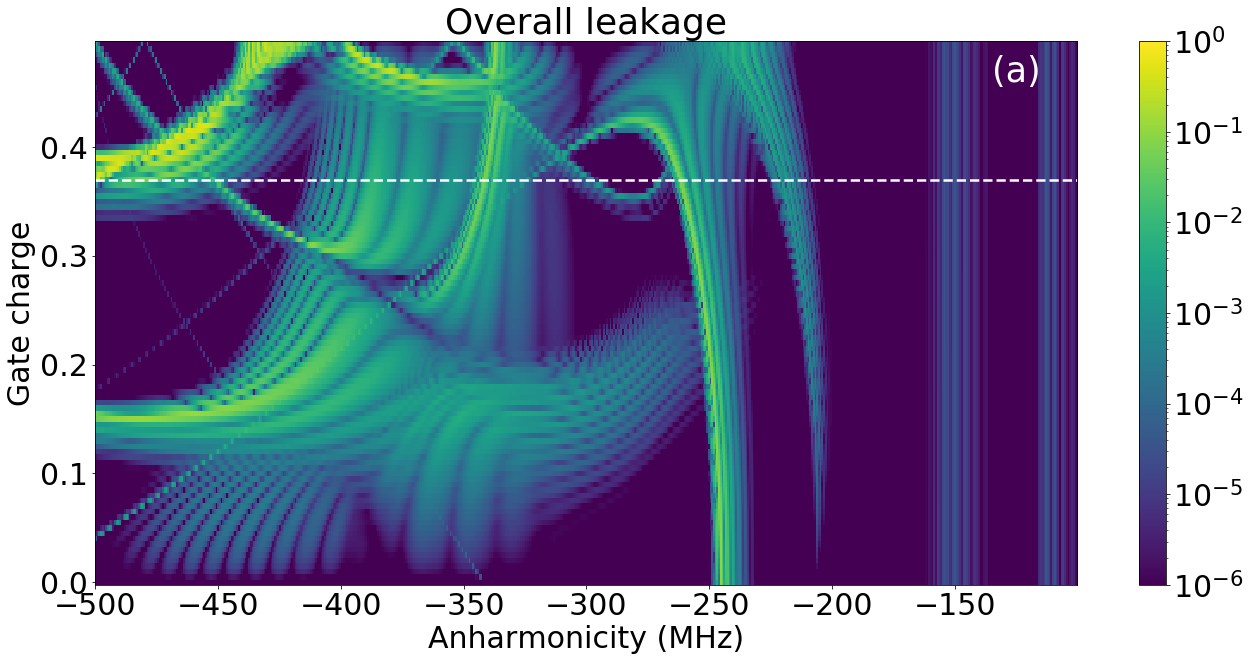}
\includegraphics[scale=0.205]{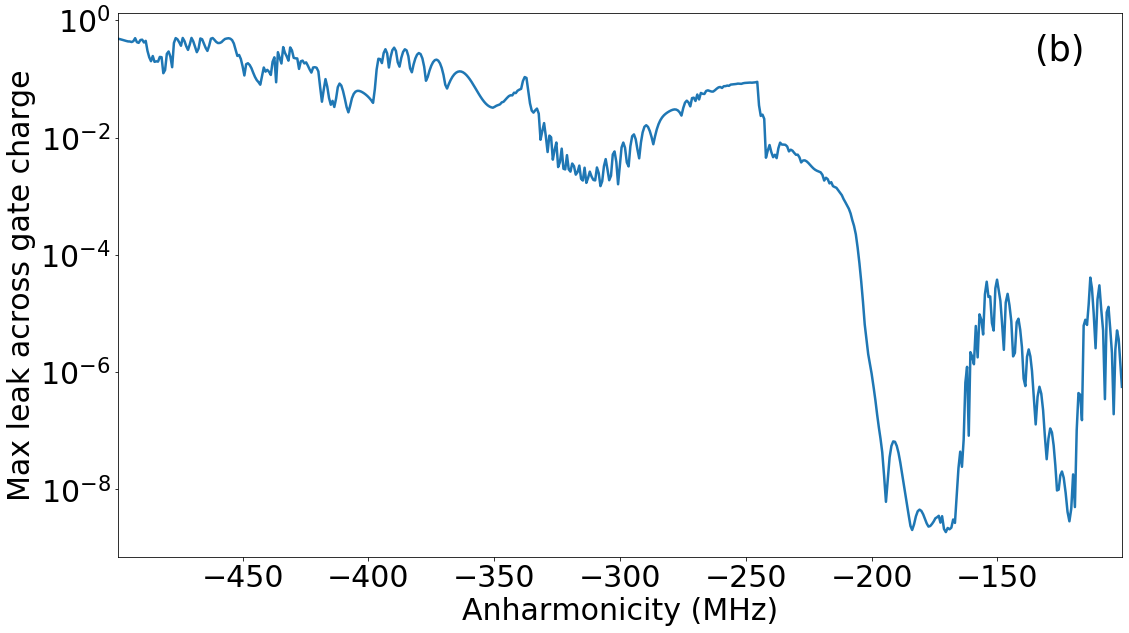}
\caption{Qubit-resonator case --- Dependence of overall leakage with DRAG on the gate charge for the same parameters as in Fig.~\ref{fig:Leak-QuResLeakage}, resonator-drive detuning set to -50 MHz, and maximum photon number set to 16. (a) 2D sweep of overall leakage as a function of gate charge and anharmonicity. The result is \textit{approximately} periodic and symmetric under change of sign for the gate charge, hence the approximately unique interval of [0,0.5] is shown. The white dashed line shows the value of gate charge at 0.37 used in Figs.~\ref{fig:Leak-2DSweepWithDRAG}--\ref{fig:Leak-HighLowStateCollision}. (b) Maximum overall leakage across the gate charge, which accounts for the worst case scenario at each anharmonicity value.} 
\label{fig:Leak-2DSweepDepOnGateCharge}
\end{figure}
%%%%%%%%%%%%%%%%%%%%%%%%%%%%%%%%%%%%%%%%%%%%%%%%%%%%%%%%%%%%%%%%%%%%%%%%%%%%%%%%%  

Given the correspondence between qubit-resonator leakage clusters and frequency collisions involving \textit{high-excitation} qubit states, it is also crucial to quantify and optimize the dependence of leakage on gate charge for two main reasons. Firstly, higher-excitation eigenenergies of the transmon qubit depend more strongly on the gate charge \cite{Koch_Charge_2007}, causing non-negligible shifts of the leakage clusters in the parameter space. Secondly, and more importantly, the gate charge is neither controllable nor predictable in the experiment. A more realistic measure is then the maximum leakage over one period of gate charge for each parameter set. Figure~\ref{fig:Leak-2DSweepDepOnGateCharge} shows the dependence of qubit-resonator overall leakage on the qubit anharmonicity and the gate charge, over the approximately unique interval of $n_{ga}\in [0,0.5]$ \footnote{Spectrum of an isolated transmon, based on $\HO_a=4E_{Ca}(\hat{n}_a-n_{ga})^2-E_{Ja}\cos(\hat{\phi}_a)$, is periodic with respect to the gate charge $n_{ga}$ \cite{Koch_Charge_2007}. However, a charge-charge coupling of the form $g_{a}\hat{n}_a\hat{n}_c$ to a resonator mode breaks such a translational symmetry. Our numerical simulations show that for experimentally relevant values of qubit-resonator coupling (full dispersive shift of the order of -5 MHz), the deviation in the spectrum and also the corresponding deviation in the qubit-resonator leakage is small under $n_{ga} \rightarrow n_{ga}+1$. Moreover, the result is approximately symmetric with respect to $n_{ga} \rightarrow -n_{ga}$, Therefore, in Fig.~\ref{fig:Leak-2DSweepDepOnGateCharge}, we have presented the dependence of leakage on the approximately unique interval of $n_{ga}\in [0,0.5]$}. We observe that smaller anharmonicity, approximately below -200 MHz, leads to significantly less leakage cluster density, less leakage amplitude and less dependence of leakage on the gate charge.  

\begin{figure}[t!]
\centering
\includegraphics[scale=0.144]{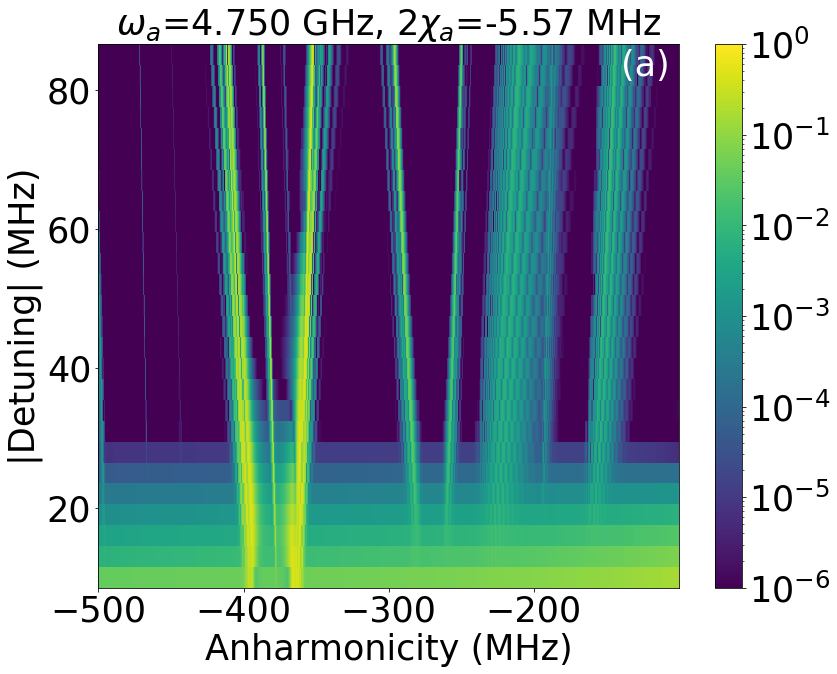}
\includegraphics[scale=0.144]{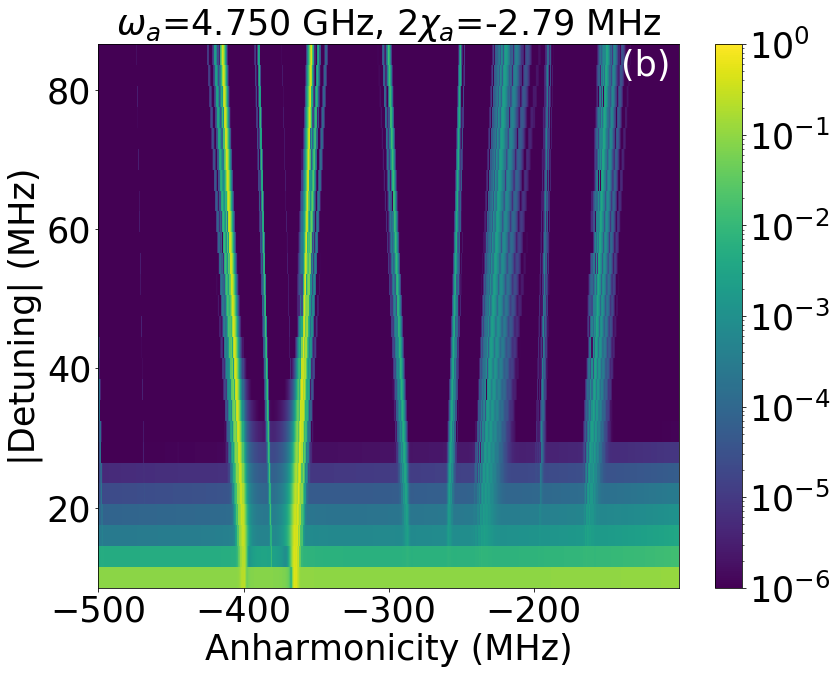}\\
\includegraphics[scale=0.144]{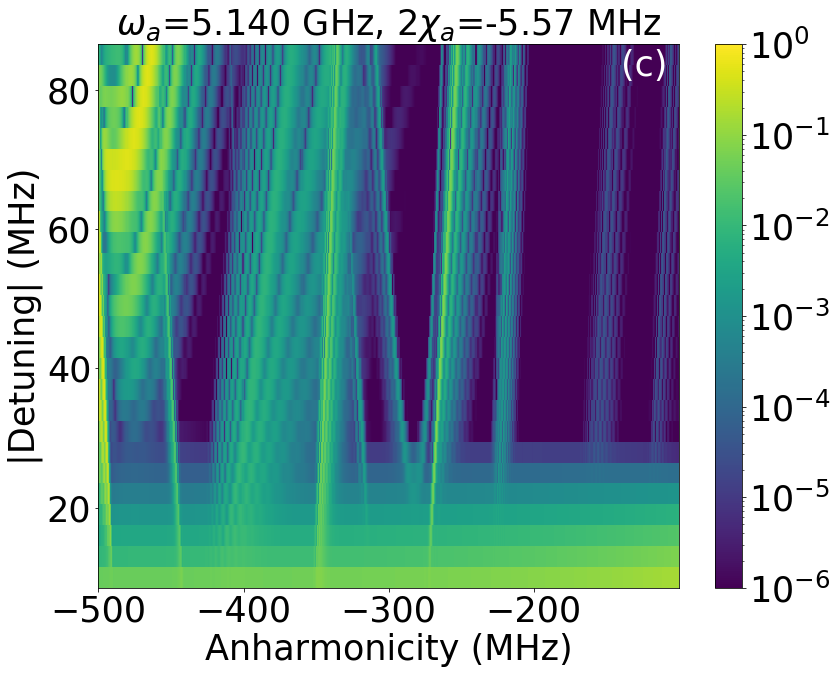}
\includegraphics[scale=0.144]{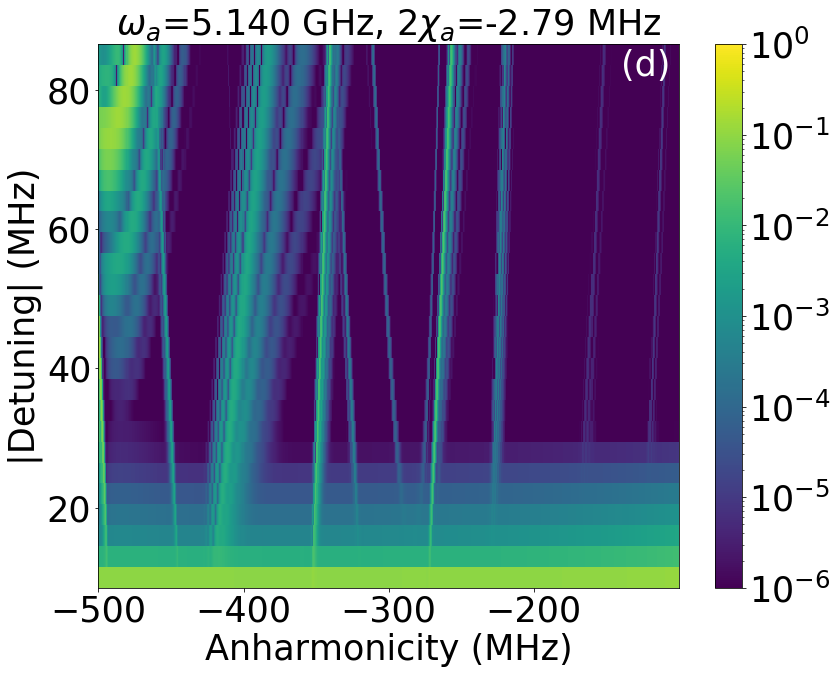}\\
\includegraphics[scale=0.144]{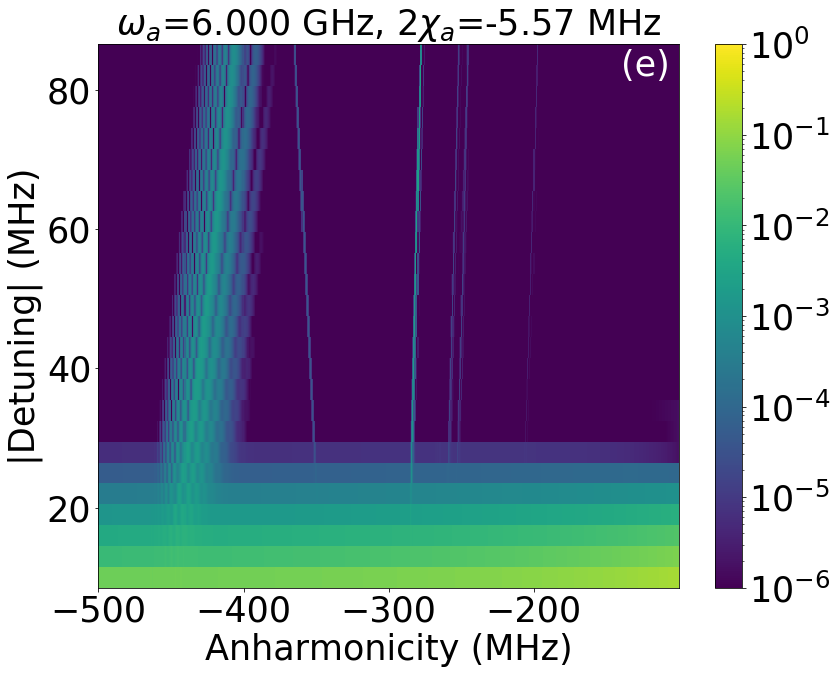}
\includegraphics[scale=0.144]{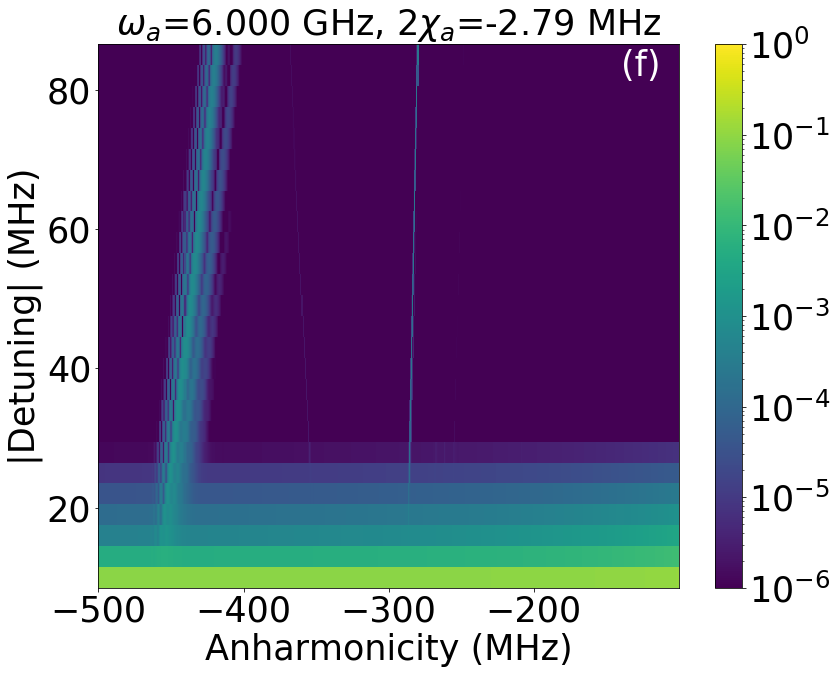}
\caption{Qubit-resonator case --- Overall leakage with DRAG for three different qubit frequencies $\omega_a /2\pi \approx 4750$ [(a), (b)], $5140$ [(c), (d)], $6000$ [(e), (f)] MHz, and two distinct dispersive shifts $2\chi_{ac} /2\pi \approx -5.57$ MHz (left column) and $-2.79$ MHz (right column), with maximum photon number set to 16. Other parameters are the same as Fig.~\ref{fig:Leak-2DSweepWithDRAG}.}
\label{fig:Leak-2DSweepDepOnFreqAndChi}
\end{figure}

The results so far were based on fixed $\omega_a/2\pi \approx 5140$, $\omega_c/2\pi \approx 6971$ and $2\chi_{ac}/2\pi \approx -5.57$ MHz. According to Table~\ref{tab:Leak-ReadoutCollisions}, the collision conditions depend also strongly on these parameters. In particular, if the qubit frequency is increased, there is \textit{less} chance of qubit-resonator collisions in the anharmonicity range relevant for experiment. For instance, consider the $\ket{6_a,0_c} \sim \ket{1_a,3_c}$ collision, with approximate Kerr condition $6\omega_a+15\alpha_a \approx \omega_a+3\omega_c+6\chi_{ac}$. Setting $\omega_a/2\pi \approx 6000$ MHz, keeping other parameters the same, pushes the collision to $\alpha_a/2\pi \approx -606.914$ MHz, away from the transmon regime. This behavior holds for qubit-resonator collisions in general. Moreover, the dispersive shift $2\chi_{ac}$ determines the number-splitting span in each cluster. Hence there is a trade-off between large $2\chi_{ac}$, desired for large dynamic $ZZ$, and the width of collision-free anharmonicity intervals. Figure~\ref{fig:Leak-2DSweepDepOnFreqAndChi} compares the qubit-resonator leakage of three distinct qubit frequencies $\omega_a /2\pi\approx 4750$, $5140$, $6000$ MHz and two dispersive shifts $2\chi_{ac} /2\pi \approx -5.57$, $-2.79$ MHz, and confirms the above-mentioned trends. Importantly, based on Figs.~\ref{fig:Leak-2DSweepDepOnFreqAndChi}(e)--\ref{fig:Leak-2DSweepDepOnFreqAndChi}(f), working with the 6000 MHz frequency qubit removes most of the leakage clusters from the considered anharmonicity range and mitigates the amplitude of the remaining ones.    

%%%%%%%%%%%%%%%%%%%%%%% Fig: 2D Sweep Leakage 2 %%%%%%%%%%%%%%%%%%%%%%%%%%%%%%%%
%%%%%%%%%%%%%%%%%%%%%%%%%%%%%%%%%%%%%%%%%%%%%%%%%%%%%%%%%%%%%%%%%%%%%%%%%%%%%%%%%

\subsection{Three-body leakage}
\label{SubSec:RIPLikeLeak}

The main physics of qubit-resonator collisions and leakage can in principle be extended to understand more complex three-body collisions. Compared to the qubit-resonator case, which results in leakage to relatively high-excitation qubit states (5th--9th), having both qubits participating in exchange interactions allows for leakage to low-excitation qubit states as well. For instance, a single excitation of the low-frequency qubit and one resonator photon can provide the energy to leak to the 2nd excited state of the high-frequency qubit, i.e. a collision of the form $\ket{1_a,0_b,1_c} \sim \ket{0_a,2_b,0_c}$. For example, this collision can be satisfied if $\omega_a /2\pi \approx 4750$, $\omega_b /2\pi \approx 6000$, $\omega_c /2\pi \approx 6971$ and $\alpha_b /2\pi \approx -279$ MHz. We refer to this frequency configuration as the high-low RIP pair.

%%%%%%%%%%%%%%%%%%% Table: Three-body collisions %%%%%%%%%%%%%%%%%%%%%%%%%%%%%%%%
\begin{table*}[t!]
\begin{tabular}{|c|c|c|}
\hline
States & Condition (Kerr) & Instance\\
\hline\hline
$\ket{3_a,0_b,n_c} \sim \ket{0_a,1_b,n_c+1}$ & $3\omega_a+3\alpha_a+6n_c\chi_{ac} \approx \omega_b+\omega_c+2(n_c+1)\chi_{bc}$ & high-low\\
\hline
$\ket{4_a,0_b,n_c} \sim \ket{0_a,1_b,n_c+2}$ & $4\omega_a+6\alpha_a+8n_c\chi_{ac} \approx \omega_b+2\omega_c+2(n_c+2)\chi_{bc}$ & high-high\\
\hline
$\ket{5_a,0_b,n_c} \sim \ket{0_a,1_b,n_c+2}$ & $5\omega_a +10\alpha_a	+10 n_c \chi_{ac}\approx \omega_b + 2 \omega_c +2(n_c+2)\chi_{bc}$ & high-low\\
\hline
$\ket{5_a,0_b,n_c} \sim \ket{1_a,1_b,n_c+1}$ & $4 \omega_a +10\alpha_a +10 n_c \chi_{ac} \approx \omega_b + \omega_c +2(n_c+1)(\chi_{ac}+\chi_{bc})$ & high-low\\
\hline
$\ket{6_a,0_b,n_c} \sim \ket{1_a,1_b,n_c+2}$ & $5 \omega_a +15\alpha_a +12 n_c \chi_{ac} \approx \omega_b + 2 \omega_c +2(n_c+2)(\chi_{ac}+\chi_{bc})$ & high-low\\
\hline
$\ket{7_a,0_b,n_c} \sim \ket{1_a,1_b,n_c+2}$ & $6\omega_a+21\alpha_{a}+14n_c\chi_{ac}\approx \omega_b + 2\omega_c +2(n_c+2)(\chi_{ac}+\chi_{bc})$ & high-low\\
\hline
$\ket{7_a,0_b,n_c} \sim \ket{1_a,1_b,n_c+3}$ & $6\omega_a+21\alpha_{a}+14n_c\chi_{ac}\approx \omega_b + 3\omega_c +2(n_c+3)(\chi_{ac}+\chi_{bc})$ & high-high\\
\hline
$\ket{8_a,0_b,n_c} \sim \ket{1_a,1_b,n_c+2}$ & $7\omega_a+28\alpha_{a}+16n_c\chi_{ac}\approx \omega_b + 2 \omega_c +2(n_c+2)(\chi_{ac}+\chi_{bc})$ & high-low\\
\hline \hline
$\ket{0_a,2_b,n_c} \sim \ket{1_a,0_b,n_c+1}$ & $2\omega_b+\alpha_b+4n_c\chi_{bc} \approx \omega_a+\omega_c+2(n_c+1)\chi_{ac}$ & high-low\\
\hline
$\ket{0_a,5_b,n_c} \sim \ket{1_a,1_b,n_c+2}$ & $4\omega_b+10\alpha_b+10n_c\chi_{bc}\approx \omega_a+2\omega_c+2(n_c+2)(\chi_{ac}+\chi_{bc})$ & high-high\\
\hline
$\ket{0_a,7_b,n_c} \sim \ket{1_a,1_b,n_c+3}$ & $6\omega_b+21\alpha_b+14n_c\chi_{bc}\approx \omega_a+3\omega_c+2(n_c+3)(\chi_{ac}+\chi_{bc})$ & high-high\\
\hline\hline
$\ket{2_a,3_b,n_c} \sim \ket{1_a,0_b,n_c+3}$ & $\omega_a+\alpha_a+3\omega_b+3\alpha_b + 4n_c\chi_{ac}+6n_c\chi_{bc}\approx 3\omega_c+2(n_c+3)\chi_{ac}$ & \text{high-low}\\
\hline
$\ket{2_a,5_b,n_c} \sim \ket{1_a,0_b,n_c+4}$ & $\omega_a+\alpha_a+5\omega_b+10\alpha_b + 4n_c\chi_{ac}+10n_c\chi_{bc}\approx 4\omega_c+2(n_c+4)\chi_{ac}$ & \text{high-low}\\
\hline
$\ket{3_a,1_b,n_c} \sim \ket{1_a,0_b,n_c+2}$ & $2\omega_a+3\alpha_a+\omega_b+ 6n_c\chi_{ac}+2n_c\chi_{bc}\approx 2\omega_c+2(n_c+2)\chi_{ac}$ & \text{high-low}\\
\hline
$\ket{3_a,2_b,n_c} \sim \ket{1_a,1_b,n_c+2}$ & $2\omega_a+3\alpha_a+\omega_b+\alpha_b + 6n_c\chi_{ac}+4n_c\chi_{bc}\approx 2\omega_c+2(n_c+2)(\chi_{ac}+\chi_{bc})$ & \text{high-low}\\
\hline
$\ket{5_a,2_b,n_c} \sim \ket{1_a,1_b,n_c+3}$ & $4\omega_a+10\alpha_a+\omega_b+\alpha_b + 10n_c\chi_{ac}+4n_c\chi_{bc}\approx 3\omega_c+2(n_c+3)(\chi_{ac}+\chi_{bc})$ & \text{high-low}\\
\hline
\end{tabular}
\caption{Examples of dominant three-body collisions observed in Figs.~\ref{fig:RIPLikeLeak-HighLowPair}--\ref{fig:RIPLikeLeak-HighHighPair}. The first column describes the colliding states, the second provides an \textit{approximate} collision condition based on the Kerr model, and the last column shows in which frequency configuration the collision was observed. The list here is derived from Figs.~\ref{fig:RIPLikeLeak-HighLowPair}--\ref{fig:RIPLikeLeak-HighHighPair} with a leakage cut-off of $10^{-5}$ for each collision assuming 4 maximum resonator photons. In principle, however, there are numerous possibilities.} 	
\label{tab:Leak-3bdCollisions}
\end{table*}
%%%%%%%%%%%%%%%%%%%%%%%%%%%%%%%%%%%%%%%%%%%%%%%%%%%%%%%%%%%%%%%%%%%%%%%%%%%%%%%%

%%%%%%%%%%%%%%%%%%%%%%% Fig: HighLow_RIPLIKE%%%%%%%%%%%%%%%%%%%%%%%%
\begin{figure}[h!]
\centering
\includegraphics[scale=0.213]{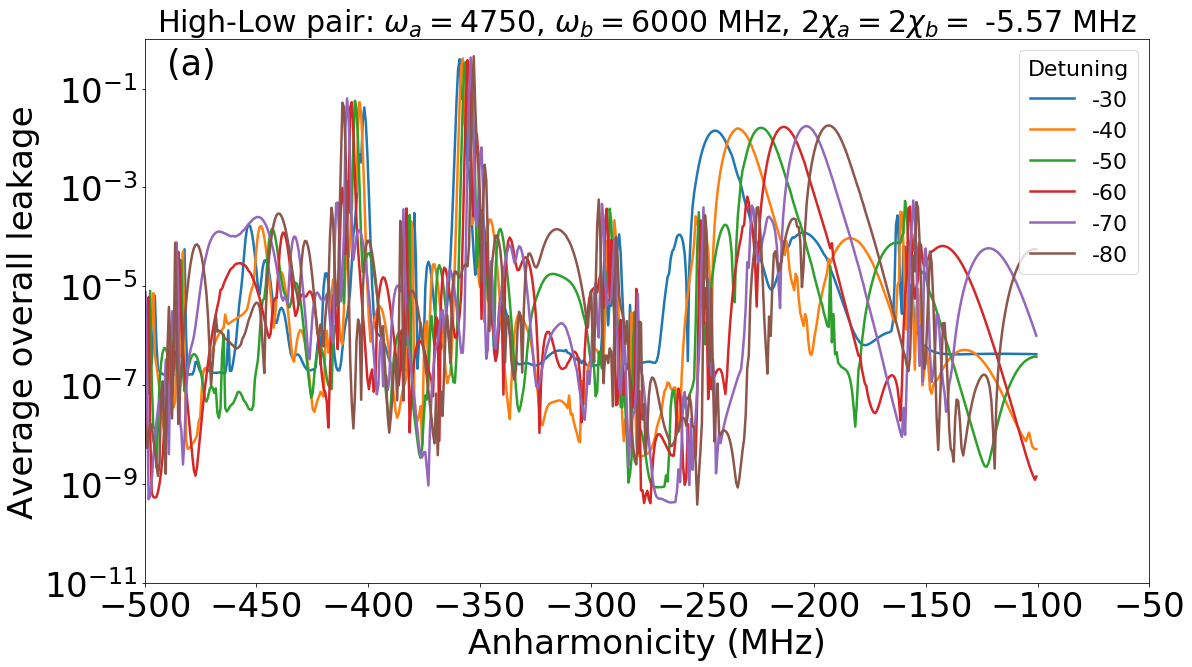}\\
\includegraphics[scale=0.213]{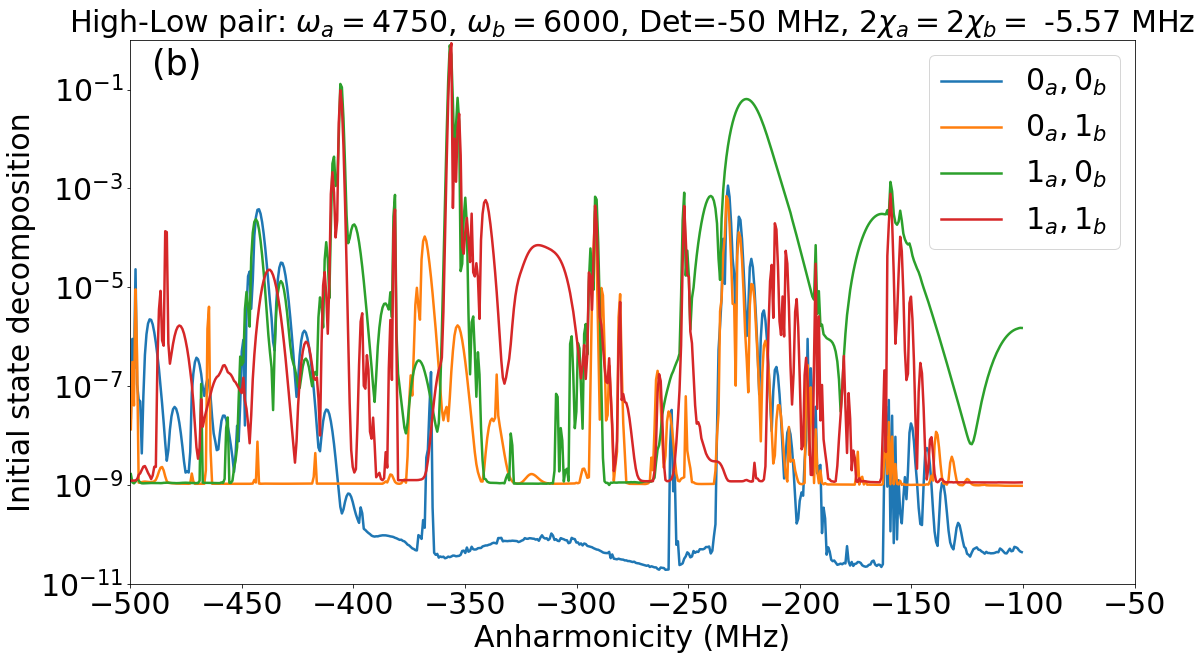}\\
\includegraphics[scale=0.1625]{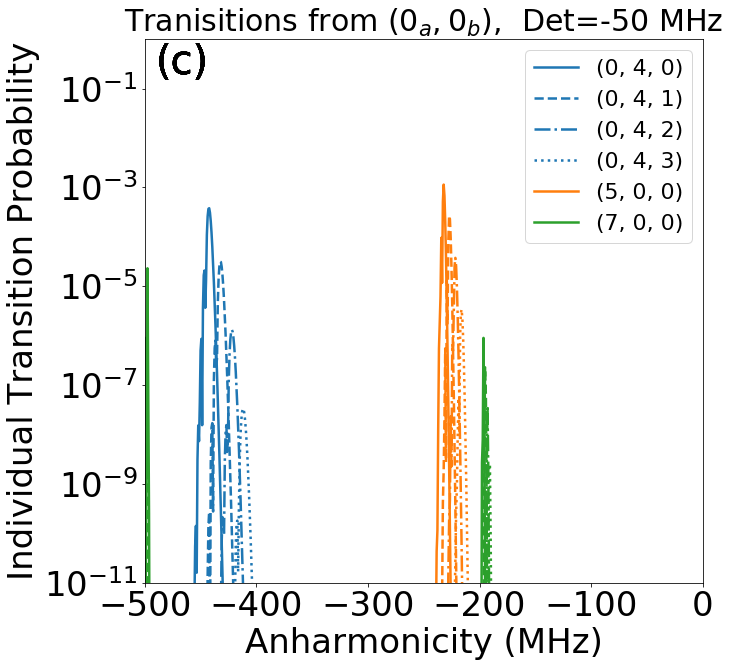}
\includegraphics[scale=0.1625]{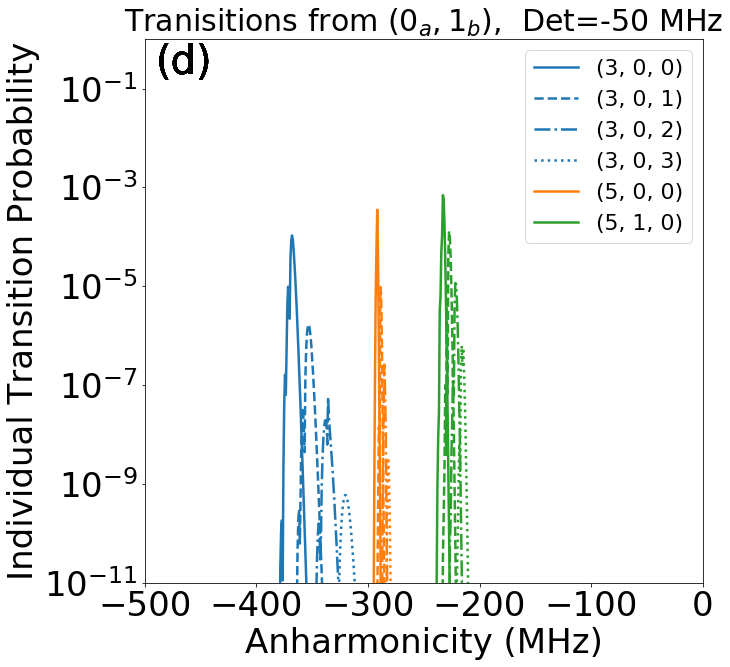}\\
\includegraphics[scale=0.1625]{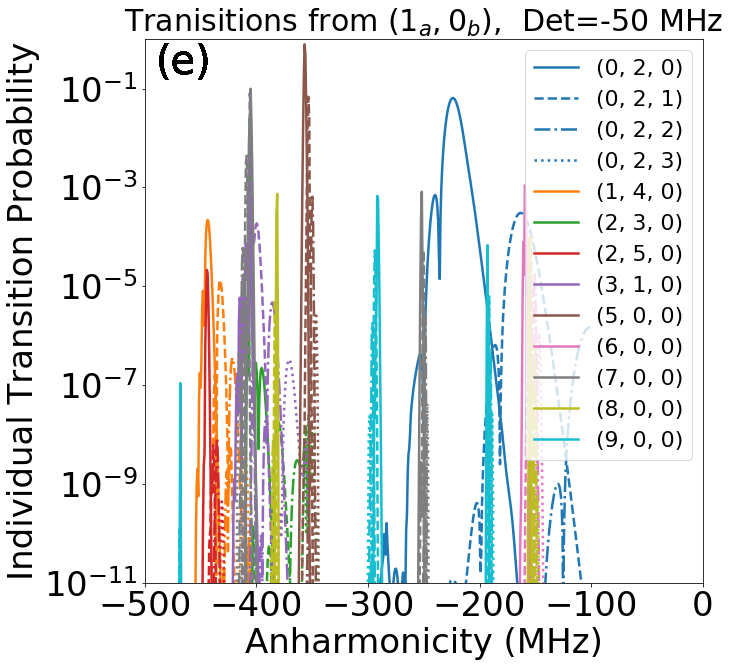}
\includegraphics[scale=0.1625]{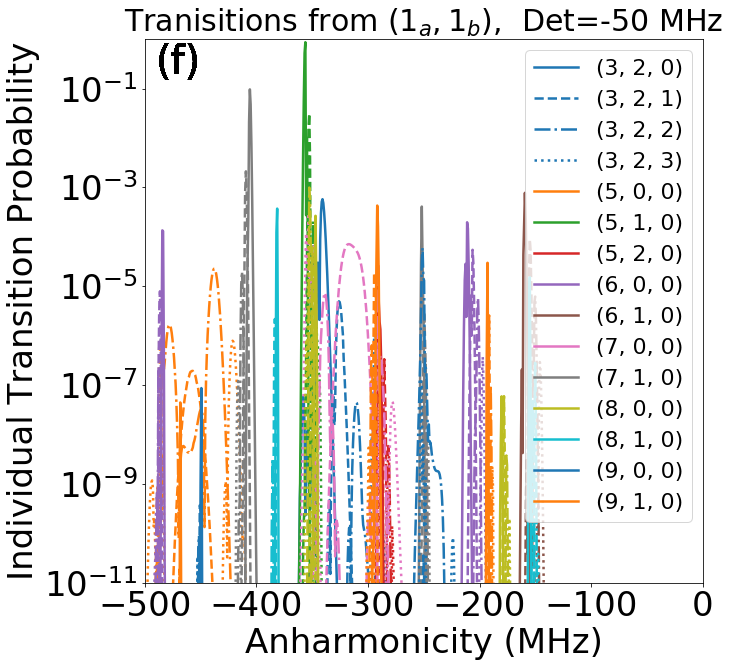}
\caption{Two-qubit simulation of leakage for the high-low pair with $\omega_a/2\pi\approx 4750$, $\omega_b/2\pi \approx 6000$, $\omega_c/2\pi\approx 6971$, $2\chi_{ac} /2\pi\approx 2\chi_{bc}/2\pi\approx -5.57$ MHz using a $200$ ns nested cosine pulse equivalent to 4 photons at the pulse maximum. (a) Overall leakage with DRAG, averaged over the four initial two-qubit states, $\ket{0_a,0_b,0_c}$, $\ket{0_a,1_b,0_c}$, $\ket{1_a,0_b,0_c}$ and $\ket{1_a,1_b,0_c}$, as a function of $\alpha_a\approx \alpha_b$ and $\Delta_{cd}$. (b) Overall leakage for each initial computational state at fixed $\Delta_{cd}/2\pi=-50$ MHz. (c)--(f) Individual leakage transition probabilities starting from the four computational states, respectively. Compared to the qubit-resonator simulations of Sec.~\ref{SubSec:READLikeLeak}, 10 energy eigenstates and 22 resonator states are kept.} 	
\label{fig:RIPLikeLeak-HighLowPair}
\end{figure}
%%%%%%%%%%%%%%%%%%%%%%%%%%%%%%%%%%%%%%%%%%%%%%%%%%%%%%%%%%%%%%%%%%%%%%%%%%%%%%%%%        
%%%%%%%%%%%%%%%%%%%%%%% Fig: HighHigh_RIPLIKE%%%%%%%%%%%%%%%%%%%%%%%%
\begin{figure}[h!]
\centering
\includegraphics[scale=0.213]{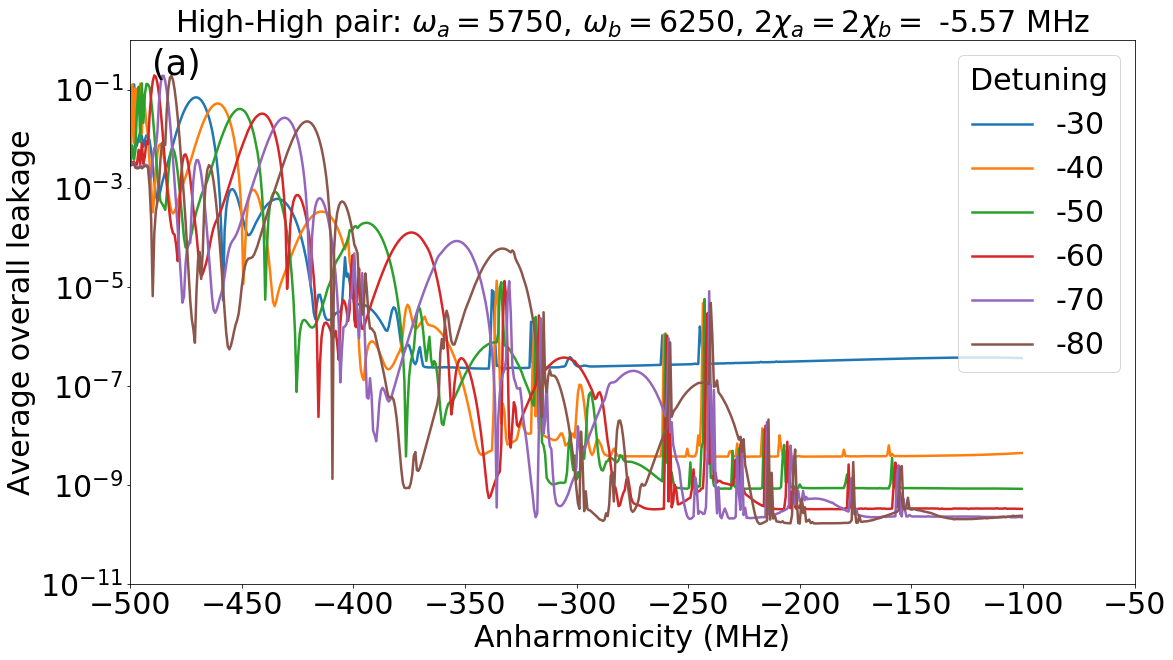}\\
\includegraphics[scale=0.213]{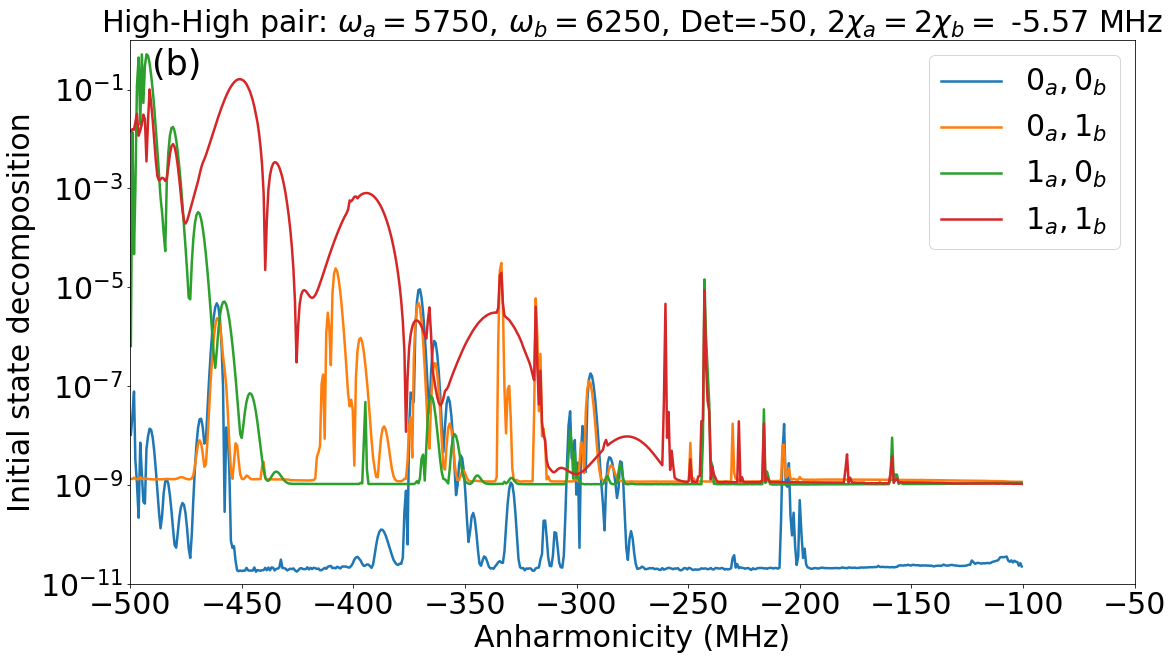}\\
\includegraphics[scale=0.160]{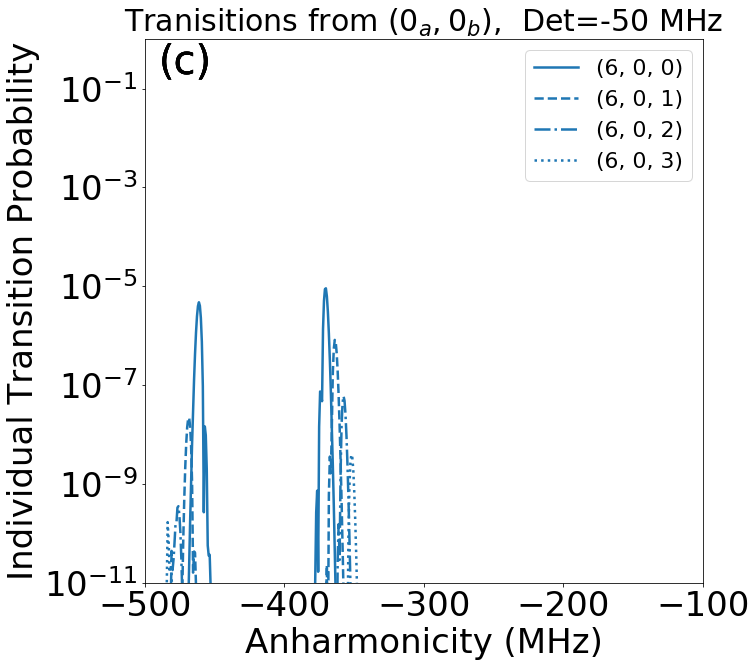}
\includegraphics[scale=0.160]{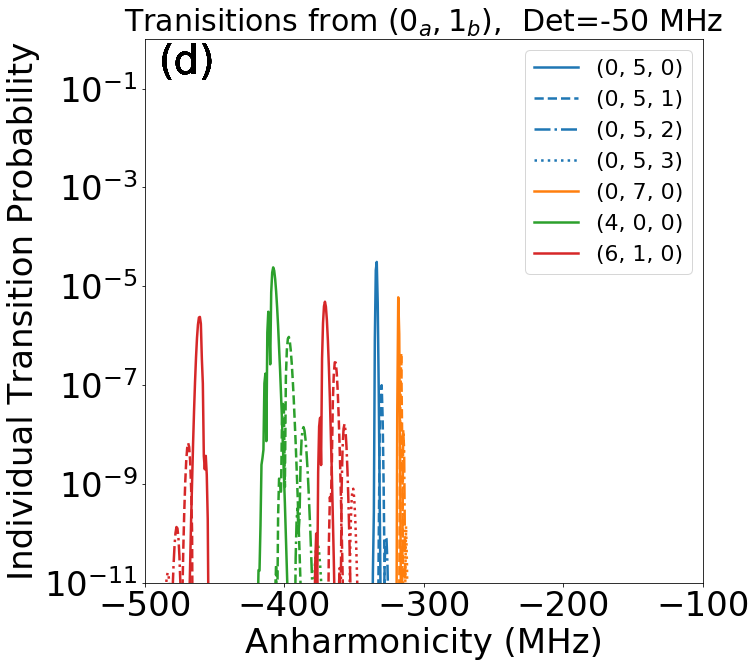}\\
\includegraphics[scale=0.160]{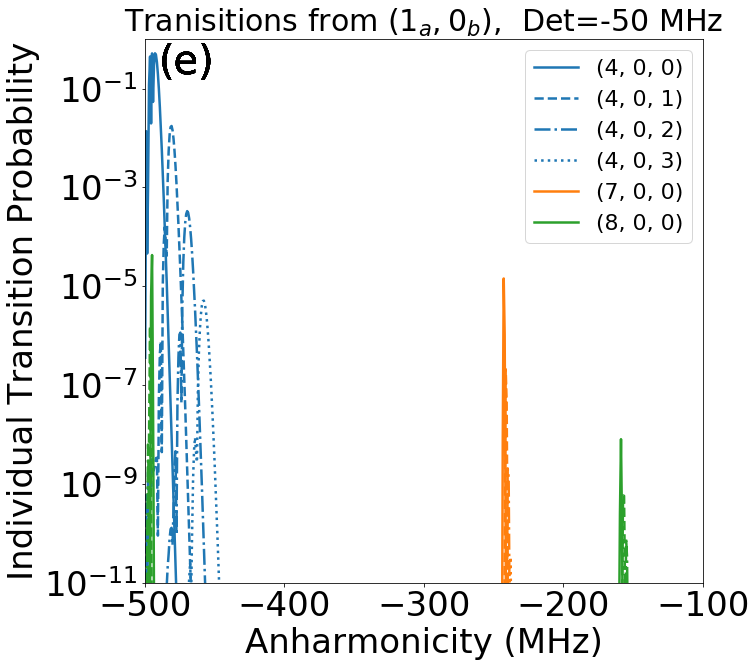}
\includegraphics[scale=0.160]{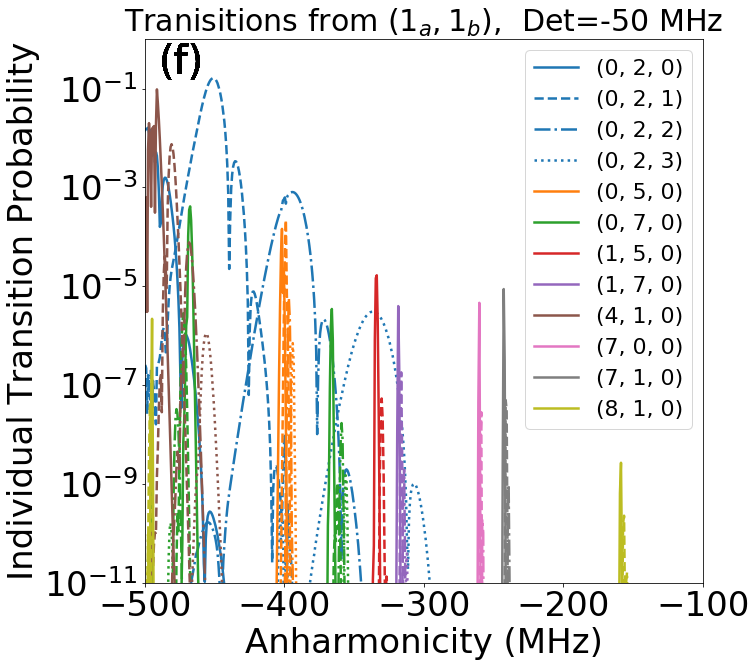}
\caption{Two-qubit simulation of leakage for the high-high pair with $\omega_a/2\pi\approx 5750$, $\omega_b/2\pi \approx 6250$ MHz and other parameters the same as Fig.~\ref{fig:RIPLikeLeak-HighLowPair}. Compared to the high-low configuration, we observe a significant suppression of leakage especially for $\alpha_a /2\pi \approx \alpha_b /2\pi \lessapprox -200$ MHz.} 	
\label{fig:RIPLikeLeak-HighHighPair}
\end{figure}
%%%%%%%%%%%%%%%%%%%%%%%%%%%%%%%%%%%%%%%%%%%%%%%%%%%%%%%%%%%%%%%%%%%%%%%%%%%%%%%%%        

We then compare the high-low configuration given above to an improved RIP pair. Based on the preceding discussion of qubit-resonator leakage (Sec.~\ref{SubSec:READLikeLeak} and Fig.~\ref{fig:Leak-2DSweepDepOnFreqAndChi}), both qubit frequencies should be set as close to the resonator frequency as is allowed by other constraints on the parameters. As one of the advantages of the RIP gate is a wider range of allowed qubit frequency values, we choose to keep the qubits outside of the straddling regime. When qubits are designed to operate in the straddling regime, the design, subject to sample-to-sample variation in fabrication, is more susceptible to qubit-qubit collisions of the form $\ket{1_a,1_b,n_c}\sim \ket{2_a,0_b,n_c}$ and $\ket{1_a,1_b,n_c}\sim \ket{0_a,2_b,n_c}$, with approximate collision conditions $\Delta_{ab}\approx \alpha_b$ and $\Delta_{ab}\approx -\alpha_a$. Moreover, since our model does not include a cancellation coupler, small qubit-qubit detuning leads to a large static $ZZ$, reducing the on/off ratio of the RIP gate. Therefore, as a contrast to the high-low configuration given above, we pick a second ``high-high'' configuration with $\omega_a /2\pi \approx 5750$, $\omega_b /2\pi \approx 6250$, and $\omega_c /2\pi \approx 6971$ MHz, corresponding to a $500$ MHz qubit-qubit detuning. We find that the high-high pair reduces three-body leakage compared to the high-low configuration while also having low qubit-resonator leakage.

For the two-qubit simulation, the ab-initio system parameters were selected to produce the qubit and resonator frequency values of the high-high and high-low scenarios while fixing qubit-resonator dispersive shifts at $2\chi_{ac} /2\pi\approx 2\chi_{bc} /2\pi \approx -5.57$~MHz. Such parameters were found for a range of qubit anharmonicities from -500 to -100~MHz, with the anharmonicity of the two qubits kept approximately equal, $\alpha_a \approx\alpha_b$, for each case. The pulse parameters were set to produce a maximum resonator photon number of 4 over multiple traces of constant resonator-drive detuning. Compared to the single-qubit simulations of Sec.~\ref{SubSec:READLikeLeak}, these simulations used 22 resonator states and 10 energy eigenstates for each qubit.

Figures~\ref{fig:RIPLikeLeak-HighLowPair} and~\ref{fig:RIPLikeLeak-HighHighPair} provide the leakage analysis for the high-low and high-high RIP pairs, respectively. We note that in such two-qubit simulations all leakage categories, as described in Fig.~\ref{fig:Leak-CollisionUniverse}, can be driven in principle. Therefore, to better understand each case, the observed leakage peaks in panel (a) are decomposed into separate transitions in terms of initial computational states in panel (b) and final leaked states in panels (c)--(f). We find that the high-high allocation produces regions of overall leakage below $10^{-6}$ especially at weak qubit anharmonicity less than -200 MHz. However, for the high-low case, the clusters are spread over the considered anharmonicity range and no discernible suppression is observed at weaker anharmonicity. This is in part due to the strong qubit-resonator collisions for the low-frequency qubit as also observed in Fig.~\ref{fig:Leak-2DSweepDepOnFreqAndChi}(a), but decomposition of leakage into individual transitions in panels (c)--(f) also reveals a series of three-body leakage transitions in which both qubits participate in the excitation exchange. In particular, the high-low case suffers from both higher three-body leakage cluster density and stronger peaks compared to the high-high case. Table~\ref{tab:Leak-3bdCollisions} provides examples of dominant three-body collisions corresponding to the two cases. For example, below -250~MHz anharmonicity, the dominant leakage transitions for the high-low case satisfy $\ket{0_a,2_b,n_c} \sim \ket{1_a,0_b,n_c+1}$.

%%%%%%%%%%%%%%%%%%%%%%% Fig: Justifying Leakage Categories %%%%%%%%%%%%%%%%%%%%%
\begin{figure}[t!]
\centering
\includegraphics[scale=0.215]{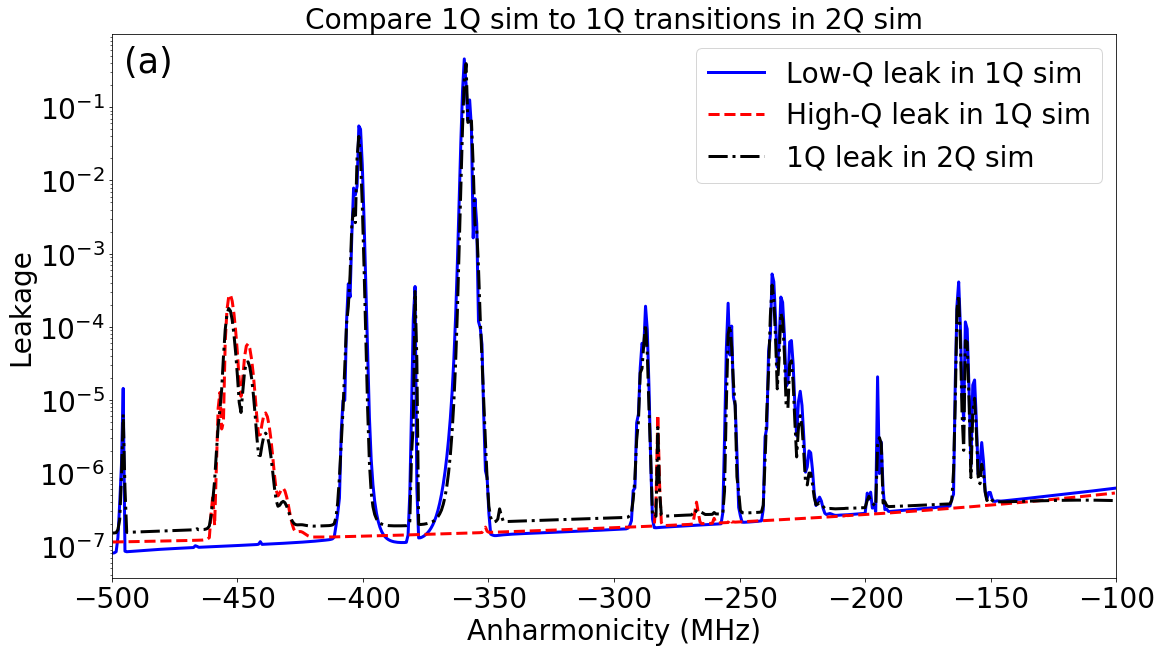}\\
\includegraphics[scale=0.215]{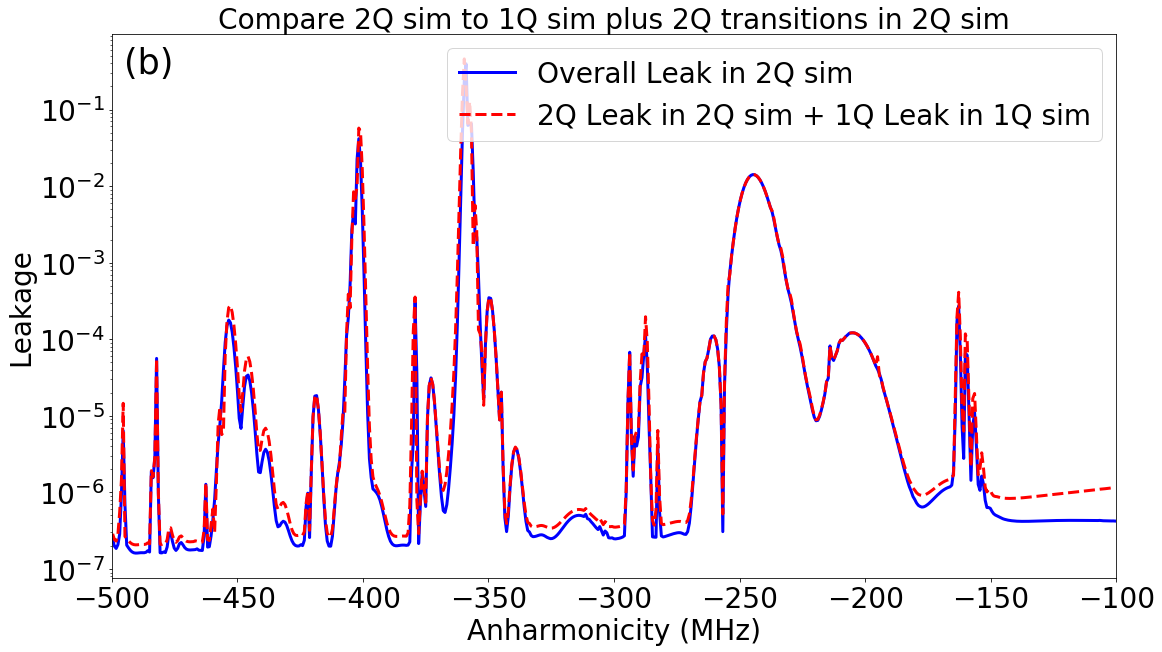}
\caption{Comparison between single- and two-qubit simulations for the high-low pair with resonator-drive detuning set to -30 MHz and 4 maximum photons --- (a) Individual qubit-resonator leakage for each qubit compared to the qubit-resonator leakage obtained from the two-qubit simulation. (b) Overall leakage obtained from the two-qubit simulation compared to the sum of qubit-resonator leakage obtained from the single-qubit simulation and three-body leakage obtained from the two-qubit simulation.}
\label{fig:Leak-JustifyingLeakCategories}
\end{figure}
%%%%%%%%%%%%%%%%%%%%%%%%%%%%%%%%%%%%%%%%%%%%%%%%%%%%%%%%%%%%%%%%%%%%%%%%%%%%%%%%%
	
To validate our characterization of the RIP gate leakage in terms of distinct categories, we show that qubit-resonator leakage computed from the single-qubit simulation in Sec.~\ref{SubSec:READLikeLeak} agrees well with the two-qubit simulation here. In particular, for the high-low pair, Fig.~\ref{fig:Leak-JustifyingLeakCategories}(a) demonstrates a good agreement between the qubit-resonator leakage from \textit{individual} single-qubit simulations and the qubit-resonator leakage from the two-qubit simulation. Furthermore, Fig.~\ref{fig:Leak-JustifyingLeakCategories}(b) shows that adding the three-body leakage from the two-qubit simulation to the qubit-resonator leakage from separate single-qubit simulations recovers the overall leakage in the two-qubit simulation. For this analysis, the two-qubit simulation results were split into qubit-resonator and three-body leakage by looking at whether the final leaked state involved a transition of one or both of the qubits. These observations confirm that, when choosing system parameters to avoid RIP gate leakage, the different leakage categories can be considered independently, since the presence of the 2nd qubit does not noticeably change the nature of the single-qubit transitions. 

In summary, our analysis of leakage has important implications in terms of optimal parameter allocation for the RIP gate. Although the background leakage can be controlled dynamically, i.e. through modifying $\omega_d$, $\tau$ and DRAG, the qubit-resonator and three-body leakage clusters result from ill-chosen static system parameters. It can be shown that average gate fidelity is limited by average leakage \cite{Wood_Quantification_2018} (see Appendix~\ref{App:FidITOLeak}). Therefore, improving fidelity requires suppressing leakage.

%%%%%%%%%%%%%%%%%%%%%%%%%%%%%%%%%%%%%%%%%%%%%%%%%%%%%%%%%%%%%%%%%%%%%%%%%%%%%

%%%%%%%%%%%%%%%%%%%%%%%%%%%%% Sec: TakeawaysForExperiment %%%%%%%%%%%%%%%%%%%%%
\section{Takeaways for control and design}
\label{Sec:Tkwys}

In this section, we summarize our findings and provide a set of instructions for optimal parameter allocation. The discussion is based on the ab-initio model in Eqs.~(\ref{eqn:Model-Starting Hs})--(\ref{eqn:Model-Starting Hd}), for which there are 12 independent system and pulse parameters: $E_{Ca}$, $E_{Cb}$, $E_{Ja}$, $E_{Jb}$, $n_{ga}$, $n_{gb}$, $g_{a}$, $g_{b}$, $\omega_c$, $\omega_d$, $\Omega_c$ and $\tau$. The nested cosine pulse shape in Eq.~(\ref{eqn:Leak-Def of NCPulse}) is fully characterized by the gate time $\tau$, while in principle there can be more pulse parameters. In what follows, we describe distinct interdependencies between certain subsets of parameters which need to be taken into account for RIP device design. In principle, these subsets are \textit{not} completely independent; however, dissecting into simpler few-parameter dependencies facilitates the search. We sort the following conditions from the most to the least trivial as:                
\begin{itemize}
\item[(i)] \textit{Resonator frequency $\omega_c$} --- The measurement equipment sets the choice for $\omega_c$ typically at 7000 MHz.
 
\item[(ii)] \textit{Drive frequency $\omega_d$ and gate time $\tau$} --- Driving above the resonator, i.e. $\omega_d>\omega_c$, leads to less background leakage, while for $\omega_d<\omega_c$ there is excessive leakage due to the other $2\chi$-shifted resonator poles [Fig.~\ref{fig:EffHam-KerrModelFreqSchematic} and Eq.~(\ref{eqn:EffHam-Def of hat(Delta)_cd})]. Pulse-resonator spectral overlap and the resulting background leakage is approximately determined by the product $\Delta_{cd} \tau$ (see Appendix~\ref{App:ResRes}). The leakage clusters tend to fan out at larger detuning (Fig.~\ref{fig:Leak-2DSweepWithDRAG}), hence $\Delta_{cd}$ should be as small as allowed by the background leakage threshold. Filtering the pulse at $\Delta_{cd}$ in general, and applying DRAG in particular [Eqs.~(\ref{eqn:Leak-Def of Omcx})--(\ref{eqn:Leak-Def of Omcy})], are beneficial in reducing $\Delta_{cd}$ for fixed leakage. Based on the simulation with DRAG, $|\Delta_{cd}/2\pi|\gtrapprox 30$ MHz is a reasonable choice for $\tau\approx 200$ ns. 

\item[(iii)] \textit{Charging energy $E_C$ and Josephson energy $E_J$} --- Qubit frequency and anharmonicity are related to $E_C$ and $E_J$ up to $O(E_C^2/E_J)$ as \cite{Koch_Charge_2007, Didier_Analytical_2018} (assuming $g=0$ and $n_g=0$)
\begin{subequations}
\begin{align}
&\quad\quad \omega \approx (8E_JE_C)^{1/2}-E_C-\frac{1}{4}\left(\frac{2E_C}{E_J}\right)^{1/2} E_C \;,
\label{eqn:Tkwys-FreqITOEcEJ}\\
&\quad\quad \alpha \approx -E_C -\frac{9}{16}\left(\frac{2E_C}{E_J}\right)^{1/2} E_C \;.
\label{eqn:Tkwys-AnhITOEcEJ}
\end{align}   
\end{subequations}
A few observations are in order based in part on our simulation results. First, operating in the transmon limit reduces the set of possible collisions by reducing the dependence of the spectrum on the gate offset charge \cite{Koch_Charge_2007}. Second, both qubit-resonator and three-body leakage \text{amplitude} are universally reduced at smaller qubit anharmonicity [Figs.~\ref{fig:Leak-2DSweepWithDRAG} and~\ref{fig:RIPLikeLeak-HighHighPair}]. Third, smaller anharmonicity leads also to less eigenenergy crowding and collisions (compare collision density at -400 to -200 MHz in Fig.~\ref{fig:Leak-2DSweepWithDRAG}). Fourth, larger qubit frequency pushes the qubit-resonator collisions to occur at larger anharmonicity values outside the range relevant to experiment (less collisions for $\omega /2\pi \approx 6000$~MHz compared to $\omega /2\pi \approx 5140$~MHz in Fig.~\ref{fig:Leak-2DSweepDepOnFreqAndChi}). All in all, it is beneficial to work with very weakly anharmonic qubits with sufficiently small anharmonicity, of the order of -200 MHz. With this choice, $\omega /2\pi \approx 5140$ and $6000$ MHz correspond to $E_J/E_C \approx 103.29$ and $136.71$, respectively. Lastly, we note that smaller anharmonicity, i.e. going from -340 MHz that is common for CR architectures to -200 MHz, can in principle enhance the $\ket{1_a}\rightarrow \ket{2_a}$ leakage during single-qubit gate operations. However, this is \textit{not} a limiting factor, since single-qubit leakage can be mitigated via a combination of DRAG \cite{Motzoi_Simple_2009, Gambetta_Analytic_2011} and \textit{slightly} longer single-qubit gate time (approximately 35 ns instead of 20 ns).           
   
\item[(iv)] \textit{Qubit-resonator coupling $g$} --- The effective dispersive coupling is approximately determined as \cite{Koch_Charge_2007} 
\begin{align}
\chi_{jc} \approx \frac{\alpha_j}{\bar{\Delta}_{jc}+\alpha_j}\frac{g_j^2}{\bar{\Delta}_{jc}} \;, 
\label{eqn:Tkwys-Expr for chi} 
\end{align}
for $j=a,b$, and the dynamic $ZZ$ behaves as $\omega_{zz}^{(2)}\propto \chi_{ac}\chi_{bc}$ [Eq.~(\ref{eqn:EffHam-w_zz^(2)})]. Working with very weakly anharmonic transmons suppresses $\chi_{jc}$, unless we compensate by keeping $g_j^2 \alpha_j$ constant or reduce $\bar{\Delta}_{jc}$. There are, however, a few trade-offs associated with large $g_j$. First, the number splitting \cite{Gambetta_Qubit-photon_2006} of leakage clusters is proportional to $\chi_{jc}$ (Tables~\ref{tab:Leak-ReadoutCollisions}--\ref{tab:Leak-3bdCollisions}); hence strong coupling leaves behind a narrower collision-free anharmonicity range (Fig.~\ref{fig:Leak-2DSweepDepOnFreqAndChi}). Second, generally speaking, large $g_j$ leads to cross-talk between the pair of qubits in the gate as well as between these qubits and other spectator qubits (beyond the scope of this paper). The static $ZZ$, for instance, grows as $g_a^2g_b^2$ [Eqs.~(\ref{eqn:EffHam-Def of J})--(\ref{eqn:EffHam-w_zz^(0)})], which can be suppressed using multi-path interference couplers \cite{Mundada_Suppression_2019, Kandala_Demonstration_2020, Kumph_Novel_APS2021}. Third, although beyond the scope of our analysis, the Purcell rate \cite{Purcell_Resonance_1946, Purcell_Spontaneous_1995, Houck_Controlling_2008, Malekakhlagh_Cutoff-Free_2017} and single-qubit measurement-induced dephasing \cite{Blais_Cavity_2004, Gambetta_Qubit-photon_2006, Puri_High-Fidelity_2016} are also enhanced at strong coupling approximately as
\begin{subequations}
\begin{align}
&\gamma_{Pj} \approx \left(\frac{g_j}{\bar{\Delta}_{jc}}\right)^2 \kappa_c \;,
\label{eqn:Tkwys-Expr for gamma_P} \\
&\gamma_{\phi j}(t)\approx \frac{2\chi_{jc}^2}{\Delta_{cd}^2+\chi_{jc}^2+(\kappa_c/2)^2}|\eta_c(t)|^2\kappa_c \;. 
\label{eqn:Tkwys-Expr for gamma_phi}
\end{align}
\end{subequations} 	  
Reference~\cite{Puri_High-Fidelity_2016} demonstrated the suppression of $\gamma_{\phi j}(t)$ via mode-squeezing.
 
\item[(v)] \textit{Drive amplitude $\Omega_c$} --- The drive amplitude sets the resonator photon number $|\eta_c(t)|^2$ [Eq.~(\ref{eqn:EffHam-Cond for eta_c(t)}) and Appendix~\ref{App:ResRes}], and the dynamic $ZZ$ rate is proportional to photon number as $\omega_{zz}^{(2)}(t)\propto |\eta_c(t)|^2$ [Eq.~(\ref{eqn:EffHam-w_zz^(2)})]. On the other hand, most leakage clusters grow \textit{super-linearly} with photon number. Therefore, the maximum drive threshold is \textit{limited} by leakage threshold. Moreover, to calibrate a controlled-phase gate, there is also an interplay with steps (2) and (4) due to a fixed rotation angle given by $\int_{0}^{\tau}dt'\omega_{zz}^{(2)}(t') = \theta_{zz}$. 
\end{itemize}

Having $2\chi_{ac}/2\pi \approx 2\chi_{bc}/2\pi\approx -5.57$ MHz at $\alpha_a/2\pi\approx\alpha_b/2\pi \approx -200$ MHz for the high-high pair requires $g_a/2\pi \approx 143.69$ and $g_b/2\pi\approx 92.13$ MHz. Using the nested cosine pulse, choosing $\Delta_{cd}/2\pi=-30$ MHz with 10 maximum photons, i.e. $[\Omega_c/(2\Delta_{cd})]^2\approx 10$, we can tune a CNOT-equivalent operation ($\theta_{zz}=\pi/2$) with a total gate time of $\tau\approx 155$ ns. The corresponding Purcell and \textit{pulse-averaged} measurement-induced dephasing rates read $\gamma_{Pa}/2\pi \approx 0.096$, $\gamma_{Pb}/2\pi \approx 0.114$, $\gamma_{\phi a}/2\pi \approx \gamma_{\phi b}/2\pi \approx 0.303$ KHz, where $\gamma_{\phi j}\equiv (1/\tau)\int_{0}^{\tau}dt' \gamma_{\phi j}(t')$. The coherence limit on the average two-qubit error due to each mechanism is estimated as $\bar{E}_{\gamma_{\phi}}\approx \sum_{j=a,b}(2/5)[(\gamma_{\phi j}/2\pi)\tau]\approx 3.74\times 10^{-5}$ and $\bar{E}_{\gamma_{P}} \approx \sum_{j=a,b}(1/5)[(\gamma_{Pj}/2\pi)\tau]\approx 6.51\times 10^{-6}$. Assuming longitudinal relaxation times $T_{1a}=T_{1b}\approx 100$ $\mu$s, the overall average incoherent error is estimated as $\bar{E}_{\text{incoh}}\approx 1.28\times 10^{-3}$.    

%%%%%%%%%%%%%%%%%%%%%%%%%%%%%%%%%%%%%%%%%%%%%%%%%%%%%%%%%%%%%%%%%%%%%%%%%%%%%%%%

%%%%%%%%%%%%%%%%%%%%%%%%%%%%% Sec: Conclusion %%%%%%%%%%%%%%%%%%%%%%%%%%%%%%%%%%
\section{Conclusion}
\label{Sec:Conc}

In this work, we presented an ab-initio analysis of the RIP gate dynamics, through which we characterized qubit leakage due to a series of unwanted transitions. The physics behind such transitions cannot be correctly analyzed using the dispersive JC or Kerr models since they are by construction diagonal with respect to the subspace of the qubits. Our ab-initio theory suggests that the qubit leakage can be reduced by using very weakly anharmonic transmon qubits with $E_J/E_C > 100$ compared to $E_J/E_C \approx 50$ \cite{Koch_Charge_2007} that is common for CR architectures \cite{Sheldon_Procedure_2016, Sundaresan_Reducing_2020, Jurcevic_Demonstration_2021}. 

In particular, we achieve this limit by simultaneously increasing the qubit frequency and decreasing the anharmonicity compared to the state-of-the-art operating point of 5 GHz and -340 MHz for CR gates \cite{Hertzberg_Laser_2021, Jurcevic_Demonstration_2021}. Weaker anharmonicity mitigates leakage amplitude, density, and its dependence on gate charge, while larger qubit frequency pushes the underlying collisions to larger negative anharmonicity away from experimentally relevant range. We demonstrated the advantage of such parameter allocation for a RIP pair with qubit and resonator frequencies at 5.75, 6.25 and 7.00 GHz, respectively (Fig.~\ref{fig:RIPLikeLeak-HighHighPair}).        

Despite focusing on the RIP gate operation, we note that our analysis of leakage and frequency collisions have immediate application to similar setups in which weakly anharmonic superconducting qubits are coupled to linear resonators. Prominent examples are dispersive readout \cite{Boissonneault_Nonlinear_2008, Boissonneault_Dispersive_2009, Minev_Catch_2019, Petrescu_Lifetime_2020, Hanai_Intrinsic_2021} and Kerr-cat qubits \cite{Mirrahimi_Dynamically_2014, Leghtas_Confining_2015, Grimm_Stabilization_2020}. Although the leakage strength depends on the specifics of the control and measurement scheme for each case, the unwanted transitions outlined in this work can also be driven in these setups.          	

%%%%%%%%%%%%%%%%%%%%%%%%%%%%%%%%%%%%%%%%%%%%%%%%%%%%%%%%%%%%%%%%%%%%%%%%%%%%%%%%

%%%%%%%%%%%%%%%%%%%%%%%%%%%%% Sec: Acknowledgements %%%%%%%%%%%%%%%%%%%%%%%%%%%%%
\section{Acknowledgements}
\label{Sec:Acknow}

We appreciate helpful discussions with the IBM Quantum team especially Lev Bishop, Oliver Dial, Aaron Finck, Jay Gambetta, Abhinav Kandala, Muir Kumph, Easwar Magesan, David McKay, James Raftery, Seth Merkel, Zlatko Minev, Matthias Steffen and Ted Thorbeck. We acknowledge the work of IBM Research Hybrid Cloud services, and especially Kenny Tran, which substantially facilitated our extensive numerical analyses.  

%%%%%%%%%%%%%%%%%%%%%%%%%%%%%%%%%%%%%%%%%%%%%%%%%%%%%%%%%%%%%%%%%%%%%%%%%%%%%%%%%%%%%%%%%%%%%%%%%%%%%%

%%%%%%%%%%%%%%%%%%%%%%%%%%%%%%%%%%%%%%%%%%%%%%%%%%%%%%%%%%%%%%%%%%%%%%%%%%%%%%%%%%%%%%%%%%%%%%%%%%%%%%%%%%%%%%%%
\appendix
%%%%%%%%%%%%%%%%%%%% App: two-level model %%%%%%%%%%%%%%%%%%%%%%%%%%%%%%%%%%%%%%%
\section{Dispersive JC model}
\label{App:TLM}
In this appendix, we provide the derivation of an effective RIP Hamiltonian based on the dispersive JC model introduced in Ref.~\cite{Puri_High-Fidelity_2016}.

The starting Hamiltonian for the dispersive JC model reads [see Fig.~\ref{fig:Model-RIPDifferentModels}(c)] 
\begin{subequations}
\begin{align}
&\HO_s=\sum\limits_{j=a,b}\frac{\omega_{j}}{2}\hat{\sigma}_j^{z}+\omega_c \hat{c}^{\dag}\hat{c}+\sum\limits_{j=a,b} \chi_{jc}\hat{c}^{\dag}\hat{c}\hat{\sigma}_j^z \;,
\label{Eq:TLM-Def of Hs}\\
&\HO_d(t)=\frac{1}{2}\left[\Omega_c^*(t)\hat{c}e^{i\omega_d t}+\Omega_c(t)\hat{c}^{\dag}e^{-i\omega_d t}\right] \;.
\label{Eq:TLM-Def of Hd(t)}
\end{align}
\end{subequations}
Moving to the rotating frame of the drive, via unitary transformation $\hat{U}_{\text{rf}}(t) \equiv \exp [-i\omega_d (\hat{c}^{\dag}\hat{c}+\sum_j\hat{\sigma}^{z}_j/2)t]$, Hamiltonian~(\ref{Eq:TLM-Def of Hs})--(\ref{Eq:TLM-Def of Hd(t)}) can be rewritten as
\begin{subequations}
\begin{align}
&\HO_{s,\text{rf}}=\sum\limits_{j=a,b}\frac{\Delta_{jd}}{2}\hat{\sigma}_j^{z}+\Delta_{cd} \hat{c}^{\dag}\hat{c}+\sum\limits_{j=a,b} \chi_{jc}\hat{c}^{\dag}\hat{c}\hat{\sigma}_j^z \;,
\label{Eq:TLM-Def of Hs rotframe}\\
&\HO_{d,\text{rf}}(t)=\frac{1}{2}\left[\Omega_c^*(t)\hat{c}+\Omega_c(t)\hat{c}^{\dag}\right] \;.
\label{Eq:TLM-Def of Hd(t) rotframe}
\end{align}
\end{subequations}

We next apply a displacement transformation in order to separate the coherent part of the field that behaves \textit{classically} from the quantum fluctuations. We take the following Ansatz
\begin{subequations}
\begin{align}
\hat{D}[\hat{\eta}_c(t)]\equiv e^{\eta_c(t)\hat{c}^{\dag}-\eta_c^*(t)\hat{c}} \;,
\label{Eq:TLM-Def of D[eta_c]}
\end{align}
where $\eta_c(t)$ is the coherent displacement of the resonator mode as
\begin{align}
&\hat{D}^{\dag}[\eta_c(t)]\hat{c}\hat{D}[\eta_c(t)]=\hat{c}+\eta_c(t) \;,
\label{Eq:TLM-D^d c D}\\
&\hat{D}^{\dag}[\eta_c(t)]\hat{c}^{\dag}\hat{D}[\eta_c(t)]=\hat{c}^{\dag}+\eta_c^*(t) \;.
\label{Eq:TLM-D^d c^d D}
\end{align}
Since the displacement is time-dependent, we also need to account for the transformation of the energy operator as
\begin{align}
&\hat{D}^{\dag}[\eta_c(t)](-i\partial_t)\hat{D}[\eta_c(t)]= -i\dot{\eta}_c(t)\hat{c}^{\dag}+i\dot{\eta}_c^*(t)\hat{c}\;.
\label{Eq:TLM-D^d (-i d_t) D}
\end{align}
\end{subequations}

Therefore, the displaced-frame Hamiltonian is obtained as
\begin{align}
\HO_{s,\text{dis}}(t)\equiv \hat{D}^{\dag}[\eta_c(t)]\left[\HO_{s,\text{rf}}+\HO_{d,\text{rf}}(t)-i\partial_t\right]\hat{D}[\eta_c(t)] \;,
\label{Eq:TLM-D^d (Hs+Hd-i d_t) D}
\end{align}
which by using Eqs.~(\ref{Eq:TLM-D^d c D})--(\ref{Eq:TLM-D^d (-i d_t) D}) can be expanded as
\begin{align}
\begin{split}
\HO_{s,\text{dis}}(t)&=\sum\limits_{j=a,b}\frac{\Delta_{jd}}{2}\hat{\sigma}_j^{z}+\Delta_{cd} [\hat{c}^{\dag}+\eta_c^*(t)][\hat{c}+\eta_c(t)]\\
&+\sum\limits_{j=a,b}\chi_{jc}[\hat{c}^{\dag}+\eta_c^*(t)][\hat{c}+\eta_c(t)]\hat{\sigma}_j^{z}\\
&+\frac{1}{2}[\Omega_c^*(t)\hat{c}+\Omega_c(t)\hat{c}^{\dag}+\eta_c(t)+\eta_c^*(t)]\\
&+[-i\dot{\eta}_c(t)\hat{c}^{\dag}+i\dot{\eta}_c^*(t)\hat{c}] \;.
\end{split}
\label{Eq:TLM-displaced Hs(t) 1}
\end{align}
Next, $\eta_c(t)$ is chosen such that the coefficients of terms linear in $\hat{c}$ and $\hat{c}^{\dag}$ in Eq.~(\ref{Eq:TLM-displaced Hs(t) 1}) become zero resulting in
\begin{align}
\dot{\eta}_c(t)+i\Delta_{cd}\eta_c(t)=-\frac{i}{2}\Omega_c(t),
\label{Eq:TLM-Cond for eta_c(t)}
\end{align}
which is the response of a classical harmonic oscillator to the pulse amplitude $\Omega_c(t)$. Under condition~(\ref{Eq:TLM-Cond for eta_c(t)}), the displaced Hamiltonian takes the form	
\begin{align}
\begin{split}
\HO_{s,\text{dis}}(t)&=\sum\limits_{j=a,b}\frac{1}{2}\Delta_{jd}\hat{\sigma}_j^z+\Delta_{cd}\hat{c}^{\dag}\hat{c}+\sum\limits_{j=a,b}\chi_{jc}\hat{c}^{\dag}\hat{c}\hat{\sigma}_j^z\\
&+\sum\limits_{j=a,b}\chi_{jc}|\eta_c(t)|^2\hat{\sigma}_j^z\\
&+\sum\limits_{j=a,b}\chi_{jc}\left[\eta_c^*(t)\hat{c}+\eta_c(t)\hat{c}^{\dag}\right]\hat{\sigma}_j^z \;.
\end{split}
\label{Eq:TLM-displaced Hs(t) 2}
\end{align}
Equation~(\ref{Eq:TLM-displaced Hs(t) 2}) exhibits a dynamic frequency shift for each qubit (2$\chi$ per photon) as well as off-diagonal interaction terms of the form $\chi_{jc}\eta_c^*(t)\hat{c}\hat{\sigma}_j^z+\text{H.c.}$ for $j=a, b$.

Employing time-dependent SWPT, we obtain an effective \textit{diagonal} Hamiltonian starting from Hamiltonian~(\ref{Eq:TLM-displaced Hs(t) 2}). We first separate the terms into zeroth order
\begin{subequations}
\begin{align}
&\HO_0 \equiv \sum\limits_{j=a,b}\frac{1}{2}\Delta_{jd}\hat{\sigma}_j^z+\Delta_{cd}\hat{c}^{\dag}\hat{c} \;, 
\label{Eq:TLM-Def of H0}
\end{align}
and interaction parts 
\begin{align}
\begin{split}
\HO_{\text{int}}(t) &\equiv \sum\limits_{j=a,b}\chi_{jc}\left[\hat{c}^{\dag}\hat{c}+|\eta_c(t)|^2\right]\hat{\sigma}_j^z\\
&+\sum\limits_{j=a,b}\chi_{jc}\left[\eta_c^*(t)\hat{c}+\eta_c(t)\hat{c}^{\dag}\right]\hat{\sigma}_j^z \;.
\end{split}
\label{Eq:TLM-Def of Hint(t)}
\end{align}
\end{subequations}
Note that $\sum_j\chi_{jc}\hat{c}^{\dag}\hat{c}\sigma_j^{z}$ [the 1st term in Eq.~(\ref{Eq:TLM-Def of Hint(t)})] is time-independent and diagonal, hence should in principle be kept in $\HO_0$ (see Appendix~\ref{App:KM}). Here, however, we follow the same level of precision as in Ref.~\cite{Puri_High-Fidelity_2016} as a point of comparison. This is valid when $|\Delta_{cd}| \gg |\chi_{jc}|$. To simplify perturbation theory, it is helpful to move to the interaction frame with respect to $\HO_0$ as
\begin{align}
\begin{split}
\HO_{I}(t) &\equiv e^{i\HO_0 t}\HO_{\text{int}}(t)e^{-i\HO_0 t}\\
&= \sum\limits_{j=a,b}\chi_{jc}\left[\hat{c}^{\dag}\hat{c}+|\eta_c(t)|^2\right]\hat{\sigma}_j^z\\
&+\sum\limits_{j=a,b}\left[\chi_{jc}\eta_c^*(t)e^{-i\Delta_{cd} t}\hat{c}\hat{\sigma}_j^z+\text{H.c.}\right]\;.	
\end{split}
\label{Eq:TLM-Def of H_I(t)}
\end{align}
According to Eq.~(\ref{Eq:TLM-Def of H_I(t)}), the perturbative method is valid when $|\chi_{jc} \eta_c(t)| < \Delta_{cd}$ for $j=a,b$.

We then apply time-dependent SWPT as
\begin{align}
\HO_{I,\text{eff}}(t) \equiv e^{i\hat{G}(t)}[\HO_{I}(t)-i\partial_t]e^{-i\hat{G}(t)} \;,
\label{Eq:TLM-SW Trans}
\end{align}
where $\HO_{I,\text{eff}}(t)$ is the effective Hamiltonian in the interaction frame and $\hat{G}(t)$ is the SW generator that needs to be determined such that the effective Hamiltonian becomes diagonal. Expanding $\hat{G}(t)$ and $\HO_{I,\text{eff}}(t)$ and collecting equal powers in perturbation, a set of perturbative operator-valued ODEs can be obtained (see Appendix C of Ref.~\cite{Malekakhlagh_First-Principles_2020}). Up to the lowest order, we find
\begin{subequations}
\begin{align}
&\HO_{I,\text{eff}}^{(1)}(t)=\mathcal{S}\left(\HO_{I}(t)\right) \;,
\label{Eq:TLM-H_I,eff^(1) Cond}\\
&\dot{\hat{G}}_1(t)=\mathcal{N}\left(\HO_{I}(t)\right) \;,
\label{Eq:TLM-G1 Cond}
\end{align}
\end{subequations}
where $\mathcal{S}(\bullet)$ and $\mathcal{N}(\bullet)$ refer to the diagonal and off-diagonal parts of an operator. The result for the second order perturbation reads
\begin{subequations}
\begin{align}
&\HO_{I,\text{eff}}^{(2)}(t)=S\Big(i[\hat{G}_1(t),\HO_I(t)]-\frac{i}{2}[\hat{G}_1(t),\dot{\hat{G}}_1(t)]\Big) \;,
\label{Eq:TLM-H_I,eff^(2) Cond}\\
&\dot{\hat{G}}_2(t)= \mathcal{N}\Big(i[\hat{G}_1(t),\HO_I(t)]-\frac{i}{2}[\hat{G}_1(t),\dot{\hat{G}}_1(t)]\Big) \;.
\label{Eq:TLM-G2 Cond}
\end{align}
\end{subequations}
At each order in perturbation, we separate the contributions that are diagonal and remove the rest by solving for the generator at that order. The off-diagonal terms can produce diagonal terms via nested commutators at higher order.

Replacing Eq.~(\ref{Eq:TLM-Def of H_I(t)}) into the first-order Eqs.~(\ref{Eq:TLM-H_I,eff^(1) Cond})--(\ref{Eq:TLM-G1 Cond}), we find
\begin{subequations}
\begin{align}
&\HO_{I,\text{eff}}^{(1)}=\sum\limits_{j=a,b}\chi_{jc}\left[\hat{c}^{\dag}\hat{c}+|\eta_c(t)|^2\right]\hat{\sigma}_j^z \;,
\label{Eq:TLM-H_I,eff^(1) Sol}\\
&\hat{G}_1(t)=\sum\limits_{j=a,b}\int^{t} dt'\left[\chi_{jc}\eta_c^*(t')e^{-i\Delta_{cd}t'}\hat{c}\hat{\sigma}_j^z+\text{H.c.}\right] \;.
\label{Eq:TLM-G1 Sol}
\end{align}
\end{subequations}
Based on Eq.~(\ref{Eq:TLM-H_I,eff^(1) Sol}), each qubit adopts a dynamic frequency shift equal to $2\chi$ per resonator photon. Furthermore, the off-diagonal terms are removed via a non-zero $\hat{G}_1(t)$ according to Eq.~(\ref{Eq:TLM-G1 Sol}). Inserting the solution~(\ref{Eq:TLM-G1 Sol}) for $\hat{G}_1(t)$ into Eq.~(\ref{Eq:TLM-H_I,eff^(2) Cond}) and further simplification results in
\begin{subequations}
\begin{align} 
&\HO_{I,\text{eff}}^{(2)}(t)=-2\chi_{ac}\chi_{bc}\mathcal{A}_{\eta}(t)\hat{\sigma}_a^z\hat{\sigma}_b^z \;, 
\label{Eq:TLM-H_I,eff^(2) Sol}\\
&\mathcal{A}_{\eta}(t)\equiv \Im\left\{\int^{t}dt'\eta_c(t)\eta_c^*(t')e^{i\Delta_{cd}(t-t')}\right\} \;,
\label{Eq:TLM-Def of A_eta(t)}
\end{align}
\end{subequations}
where $\mathcal{A}_{\eta}(t)$ captures the time dependence of the effective qubit-qubit interaction due to the pulse shape.

In the case where the resonator response is adiabatic, i.e. $\Delta_{cd}\gg 1/\tau$ where $\tau$ is the pulse (rise) time, we can apply an adiabatic expansion to $\mathcal{A}_{\eta}(t)$ via integration by parts as  
\begin{subequations}
\begin{align}
\mathcal{A}_{\eta}(t) = \frac{|\eta_c(t)|^2}{\Delta_{cd}}+\frac{\Im\left\{\eta_c(t)\dot{\eta}_c^*(t)\right\}}{\Delta_{cd}^2}+O\left(\frac{\eta_c(t)\ddot{\eta}^*_c(t)}{\Delta_{cd}^3}\right)\;,
\label{Eq:TLM-A_eta Sol ConstDr}
\end{align}
with the first and the second terms denoting the dominant \textit{dynamic} and the relatively smaller \textit{geometric} contributions. Keeping the dynamic corrections recovers the known effective RIP interaction as
\begin{align}
\HO_{I,\text{eff}}^{(2)}(t)\approx-\frac{2\chi_{ac}\chi_{bc}|\eta_c(t)|^2}{\Delta_{cd}}\hat{\sigma}_a^z\hat{\sigma}_b^z \;.
\label{Eq:TLM-H_I,eff Sol ConstDr}
\end{align}
\end{subequations}
Equation~(\ref{Eq:TLM-H_I,eff Sol ConstDr}) is consistent with Ref.~\cite{Puri_High-Fidelity_2016} where $|\eta_c(t)|^2$ is the average number of resonator photons at time $t$.

In summary, Eqs.~(\ref{Eq:TLM-H_I,eff^(1) Sol}),~(\ref{Eq:TLM-H_I,eff^(2) Sol}) and~(\ref{Eq:TLM-Def of A_eta(t)}) are the main results of this section that provide the effective Hamiltonian for the RIP gate in terms of an arbitrary input pulse shape $\Omega_c(t)$. In the following appendices, we extend our method to multilevel models. First, in Appendix~\ref{App:KM}, we follow a \text{phenomenological} Kerr model similar to Ref.~\cite{Cross_Optimized_2015}. The rest of the appendices repeat the same procedure based on Josephson nonlinearity.
%%%%%%%%%%%%%%%%%%%%%%%%%%%%%%%%%%%%%%%%%%%%%%%%%%%%%%%%%%%%%%%%%%%%%%%%%%%%%%%%

%%%%%%%%%%%%%%%%%%%%%%%%%%%%%%%%%% App: Kerr Model %%%%%%%%%%%%%%%%%%%%%%%%%%%%%%
\section{Kerr Model}
\label{App:KM}
Here we consider a multilevel Kerr model for each qubit and repeat the calculation outlined in Appendix~\ref{App:TLM}. The bosonic algebra lays the groundwork for the ab-initio theory based on Josephson nonlinearity in the remaining appendices.

Consider the following starting Hamiltonian for the RIP gate
\begin{subequations}
\begin{align}
\begin{split}
\HO_s &=\sum\limits_{j=a,b}\left[\omega_j \hat{n}_j+\frac{\alpha_j}{2}\hat{n}_j(\hat{n}_j-1)\right]\\
&+\omega_c\hat{n}_c+\sum\limits_{j,k=a,b,c \atop j>k}2\chi_{jk}\hat{n}_j\hat{n}_k \;,
\label{KM-Def of Hs}
\end{split}\\
\HO_d(t)&=\frac{1}{2}\left[\Omega_c^*(t)\hat{c}e^{i\omega_d t}+\Omega_c(t)\hat{c}^{\dag}e^{-i\omega_d t}\right] \;,
\label{Eq:KM-Def of Hd(t)}
\end{align}
\end{subequations}
where the qubits are modeled as Kerr oscillators with anharmonicity $\alpha_{a,b}$. Note the additional factor of 2 in the Kerr interaction, which can be understood in terms of two-level mapping of the number operator as $\hat{n}\approx\ket{1}\bra{1}=(\hat{I}+\hat{\sigma}^{z})/2$. Following the procedure in Appendix~\ref{App:TLM}, we obtain a diagonal effective Hamiltonian for the RIP gate. 	

Upon a displacement transformation, and in the rotating-frame of the drive, we obtain
\begin{align}
\begin{split}
\HO_{s,\text{dis}}(t)&=\sum\limits_{j=a,b}\left[\Delta_{jd} \hat{n}_j+\frac{\alpha_j}{2}\hat{n}_j(\hat{n}_j-1)\right]\\
&+\Delta_{cd}\hat{n}_c+\sum\limits_{j,k=a,b,c \atop j>k}2\chi_{jk}\hat{n}_j\hat{n}_k\\
&+\sum\limits_{j=a,b}2\chi_{jc}|\eta_c(t)|^2\hat{n}_j\\
&+\sum\limits_{j=a,b}2\chi_{jc}\left[\eta_c^*(t)\hat{c}+\eta_c(t)\hat{c}^{\dag}\right]\hat{n}_j \;,
\end{split}
\label{Eq:KM-displaced Hs(t)}
\end{align}
where the off-diagonal terms $2\chi_{jc}\eta_c^*(t)\hat{n}_j\hat{c}+\text{H.c.}$, $j=a, b$ are responsible for the lowest order effective RIP interaction. Moreover, the condition for the coherent displacement $\eta_c(t)$ is the same as Eq.~(\ref{Eq:TLM-Cond for eta_c(t)}) of the dispersive JC derivation.

We next split the contributions in Eq.~(\ref{Eq:KM-displaced Hs(t)}) in terms of the zeroth-order 
\begin{subequations}
\begin{align}
\begin{split}
\HO_0&=\sum\limits_{j=a,b}\left[\Delta_{jd} \hat{n}_j+\frac{\alpha_j}{2}\hat{n}_j(\hat{n}_j-1)\right]\\
&+\Delta_{cd}\hat{n}_c+\sum\limits_{j,k=a,b,c \atop j>k}2\chi_{jk}\hat{n}_j\hat{n}_k \;,
\end{split}
\label{Eq:KM-Def of H0}
\end{align}
and interaction part as
\begin{align}
\begin{split}
\HO_{\text{int}}(t)&=\sum\limits_{j=a,b}2\chi_{jc}|\eta_c(t)|^2\hat{n}_j\\
&+\sum\limits_{j=a,b}2\chi_{jc}\left[\eta_c^*(t)\hat{c}+\eta_c(t)\hat{c}^{\dag}\right]\hat{n}_j \;,
\end{split}
\label{Eq:KM-Def of H_int(t)}
\end{align}
\end{subequations}
where all time-independent diagonal terms are kept in $\HO_0$ and the rest in $\HO_{\text{int}}(t)$. We then move to the interaction frame with respect to $\HO_0$. The interaction frame transformation of the resonator creation and annihilation operators reads 
\begin{subequations}
\begin{align}
&e^{i\HO_0 t}\hat{c}e^{-i\HO_0 t}=e^{-i\hat{\Delta}_{cd} t} \hat{c} \;,
\label{Eq:KM-U^d*c*U}\\
&e^{i\HO_0 t}\hat{c}^{\dag}e^{-i\HO_0 t}= \hat{c}^{\dag}e^{i\hat{\Delta}_{cd} t} \;,
\label{Eq:KM-U^d*c^d*U}
\end{align}
where $\hat{\Delta}_{cd}$ is the operator-valued resonator-drive detuning
\begin{align}
\hat{\Delta}_{cd} \equiv \Delta_{cd}+2\chi_{ac}\hat{n}_a+2\chi_{bc}\hat{n}_b \;,
\label{Eq:KM-Def of hat(Del)_cd}
\end{align}
accounting for number-splitting of distinct qubit states. Equations~(\ref{Eq:KM-U^d*c*U})--(\ref{Eq:KM-Def of hat(Del)_cd}) are found using identities 
\begin{align}
&f(\hat{n}_c)\hat{c}=\hat{c}f(\hat{n}_c-1) \;,
\label{Eq:KM-f(nc)c=cf(nc-1)}\\
&f(\hat{n}_c)\hat{c}^{\dag}=\hat{c}^{\dag}f(\hat{n}_c+1) \;,
\label{Eq:KM-f(nc)c^d=c^df(nc+1)}
\end{align}
on commutation of creation/annihilation operators with an arbitrary function of the number operator. Employing Eqs.~(\ref{Eq:KM-U^d*c*U})--(\ref{Eq:KM-Def of hat(Del)_cd}), the interaction frame Hamiltonian is found as
\begin{align}
\begin{split}
&\HO_I(t) \equiv e^{i\HO_0 t}\HO_{\text{int}}(t)e^{-i\HO_0 t} \\
&=\sum\limits_{j=a,b}2\chi_{jc}|\eta_c(t)|^2\hat{n}_j\\
&+\sum\limits_{j=a,b}2\chi_{jc}\left[\eta_c^*(t)e^{-i\hat{\Delta}_{cd} t}\hat{c}\hat{n}_j+\text{H.c.}\right] \;.
\end{split}
\label{Eq:KM-Def of H_I(t)}
\end{align}
\end{subequations}

Following Eqs.~(\ref{Eq:TLM-H_I,eff^(1) Cond})--(\ref{Eq:TLM-G2 Cond}), up to the first order, we find 
\begin{subequations}
\begin{align}
&\HO_{I,\text{eff}}^{(1)}=\sum\limits_{j=a,b}2\chi_{jc}|\eta_c(t)|^2\hat{n}_j \;,
\label{Eq:KM-H_I,eff^(1) Sol}\\
&\hat{G}_1(t)=\sum\limits_{j=a,b}\int^{t} dt'\left[2\chi_{jc}\eta_c^*(t')e^{-i\hat{\Delta}_{cd} t'}\hat{c}\hat{n}_j+\text{H.c.}\right] \;.
\label{Eq:KM-G1 Sol}
\end{align}
\end{subequations}
Replacing solution~(\ref{Eq:KM-G1 Sol}) for $\hat{G}_1(t)$ into expression~(\ref{Eq:TLM-H_I,eff^(2) Cond}) results in
\begin{subequations}
\begin{align}
\begin{split}
\HO_{I,\text{eff}}^{(2)}(t)=&-8\chi_{ac}\chi_{bc}\hat{\mathcal{A}}_{\eta}(t)\hat{n}_a\hat{n}_b\\
&-4\chi_{ac}^2\hat{\mathcal{A}}_{\eta}(t)\hat{n}_a^2\\ 
&-4\chi_{bc}^2\hat{\mathcal{A}}_{\eta}(t)\hat{n}_b^2 \;,\\ 
\end{split}
\label{Eq:KM-H_I,eff^(2) Sol}
\end{align}
where $\hat{\mathcal{A}}_{\eta}(t)$ is the operator generalization of Eq.~(\ref{Eq:TLM-Def of A_eta(t)}) as
\begin{align}
\hat{\mathcal{A}}_{\eta}(t) \equiv \Im\left\{\int^{t}dt'\eta_c(t)\eta_c^*(t')e^{i\hat{\Delta}_{cd}(t-t')}\right\} \;. 
\label{Eq:KM-Def of A_eta(na,nb)}
\end{align}
\end{subequations}
Besides the intended RIP interaction, one finds a dynamic anharmonic shift given by the last two lines of Eq.~(\ref{Eq:KM-H_I,eff^(2) Sol}). Such a correction is irrelevant in the two-level description since $(\hat{\sigma}_j^z)^2=\hat{I}_j$ for $j=a,b$. 

In the adiabatic limit, $\hat{\mathcal{A}}_{\eta}(t)$ simplifies to
\begin{align}
\begin{split}
\hat{\mathcal{A}}_{\eta} (t)=\frac{|\eta_c(t)|^2}{\hat{\Delta}_{cd}}+\frac{\Im\left\{\eta_c(t)\dot{\eta}_c^*(t)\right\}}{\hat{\Delta}_{cd}^2}+O\left(\frac{\eta_c(t)\ddot{\eta}_c^*(t)}{\hat{\Delta}_{cd}^3}\right), 
\end{split}
\label{Eq:KM-A_eta Sol ConstDr}
\end{align}
where the integration by parts holds as in Eq.~(\ref{Eq:TLM-A_eta Sol ConstDr}) given that $\hat{\Delta}_{cd}$ is a diagonal operator. Keeping the dynamic contribution in Eq.~(\ref{Eq:KM-A_eta Sol ConstDr}) and replacing it into the solution~(\ref{Eq:KM-H_I,eff^(2) Sol}) for $\HO_{I,\text{eff}}^{(2)}(t)$ we find
\begin{align} 
\begin{split}
\HO_{I,\text{eff}}^{(2)}(t) 	\approx &-\frac{8\chi_{ac}\chi_{bc}\hat{n}_a\hat{n}_b}{\Delta_{cd}+2\chi_{ac}\hat{n}_a+2\chi_{bc}\hat{n}_b}|\eta_c(t)|^2\\
&-\frac{4\chi_{ac}^2\hat{n}_a^2}{\Delta_{cd}+2\chi_{ac}\hat{n}_a+2\chi_{bc}\hat{n}_b}|\eta_c(t)|^2\\
&-\frac{4\chi_{bc}^2\hat{n}_b^2}{\Delta_{cd}+2\chi_{ac}\hat{n}_a+2\chi_{bc}\hat{n}_b}|\eta_c(t)|^2 \;.\\
\end{split}
\label{Eq:KM-H_I,eff^(2) AdiabSol}
\end{align}
%%%%%%%%%%%%%%%%%%%%%%%%%%%%%%%%%%%%%%%%%%%%%%%%%%%%%%%%%%%%%%%%%%%%%%%%%%%%%%%%%%%%%%%%%%%%

In summary, the effective RIP Hamiltonian~(\ref{Eq:KM-H_I,eff^(2) Sol}) along with the operator-valued drive contribution $\hat{A}_{\eta}(t)$ in Eq.~(\ref{Eq:KM-Def of A_eta(na,nb)}) are the main result of this appendix. Projecting the adiabatic Hamiltonian~(\ref{Eq:KM-H_I,eff^(2) AdiabSol}) onto the computational basis results in expressions for gate parameters that agree with Refs.~\cite{Cross_Optimized_2015, Paik_Experimental_2016} [see Eqs.~(\ref{eqn:EffHam-w_iz^(2)})--(\ref{eqn:EffHam-w_zz^(2)}) of the main text].

%%%%%%%%%%%%%%%%%%%% App: Derivation of the dispersive Hamiltonian %%%%%%%%%%%%%%%%%%%%%%%%%%%%%%%%%%%%%%%%%%%%%%
\section{Normal mode Hamiltonian}
\label{App:NormModeHam}

Here, starting from the Josephson nonlinearity, we provide the transformation from the bare to the normal mode frames, in which the harmonic part of the system Hamiltonian~(\ref{eqn:Model-Starting Hs}) becomes diagonal. The resulting hybridization coefficients determine the strength of various possible nonlinear mixing terms between the normal modes. The procedure outlined here follows the same logic of normal mode expansion as in the black-box quantization \cite{Nigg_BlackBox_2012} and energy-participation ratio \cite{Minev_EPR_2020} methods but assumes a \textit{given} Hamiltonian.

We begin by expanding the Josephson potential in powers of the weak anharmonicity measure $\epsilon_j$ as
\begin{align}
\begin{split}
\frac{\bar{\omega}_j}{4}\left[\hat{\bar{y}}_j^2-\frac{2}{\epsilon_j}\cos(\sqrt{\epsilon_j}\hat{\bar{x}}_j)\right]=\frac{\bar{\omega}_j}{4}\left(\hat{\bar{x}}_j^2+\hat{\bar{y}}_j^2\right)\\
+\frac{\bar{\omega}_j}{2}\sum\limits_{n=2}^{\infty}(-\epsilon_j)^{n-1}\frac{\hat{\bar{x}}_j^{2n}}{(2n)!} \;, \quad j=a,b \;,
\end{split}
\label{Eq:NormModeHam-Hj Exp in eps} 
\end{align}
where we assumed zero gate charge for simplicity. Our goal is to find the normal modes of the quadratic (harmonic) Hamiltonian
\begin{align}
\HO_{2}=\sum\limits_{k=a,b,c}\frac{\bar{\omega}_k}{4}\left(\hat{\bar{x}}_k^{2}+\hat{\bar{y}}_k^{2}\right)+\sum\limits_{j=a,b}g_j\hat{\bar{y}}_j\hat{\bar{y}}_c \;.
\label{Eq:NormModeHam-Def of H2} 
\end{align}
Defining a quadrature vector $\hat{\bar{\mathbf{Q}}}^{\dag}\equiv [\hat{\bar{\mathbf{X}}}^{\dag} \ \hat{\bar{\mathbf{Y}}}^{\dag}]$, with flux (inductive) and charge (capacitive) components, Eq.~(\ref{Eq:NormModeHam-Def of H2}) can be expressed as
\begin{subequations}
\begin{align}
\HO_2 =\frac{1}{4}\hat{\bar{\mathbf{Q}}}^{\dag}\bar{\mathbf{H}}_2\hat{\bar{\mathbf{Q}}} \;.
\label{Eq:NormModeHam-QuadRep of H2} 
\end{align}
In Eq.~(\ref{Eq:NormModeHam-QuadRep of H2}), $\bar{\mathbf{H}}_2$ is the matrix representation of the quadratic Hamiltonian 
\begin{align}
\bar{\mathbf{H}}_{2}=
\begin{bmatrix}
\bar{\mathbf{H}}_{X} & \mathbf{0}\\
\mathbf{0} & \bar{\mathbf{H}}_{Y}\\
\end{bmatrix}\;,
\label{Eq:NormModeHam-H2 Mat} 
\end{align}
with $\bar{\mathbf{H}}_{X}$ and $\bar{\mathbf{H}}_{Y}$ denoting the charge and flux subspaces as
\begin{align}
&\bar{\mathbf{H}}_{X}=
\begin{bmatrix}
\bar{\omega}_a & 0 & 0\\
0 & \bar{\omega}_b & 0\\
0 & 0 & \bar{\omega}_c
\end{bmatrix} \;,
\label{Eq:NormModeHam-H2,X Mat}\\
&\bar{\mathbf{H}}_{Y}=
\begin{bmatrix}
\bar{\omega}_a & 0 & 2g_a\\
0 & \bar{\omega}_b & 2g_b\\
2g_a & 2g_b & \bar{\omega}_c
\end{bmatrix} \;.
\label{Eq:NormModeHam-H2,Y Mat}
\end{align}
\end{subequations}

We then write $\hat{\bar{\mathbf{Q}}}\equiv \mathbf{C} \hat{\mathbf{Q}}$, where $\mathbf{C}$ is a canonical transformation to the normal mode frame shown without a bar. Since the couplings are capacitive in nature, and hence there is no flux-charge mixing, we take the following Ansatz for $\mathbf{C}$ 
\begin{align}
\mathbf{C}=
\begin{bmatrix}
\mathbf{U} & \mathbf{0} \\
\mathbf{0} & \mathbf{V}
\end{bmatrix}\;,
\label{Eq:NormModeHam-Canonical Trans}
\end{align}
with $\mathbf{U}$ and $\mathbf{V}$ being the transformations in the flux and charge subspaces, respectively. To conserve commutation relations, the canonical transformation must satisfy the symplectic condition \cite{DeGosson_Symplectic_2006, DeGosson_Symplectic_2011}
\begin{align}
\mathbf{C}^{T}\mathbf{J}\mathbf{C}=\mathbf{J} \;,
\label{Eq:NormModeHam-Symplectic Cond 1}
\end{align}
where $\mathbf{J}$ is the symplectic matrix
\begin{align}
\mathbf{J}=
\begin{bmatrix}
\mathbf{0} & \mathbf{I}\\
-\mathbf{I} & \mathbf{0}
\end{bmatrix} \;.
\label{Eq:NormModeHam-Def of J}
\end{align}
Replacing Ansatz~(\ref{Eq:NormModeHam-Canonical Trans}) into Eq.~(\ref{Eq:NormModeHam-Symplectic Cond 1}), we find 
\begin{align}
\mathbf{U}^{T}\mathbf{V}=\mathbf{I} \;.
\label{Eq:NormModeHam-Symplectic Cond 2}
\end{align}
In practice, a canonical transformation can be decomposed in terms of a set of consecutive \textit{scaling} and \textit{orthogonal} transformations. In the following, we implement this procedure for the RIP gate, which consists of three capacitively coupled harmonic oscillators.

\textbf{(i) Scaling transformation}--- We first scale the quadratures such that the Hamiltonian becomes proportional to the identity matrix in the flux subspace. This scaling allows for the removal of off-diagonal elements in the charge subspace without effecting the flux subspace. We consider the following scaling transformation:
\begin{align}
\begin{bmatrix}
\hat{\bar{X}}\\
\hat{\bar{Y}}
\end{bmatrix}=
\begin{bmatrix}
\mathbf{S}_X & 0\\
0 & \mathbf{S}_X^{-1}
\end{bmatrix}
\begin{bmatrix}
\hat{X}'\\
\hat{Y}'
\end{bmatrix} \;,
\label{Eq:NormModeHam-Def [X' Y']}
\end{align}
where we assume a diagonal scaling matrix of the form
\begin{align}
\mathbf{S}_X=
\begin{bmatrix}
s_{xa} & 0 & 0\\
0 & s_{xb} & 0\\
0 & 0 & s_{xc}\\
\end{bmatrix} \;.
\label{Eq:NormModeHam-Def of Sx Mat}	
\end{align}
In writing Eq.~(\ref{Eq:NormModeHam-Def [X' Y']}), we have employed the symplectic condition~(\ref{Eq:NormModeHam-Symplectic Cond 2}) to simplify the scaling matrix as $\mathbf{S}_Y=(\mathbf{S}_X^{T})^{-1}=\mathbf{S}_X^{-1}$. In the new basis, the harmonic Hamiltonian takes the form
\begin{align}
\mathbf{H}_2'=
\begin{bmatrix}
\mathbf{S}_X\bar{\mathbf{H}}_{X}\mathbf{S}_X & 0\\
0 & \mathbf{S}_X^{-1}\bar{\mathbf{H}}_{Y}\mathbf{S}_X^{-1} 
\end{bmatrix}\;.
\label{Eq:NormModeHam-H'2}	
\end{align}
We then impose the following condition for the elements of the scaling matrix 
\begin{align}
s_{xa}^2\bar{\omega}_a=s_{xb}^2\bar{\omega}_b=s_{xc}^2\bar{\omega}_c=(\bar{\omega}_a\bar{\omega}_b\bar{\omega}_c)^{1/3}\equiv \omega'_{abc} \;,
\label{Eq:NormModeHam-PropConst Cond}	
\end{align}
in order to make the harmonic Hamiltonian in the flux subspace proportional to the identity matrix. The proportionality constant is set to be the geometric mean of the three oscillator frequencies as $\omega'_{abc}\equiv(\bar{\omega}_a\bar{\omega}_b\bar{\omega}_c)^{1/3}$ for simplicity; the choice is irrelevant in the determination of the \textit{overall} canonical transformation. Consequently, the flux and charge subspaces of the Hamiltonian in the new basis are obtained as
\begin{subequations}
\begin{align}
&\mathbf{H}_X'\equiv \mathbf{S}_X\bar{\mathbf{H}}_{X}\mathbf{S}_X=
\begin{bmatrix}
\omega_{abc}' & 0 & 0 \\
0 & \omega_{abc}' & 0 \\
0 & 0 & \omega_{abc}' 
\end{bmatrix}\;,
\label{Eq:NormModeHam-H'_X}\\
&\mathbf{H}_Y'\equiv \mathbf{S}_X^{-1}\bar{\mathbf{H}}_{Y}\mathbf{S}_X^{-1}=
\begin{bmatrix}
\omega_a' & 0 & 2g_a' \\
0 & \omega_b' & 2g_b' \\
2g_a' & 2g_b' & \omega_c' 
\end{bmatrix} \;,
\label{Eq:NormModeHam-H'_Y}
\end{align}
\end{subequations}
where the prime parameters in the charge subspace are defined as
\begin{align}
&\omega_j'\equiv s_{xj}^{-1}\bar{\omega}_j s_{xj}^{-1}, \quad j=a,b,c \;,
\label{Eq:NormModeHam-Def of w'_j}\\
&g_a'\equiv s_{xa}^{-1}g_a s_{xc}^{-1} \;,
\label{Eq:NormModeHam-Def of s_xa}\\
&g_b'\equiv s_{xb}^{-1}g_b s_{xc}^{-1} 
\label{Eq:NormModeHam-Def of s_ya}\;.	
\end{align}

\textbf{(ii) Orthogonal transformation}---Next, we apply an orthogonal transformation (rotation) to make matrix~(\ref{Eq:NormModeHam-H'_Y}) diagonal. We take the following Ansatz for the transformation
\begin{align}
\begin{bmatrix}
\hat{X}'\\
\hat{Y}'
\end{bmatrix}=
\begin{bmatrix}
\mathbf{O}_Y & 0\\
0 & \mathbf{O}_Y
\end{bmatrix}
\begin{bmatrix}
\hat{X}''\\
\hat{Y}''
\end{bmatrix} \;,
\label{Eq:NormModeHam-Def of [X" Y"]}
\end{align}
where $\mathbf{O}_Y$ is orthogonal and thus $\mathbf{O}_X=\left(\mathbf{O}_Y^{T}\right)^{-1}=\mathbf{O}_Y$. At this stage, the harmonic Hamiltonian reads
\begin{align}
\mathbf{H}_2''=
\begin{bmatrix}
\mathbf{O}_Y^{T}\mathbf{H}'_{X}\mathbf{O}_Y & 0\\
0 & \mathbf{O}_Y^{T}\mathbf{H}'_{Y}\mathbf{O}_Y
\end{bmatrix} \;.
\label{Eq:NormModeHam-H"2}
\end{align}
The flux subspace remains intact
\begin{align}
\mathbf{H}_X''\equiv \mathbf{O}_Y^{T}\mathbf{H}'_{X}\mathbf{O}_Y=\omega'_{abc}\mathbf{I} \;,
\label{Eq:NormModeHam-H_X"2}
\end{align}
while the charge subspace becomes diagonal as
\begin{align}
\mathbf{H}_Y''\equiv \mathbf{O}_Y^{T}\mathbf{H}'_{Y}\mathbf{O}_Y \equiv
\begin{bmatrix}
\omega''_{a} & 0 & 0\\
0 & \omega''_{b} & 0\\
0 & 0 & \omega''_{c}
\end{bmatrix} \;.
\label{Eq:NormModeHam-H_Y"2}
\end{align}

\textbf{(iii) Scaling transformation}---Lastly, we apply another scaling transformation to compensate for the differences between the diagonal entries of the flux and charge subspaces for each mode. We define the normal modes of the system according to 
\begin{align}
\begin{bmatrix}
\hat{X}''\\
\hat{Y}''
\end{bmatrix}=
\begin{bmatrix}
\mathbf{S} & 0\\
0 & \mathbf{S}^{-1}
\end{bmatrix}
\begin{bmatrix}
\hat{X}\\
\hat{Y}
\end{bmatrix} \;,
\label{Eq:NormModeHam-Def of [X Y]}
\end{align}
with the Anstaz for $\mathbf{S}$
\begin{align}
\mathbf{S}=
\begin{bmatrix}
s_a & 0 & 0 \\
0 & s_b & 0 \\
0 & 0 & s_c 
\end{bmatrix} \;.
\label{Eq:NormModeHam-Def of S}
\end{align}
At this step, the harmonic Hamiltonian reads
\begin{align}
\mathbf{H}_2=
\begin{bmatrix}
\mathbf{S}\mathbf{H}''_{X}\mathbf{S} & 0\\
0 & \mathbf{S}^{-1}\mathbf{H}''_{Y}\mathbf{S}^{-1}
\end{bmatrix} \;.
\label{Eq:NormModeHam-H2}
\end{align}
Requiring the two subspaces to have the same frequency for each mode sets the scaling parameters as
\begin{align}
s_j =\left(\frac{\omega''_j}{\omega'_{abc}}\right)^{1/4}, \quad j=a,b,c \;.
\label{Eq:NormModeHam-Cond for s_j}
\end{align}
Furthermore, the normal mode frequencies are found as
\begin{align}
\tilde{\omega}_j \equiv (\omega'_{abc}\omega''_j)^{1/2}, \quad j=a,b,c \;.
\label{Eq:NormModeHam-Def of tilde(omega)_j}
\end{align}

\textbf{Overall canonical transformation}---Putting together the three transformations, we find the hybridization matrices $\mathbf{U}$ and $\mathbf{V}$ that construct the desired canonical transformation~(\ref{Eq:NormModeHam-Canonical Trans}) as
\begin{subequations}
\begin{align} 
&\mathbf{U}=\mathbf{S}_X \mathbf{O}_Y \mathbf{S} \;,
\label{Eq:NormModeHam-U Sol}\\
&\mathbf{V}=\mathbf{S}_X^{-1} \mathbf{O}_Y \mathbf{S}^{-1} \;,
\label{Eq:NormModeHam-V Sol}
\end{align}
\end{subequations}
with $\mathbf{S}_X$, $\mathbf{O}_Y$ and $\mathbf{S}$ given in Eqs.~(\ref{Eq:NormModeHam-PropConst Cond}), (\ref{Eq:NormModeHam-H_Y"2}) and (\ref{Eq:NormModeHam-Cond for s_j}), respectively.

Employing Eqs.~(\ref{Eq:NormModeHam-U Sol})--(\ref{Eq:NormModeHam-V Sol}), we obtain a normal mode representation of the RIP gate Hamiltonian as
\begin{subequations}
\begin{align} 
\begin{split}
\HO_s &=\sum\limits_{k=a,b,c}\tilde{\omega}_k\hat{k}^{\dag}\hat{k}+\sum\limits_{j=a,b}\sum\limits_{n=2}^{\infty} \frac{\bar{\omega}_j}{2}(-\epsilon_j)^{n-1}\\
&\times \frac{\left[\left(\sum\limits_{k=a,b,c}u_{jk}\hat{k}\right)+\text{H.c.}\right]^{2n}}{(2n)!} \;,
\end{split}
\label{Eq:NormModeHam-NormMode Hs}
\end{align}
Furthermore, the RIP drive Hamiltonian transforms to
\begin{align}
\begin{split}
\HO_d(t)&=[\Omega_{cx}(t)\cos(\omega_d t)+\Omega_{cy}(t)\sin(\omega_d t)]\\
& \times i[\sum\limits_{k=a,b,c}v_{ck}\hat{k}-\text{H.c.}]	
\label{Eq:NormModeHam-NormMode Hd}
\end{split}
\end{align}
\end{subequations}	

%%%%%%%%%%%%%%% Figure: Normal mode transformation %%%%%%%%%%%%%%%%%%%%%%%%%%%%%
\begin{figure}
\centering
\includegraphics[scale=0.209]{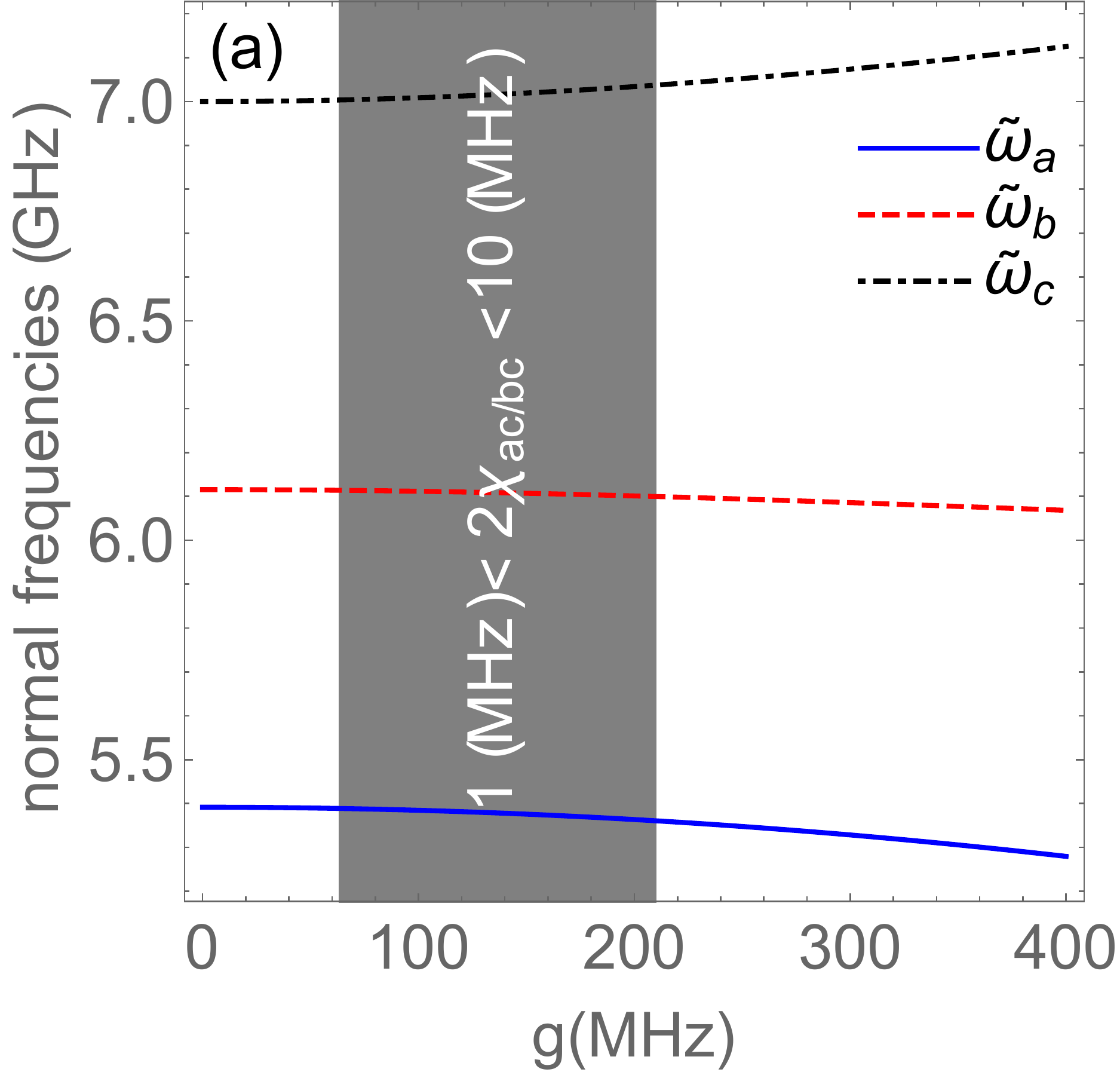}
\includegraphics[scale=0.222]{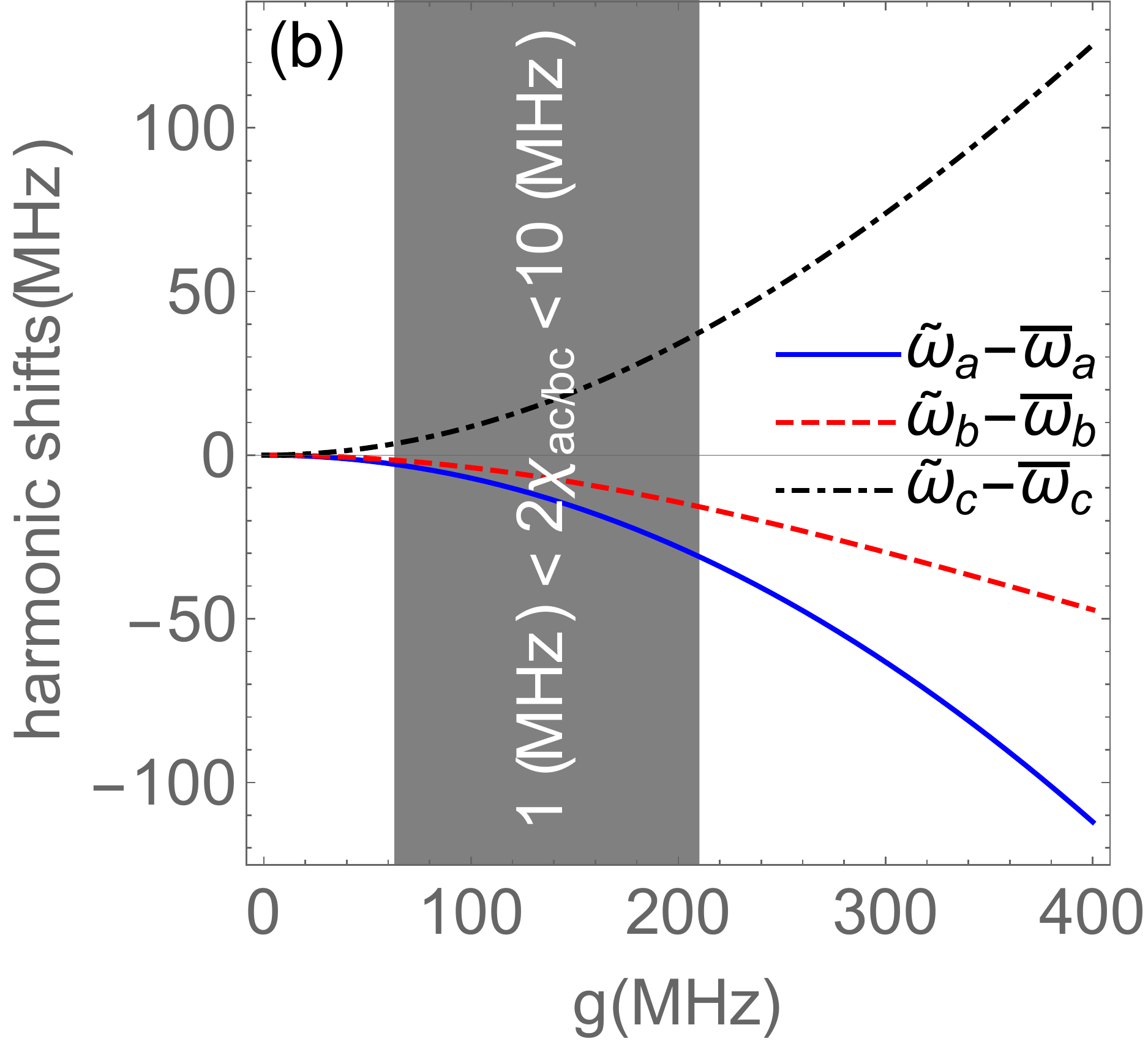}\\
\includegraphics[scale=0.220]{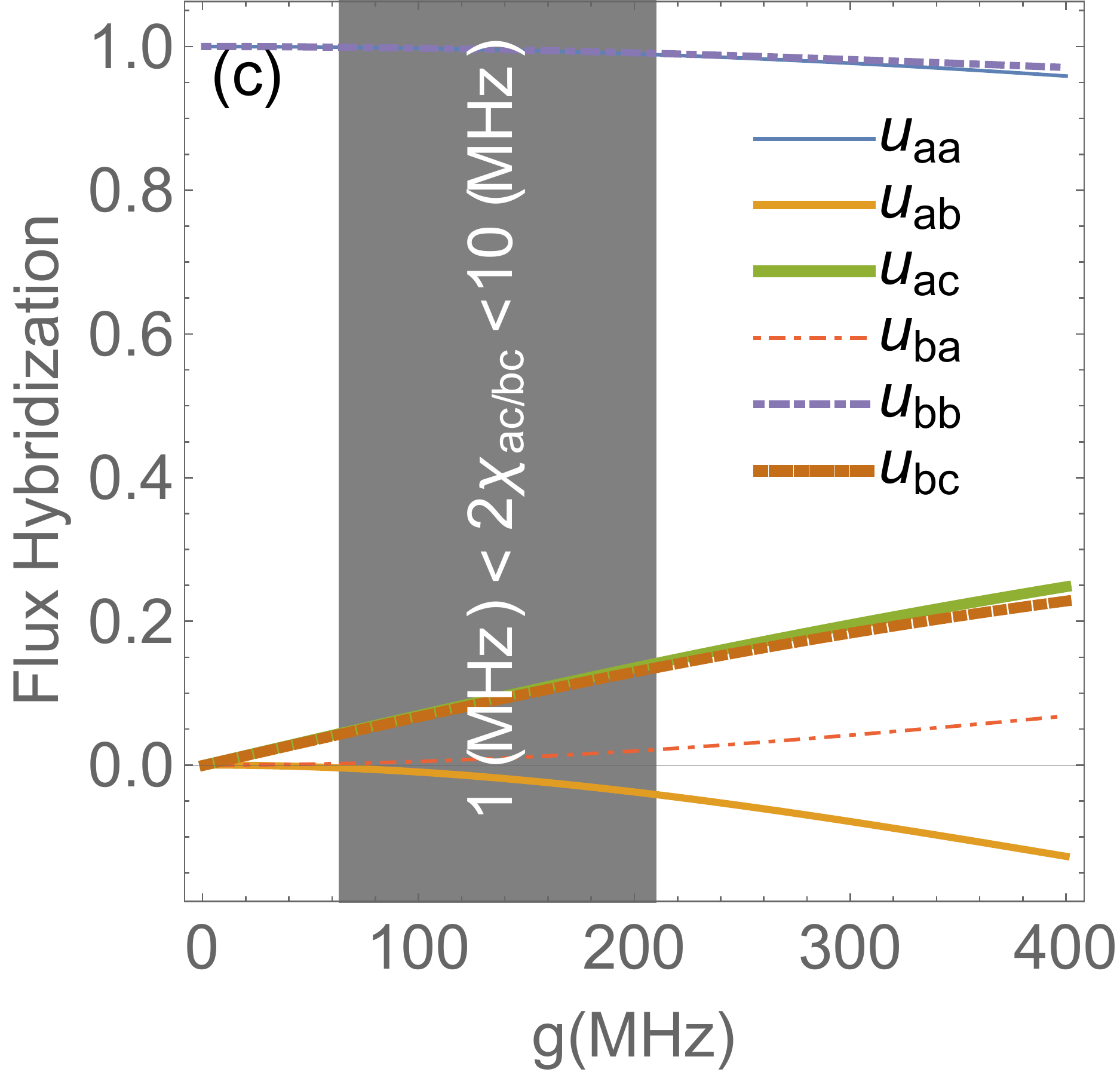}
\includegraphics[scale=0.220]{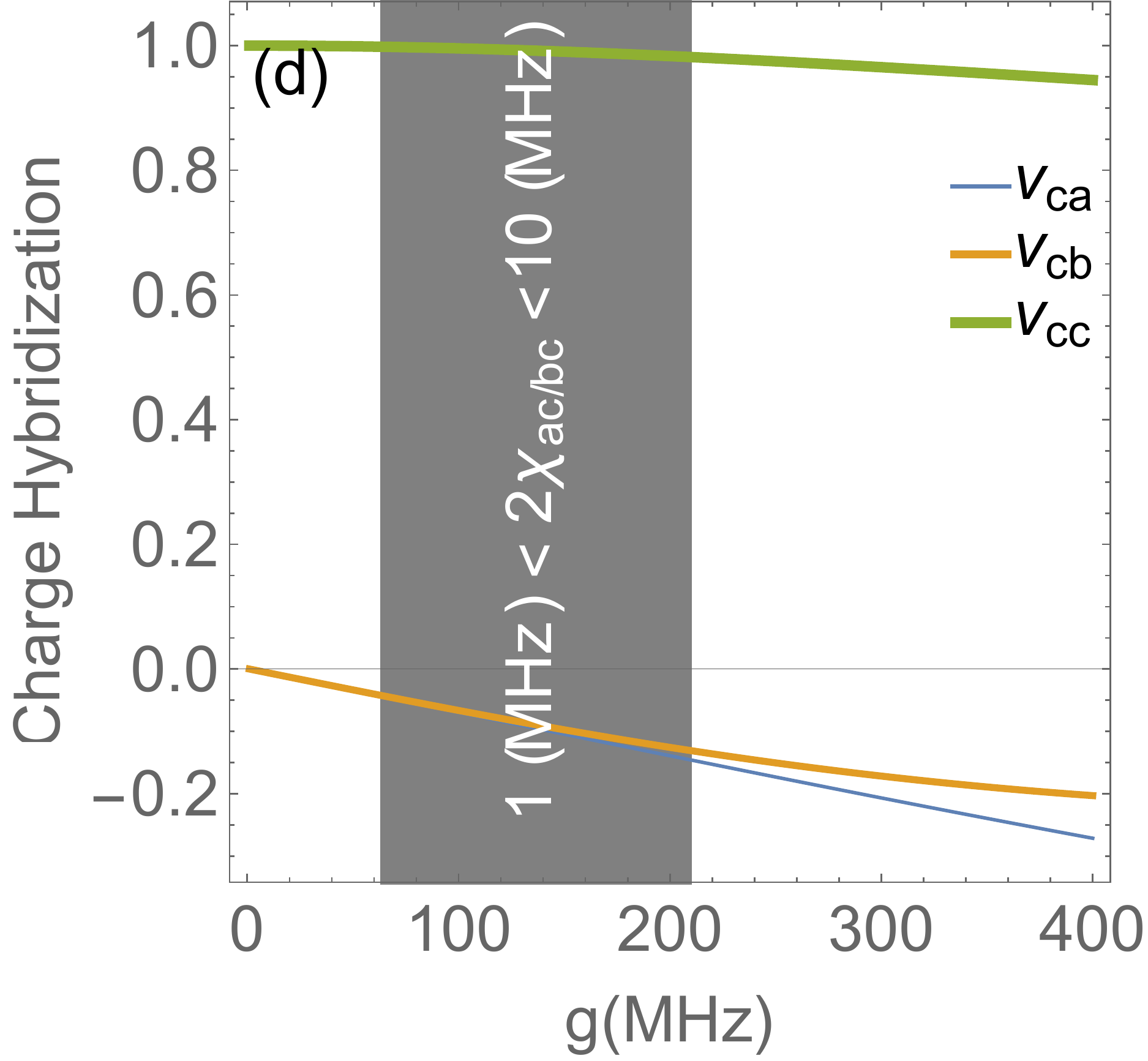}
\caption{Normal mode quantities as a function of qubit-resonator coupling: (a) \textit{harmonic} frequencies, (b) harmonic frequency shifts, (c) projection of bare qubit flux quadrature onto the normal flux quadratures, and (d) projection of the bare resonator charge quadrature that quantifies the direct drive on each normal mode. System parameters are the same as Fig.~\ref{fig:EffHam-GateParamsComparison}. Here, the resonator-qubit couplings are tuned according to $1.764g_b \approx g_a\equiv g$ such that the dispersive shifts are almost equal. The gray region provides the coupling range for which the full dispersive shift for the qubits range approximately between 1 to 10 MHz that is relevant to RIP gate implementation.}
\label{fig:Model-NormNormQuantities}
\end{figure}
%%%%%%%%%%%%%%%%%%%%%%%%%%%%%%%%%%%%%%%%%%%%%%%%%%%%%%%%%%%%%%%%%%%%%%%%%%%%%%%%

In Eqs.~(\ref{Eq:NormModeHam-NormMode Hs})--(\ref{Eq:NormModeHam-NormMode Hd}), $u_{kk'}$ and $v_{kk'}$ are matrix elements of $\mathbf{U}$ and $\mathbf{V}$ for $k,k'=a,b,c$. Normal mode \textit{harmonic} frequencies are denoted by a tilde to distinguish them from the renormalized (Lamb-shifted) normal frequencies (see Appendix~\ref{App:MLM}). Note that due to the inductive nature of the Josephson potential, nonlinear processes appear \textit{only} via mixing of flux hybridization coefficients. On the other hand, a fraction of the RIP drive acts on the normal qubit modes through charge hybridization. Figure~\ref{fig:Model-NormNormQuantities} shows the dependence of normal mode quantities on qubit-resonator couplings $g_{a,b}$. 
%%%%%%%%%%%%%%%%%%%%%%%%%%%%%%%%%%%%%%%%%%%%%%%%%%%%%%%%%%%%%%%%%%%%%%%%%%%%%%%%%

%%%%%%%%%%%%%%%%%%% App: Displacement transformation %%%%%%%%%%%%%%%%%%%%%%%%%%%%
\section{Displacement transformation of the resonator mode}
\label{App:DispTrans}
Here, we apply a displacement transformation on the normal mode Hamiltonian~(\ref{Eq:NormModeHam-NormMode Hs})--(\ref{Eq:NormModeHam-NormMode Hd}) and derive an effective classical response for the resonator mode.

We adopt the following Ansatz 
\begin{align}
\hat{D}[d_c(t)]\equiv e^{d_c(t)\hat{c}^{\dag}-d_c^{*}(t)\hat{c}} \;,	
\label{Eq:DispTrans-Def of Ud(t)}
\end{align}
where $d_c(t)$ is the \textit{time-dependent} coherent displacement. Here, we have used a distinct notation, compared to Eq.~(\ref{Eq:TLM-Def of D[eta_c]}), where $d_c(t)$ accounts also for possible counter-rotating contributions in the drive. The displaced-frame Hamiltonian can be written as
\begin{align}
\HO_s(t)\equiv \hat{D}^{\dag}[d_c(t)]\left[\HO_s+\HO_d(t)-i\partial_t\right]\hat{D}[d_c(t)] \;.
\label{Eq:DispTrans-Def of Hs(t)}
\end{align}
Expanding Eq.~(\ref{Eq:DispTrans-Def of Hs(t)}) in terms of $d_c(t)$ we find	
\begin{align}
\begin{split}
&\HO_s(t)=\sum\limits_{k=a,b,c}\tilde{\omega}_k\hat{k}^{\dag}\hat{k}+\tilde{\omega}_c [d_c^{*}(t)\hat{c}+\text{H.c.}]\\
&+\sum\limits_{j=a,b}\sum\limits_{n=2}^{\infty} \frac{\bar{\omega}_j}{2} (-\epsilon_j)^{n-1}\frac{\left[\sum\limits_{k=a,b,c}u_{jk}\hat{k}+u_{jc}d_c(t)+\text{H.c.}\right]^{2n}}{(2n)!}\\
&+i[\sum\limits_{j=a,b,c}v_{cj}\hat{j}-\text{H.c.}][\Omega_{cx}(t)\cos(\omega_d t)+\Omega_{cy}(t)\cos(\omega_d t)]\\
&+i[\dot{d}_c^{*}(t)\hat{c}-\text{H.c.}]\;.
\end{split}
\label{Eq:DispTrans-Hs(t) Trans}
\end{align}

Coherent displacement $d_c(t)$ is then set such that terms \textit{linear} in $\hat{c}$ and $\hat{c}^{\dag}$ are canceled out in the displaced Hamiltonian~(\ref{Eq:DispTrans-Hs(t) Trans}). Note that such terms can also emerge from the anharmonic part of Hamiltonian~(\ref{Eq:DispTrans-Hs(t) Trans}) due to the non-commuting bosonic algebra. Following this procedure up to the quartic expansion we find
\begin{align}
\begin{split}
\dot{d}_c(t)=&-i\tilde{\omega}_c d_c(t)-v_{cc}[\Omega_{cx}(t)\cos(\omega_d t)+\Omega_{cy}(t)\sin(\omega_d t)]\\
&-i\delta_c[d_c(t)+d_c^{*}(t)]\\
&-\frac{i}{3}\alpha_c[d_c(t)+d_c^{*}(t)]^3+O(\epsilon^2) \;.
\end{split}
\label{Eq:DispTrans-Cond for d_c(t)}
\end{align}
The first line of Eq.~(\ref{Eq:DispTrans-Cond for d_c(t)}) contains the terms coming from the harmonic Hamiltonian. The second and the third lines, however, contain nonlinear corrections, from which we find a \textit{static} frequency shift as well as an effective anharmonicity for the resonator mode as
\begin{align}
\begin{split}
\delta_{Sc}\equiv &-\frac{1}{4}\epsilon_a\bar{\omega}_a u_{ac}^2\left(u_{aa}^2+u_{ab}^2+u_{ac}^2\right)\\
&-\frac{1}{4}\epsilon_b\bar{\omega}_b u_{bc}^2\left(u_{ba}^2+u_{bb}^2+u_{bc}^2\right) +O(\epsilon^2) \;,
\end{split}
\label{Eq:DispTrans-Def of delta_Sc}\\
\alpha_c\equiv &-\frac{1}{4}\left(\epsilon_a\bar{\omega}_au_{ac}^4+\epsilon_b\bar{\omega}_bu_{bc}^4\right) +O(\epsilon^2)\;.
\label{Eq:DispTrans-Def of alpha_c}
\end{align}
Equation~(\ref{Eq:DispTrans-Cond for d_c(t)}) determines $d_c(t)$ as the response of a driven classical Duffing oscillator. Resorting to Eq.~(\ref{Eq:DispTrans-Cond for d_c(t)}) is needed when the RIP drive is comparatively strong. In the rotating frame of the drive and under the RWA, a simplified equation for the slowly varying amplitude $d_c(t)\equiv \eta_c(t)e^{-i\omega_d t}$ is obtained as
\begin{align}
\dot{\eta}_c(t)+i\Delta_{cd}\eta_c(t)+i\alpha_c|\eta_c(t)|^2\eta_c(t)=-\frac{i}{2}\Omega_{c}(t)\;,
\label{Eq:DispTrans-Cond for eta_c(t)}
\end{align}
where $\omega_c\equiv \tilde{\omega}_c+\delta_{Sc}$, $\Delta_{cd}\equiv \omega_c-\omega_d$ and $\Omega_c(t)=\Omega_{cy}(t)-i\Omega_{cx}(t)$. 
We note that higher order terms, e.g. coming from the sextic nonlinearity, can become relevant in the strong drive regime. For typical RIP parameters, $v_{cc}\approx 0.99$, leading to a negligible renormalization of the intended RIP drive amplitude which is neglected for simplicity. 
%%%%%%%%%%%%%%%%%%%%%%%%%%%%%%%%%%%%%%%%%%%%%%%%%%%%%%%%%%%%%%%%%%%%%%%%%%%%%%%%%

%%%%%%%%%%%%%%%%%%%%% Sec: Resonator Response %%%%%%%%%%%%%%%%%%%%%%%%%%%%%%%%%%%
\section{Resonator response and leakage}
\label{App:ResRes}

A possible source of coherent error for the RIP gate is residual resonator photon population at the end of the pulse leading to a faulty $ZZ$ interaction. Here, we analyze the resonator response based on condition~(\ref{Eq:DispTrans-Cond for eta_c(t)}) and discuss parameter choices that suppress leakage. We provide a classical leakage measure [Eq.~(\ref{Eq:ResRes-Def of L(theta,theta_G)})], which relates the \textit{relative} resonator leakage to the resonator-drive detuning and pulse duration. Furthermore, we demonstrate the advantage of adding DRAG to the resonator drive in suppressing leakage.   	

%%%%%%%%%%%%%%%%%%%% Figure: Dependence of Res Response on Pulse Width  %%%%%%%%%
\begin{figure}[t!]
\centering
\includegraphics[scale=0.120]{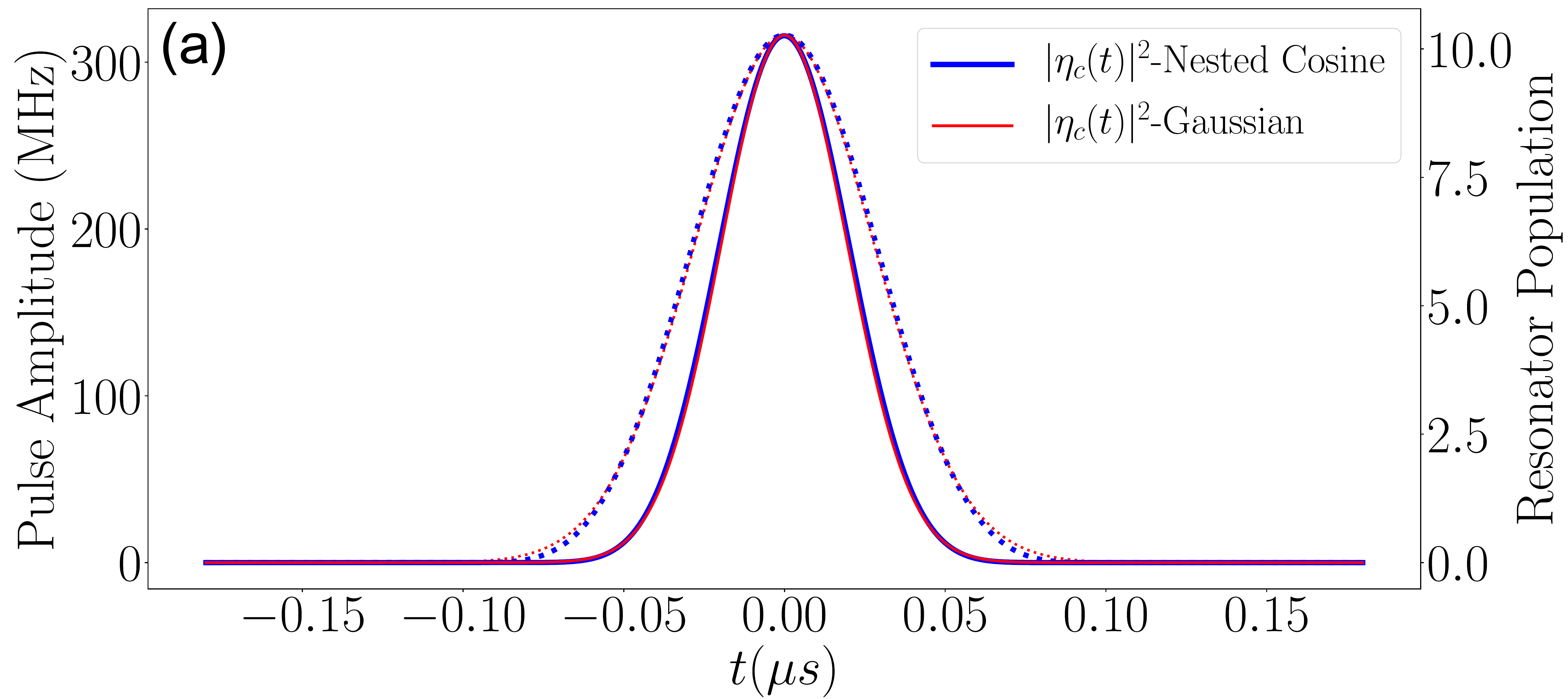} \\
\includegraphics[scale=0.123]{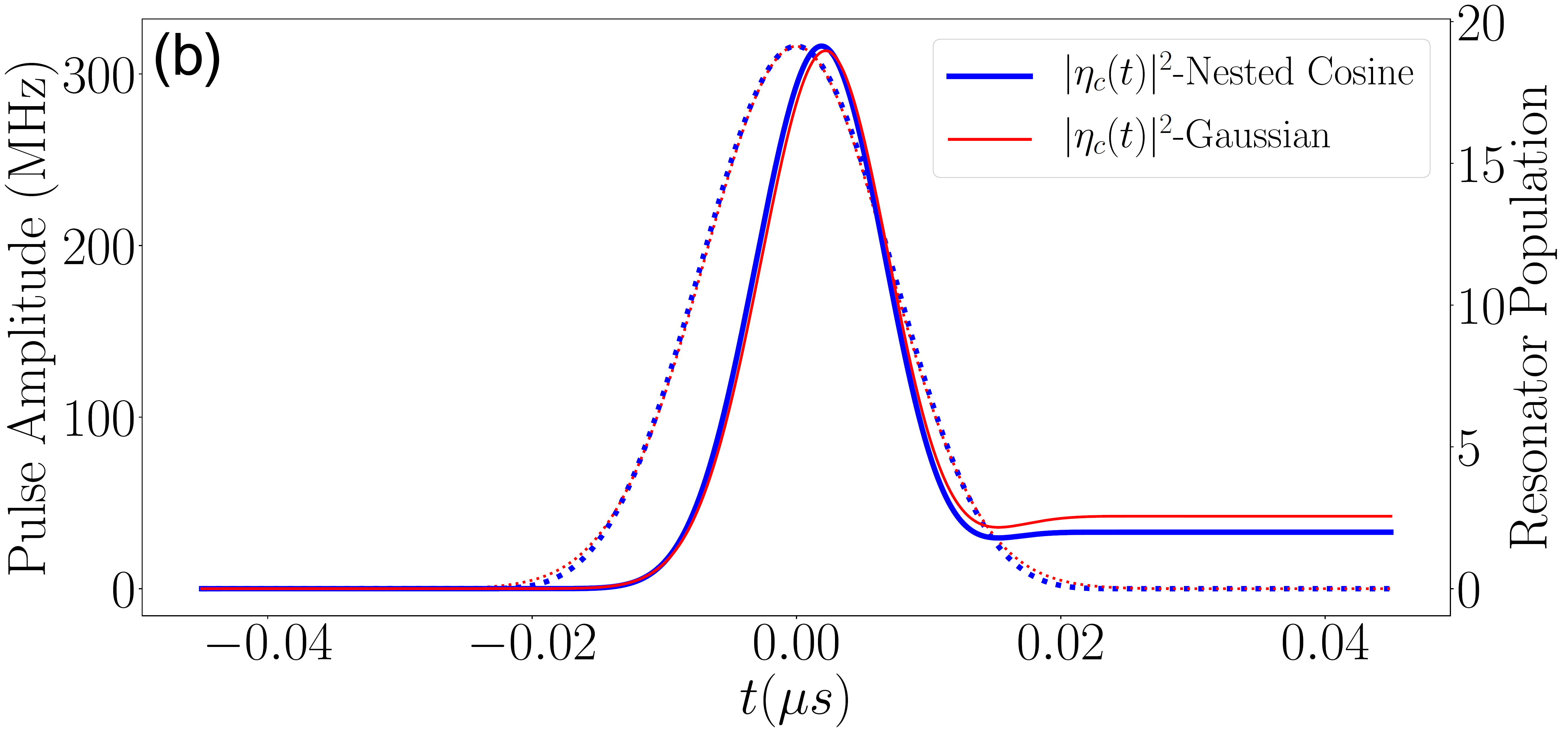}\\ 
\includegraphics[scale=0.253]{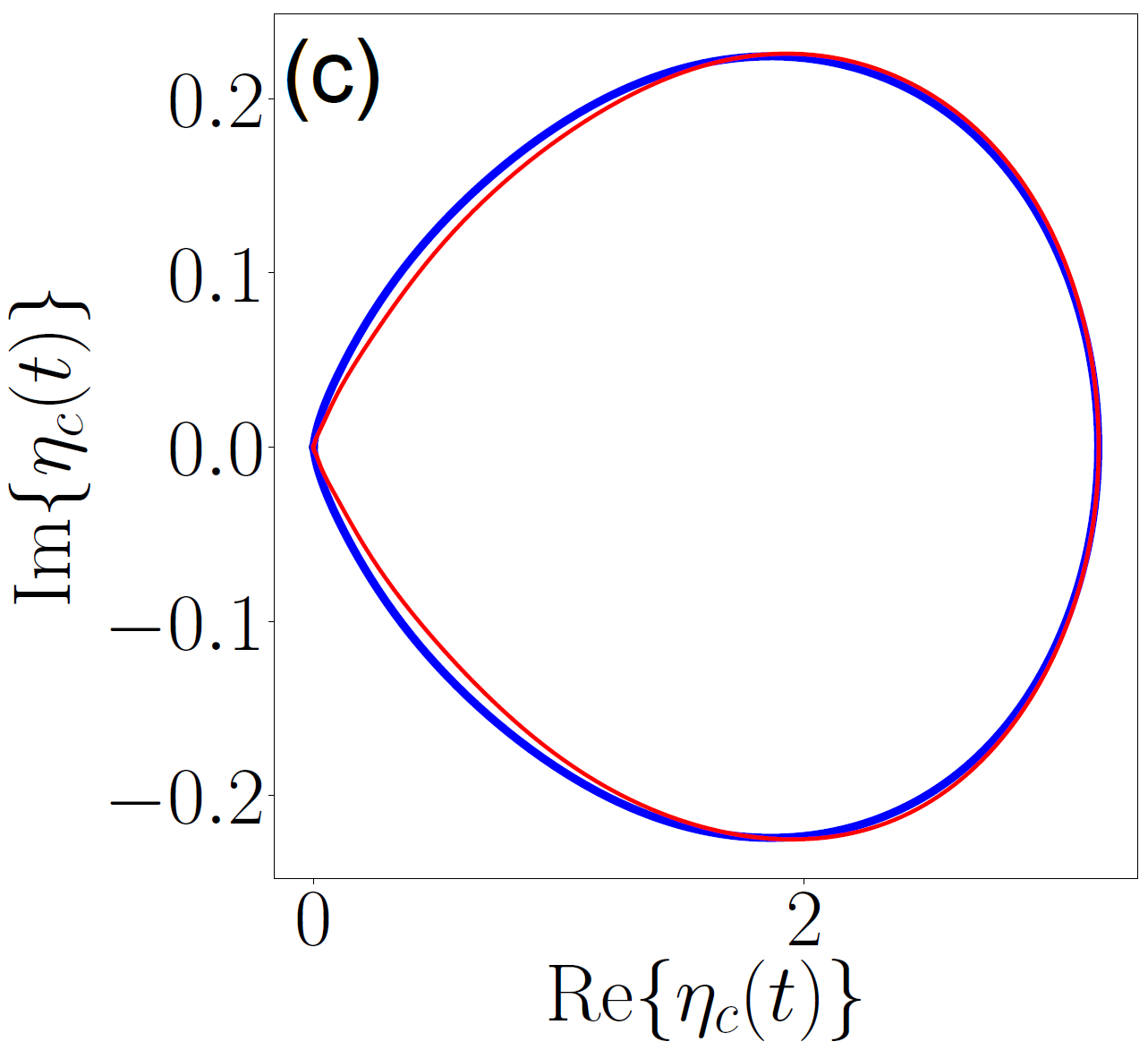}
\includegraphics[scale=0.253]{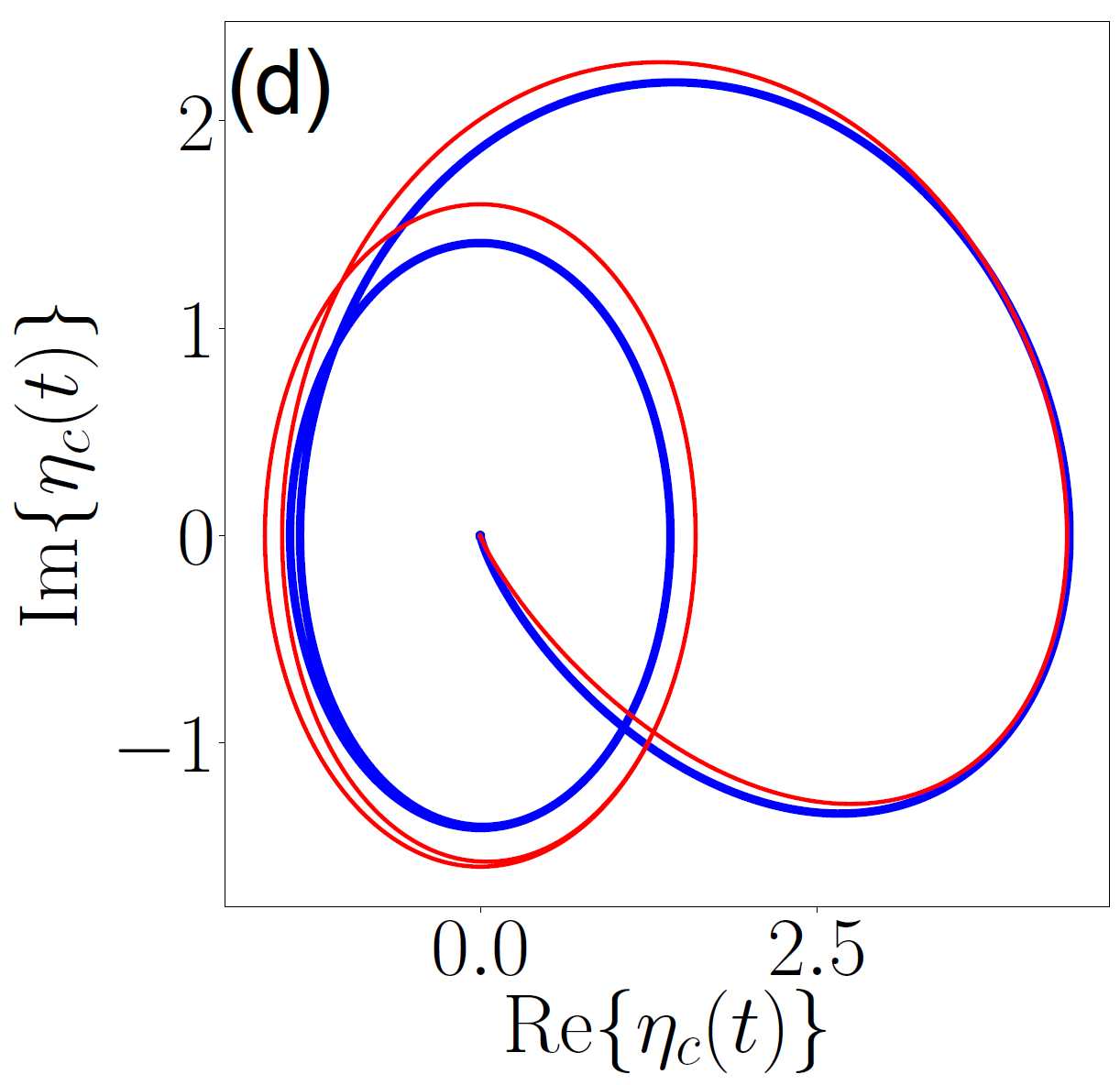}
\caption{Dependence of residual photon population on pulse time $\mathbf{\tau}$ based on Eq.~(\ref{Eq:DispTrans-Cond for eta_c(t)}). Resonator and drive parameters are set as $\Delta_{cd}/2\pi=-50$ MHz, $\Omega_c/2\pi=316.227$ MHz and $\alpha_c/2\pi=0$. The pulse time is set to $\tau=200$ ns for (a) and (c) and to $\tau=50$ ns for (b) and (d). The input pulse and the resonator response are shown in dotted and solid curves, respectively. Shorter pulse time leads to larger residual photon population, which is seen as a steady-state orbit around the origin in phase space.}	
\label{Fig:ResRes-DepOnPulseWidth}
\end{figure}
%%%%%%%%%%%%%%%%%%%%%%%%%%%%%%%%%%%%%%%%%%%%%%%%%%%%%%%%%%%%%%%%%%%%%%%%%%%%%%%%

For comparison, we consider two pulse shapes. Firstly, a truncated Gaussian pulse as 
\begin{subequations}
\begin{align}
&\Omega_{cy}(t)=\Omega_c P_G(t;t_{\text{G}},\sigma,\tau) \;, 
\label{Eq:ResRes-Def of Omega_G(t)}\\
&P_{\text{G}}(t;t_G,\sigma,\tau)\equiv \frac{e^{-\frac{(t-t_G)^2}{2\sigma^2}}-e^{-\frac{\tau^2}{8\sigma^2}}}{1-e^{-\frac{\tau^2}{8\sigma^2}}} \;,
\label{Eq:ResRes-Def of P_G(t,t_G,sigma)}
\end{align}
\end{subequations}
for $-\tau/2<t-t_{\text{G}}<\tau/2$, where $\tau$ is the pulse time, $t_{\text{G}}$ is the center, $\sigma$ is the standard deviation and the pulse is forced to zero at the tails. The Gaussian pulse serves as a standard point of reference for which simple analytical solutions exist. In our numerical simulations, as well as earlier experiments \cite{Paik_Experimental_2016}, we use the nested cosine pulse defined as
\begin{subequations}
\begin{align}
&\Omega_{cy}(t)\equiv\Omega_{c} P_{\text{nc}}(t;t_{\text{nc}},\tau) \;,
\label{Eq:ResRes-Def of Omega_nc(t)}\\
&P_{\text{nc}}(t;t_{\text{nc}},\tau)\equiv \frac{1}{2}\left\{\cos\left[\pi\cos\left(\pi\frac{t-t_{\text{nc}}}{\tau}\right)\right]+1\right\} \;,
\label{Eq:ResRes-Def of P_nc(t,t_nc,tau)}
\end{align}
\end{subequations}
for $0<t-t_{\text{nc}}<\tau$. The advantage of the latter is its smoother rise and fall (zero first-order derivative), which mitigates residual photons \cite{Cross_Optimized_2015}. For a fair comparison, we set the area under the pulse to be equal as
\begin{align}
\int_{t_{\text{nc}}}^{t_{\text{nc}}+\tau}dt P_{\text{nc}}(t;t_{\text{nc}},\tau)=\int_{t_{\text{G}}-\tau/2}^{t_{\text{G}}+\tau/2}dt P_{\text{G}}(t;t_\text{G},\sigma,\tau) \;,
\label{Eq:ResRes-EqualAreaUnderCurve}
\end{align}
resulting in $\tau\approx 7.182 \sigma$. Furthermore, to align the pulse centers, we set $t_{\text{nc}}+\tau/2=t_{\text{G}}$.

The solution to Eq.~(\ref{Eq:DispTrans-Cond for eta_c(t)}) depends on \textit{four} independent parameters: resonator-drive detuning $\Delta_{cd}$, pulse amplitude $\Omega_c$, pulse time $\tau$ and resonator anharmonicity $\alpha_c$. In the following, we discuss two separate interplays between these parameters. Firstly, the ratio of residual photons to the intended maximum photon number depends \textit{primarily} on the product $\Delta_{cd} \tau$. To focus on this, we first set $\alpha_c$ to zero. Figure~\ref{Fig:ResRes-DepOnPulseWidth} compares the resonator response for two pulse times of $\tau=200$ ns and $\tau=50$ ns and fixed $\Delta_{cd}/2\pi=-50$ MHz. We observe a significantly larger residual photon population for $\tau=50$ ns for both pulse forms, but the nested cosine pulse results in a smaller population than the Gaussian pulse. The increase in residual photons for the shorter pulse duration can be understood as the pulse becoming broader in the frequency domain leading to a larger overlap with the resonator and hence a transfer of energy to the resonator mode.

%%%%%%%%%%%% Figure: Interplay between pulse width and detuning %%%%%%%%%%%%%%%%%%%%%%%%
\begin{figure}[t!]
\centering
\includegraphics[scale=0.44]{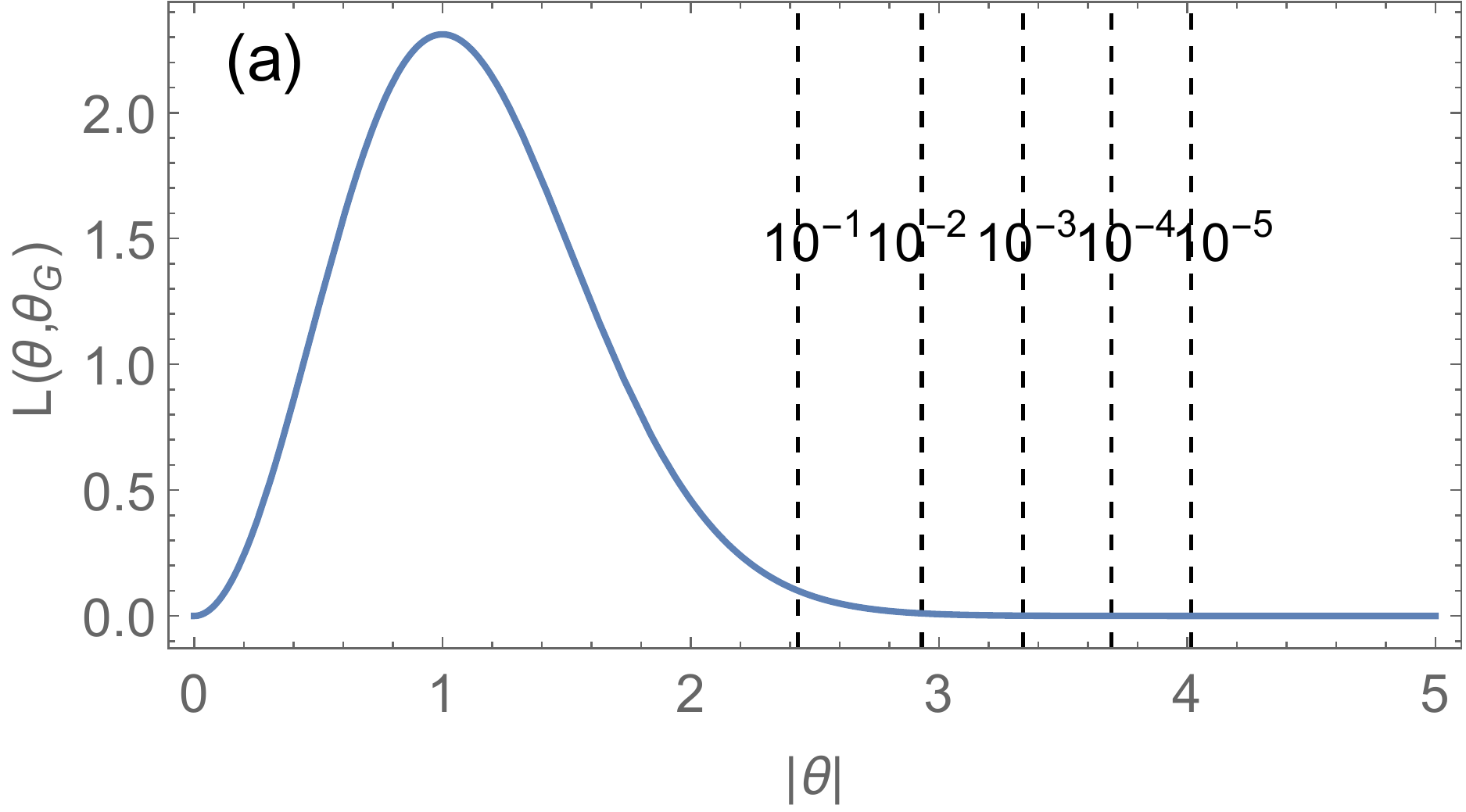}\\
\includegraphics[scale=0.45]{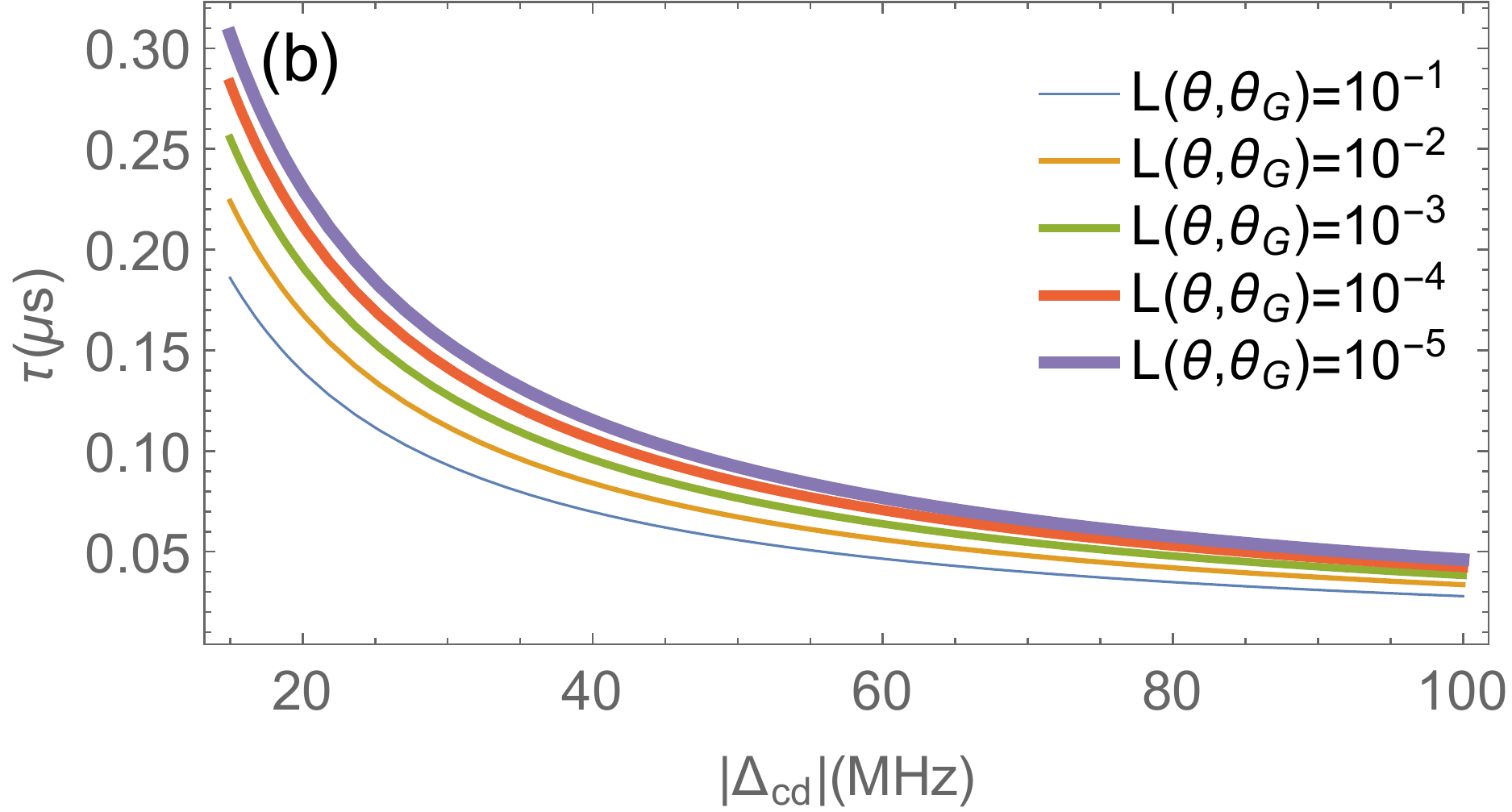}
\caption{(a) Normalized leakage function $L(\theta,\theta_{\text{G}})$ in Eq.~(\ref{Eq:ResRes-Def of L(theta,theta_G)}). Vertical dashed lines show values of $\theta\equiv \Delta_{cd} \sigma$ that $L(\theta,\theta_G)$ is fixed to in panel (b). (b) Corresponding constant leakage contours as a function of $\tau$ and $\Delta_{cd}$ using the relation $\tau\approx 7.182 \sigma$.} 
\label{Fig:ResRes-LeakageFunction}
\end{figure}
%%%%%%%%%%%%%%%%%%%%%%%%%%%%%%%%%%%%%%%%%%%%%%%%%%%%%%%%%%%%%%%%%%%%%%%%%%%%%%%%%  

In the linear case, i.e. $\alpha_c/2\pi=0$, it is possible to obtain an analytical solution to Eq.~(\ref{Eq:DispTrans-Cond for eta_c(t)}). For clarity, consider a non-truncated Gaussian input as $\Omega_c(t)=\Omega_c \exp[(t-t_{\text{G}})^2/2\sigma^2]$. The solution with initial zero resonator photons [i.e. $\eta_c(0)=0$] reads   
\begin{align}
\begin{split}
&\eta_c(t)=-\frac{i}{2}\Omega_c \sqrt{\frac{\pi}{2}}\sigma e^{-\frac{1}{2}\Delta_{cd}^2\sigma^2} e^{-i\Delta_{cd} (t-t_{\text{G}})}\\
&\times \left[\text{erf}\left(\frac{t-t_{\text{G}}}{\sqrt{2}\sigma}-i\frac{\Delta_{cd}\sigma}{\sqrt{2}}\right)+\text{erf}\left(\frac{t_G}{\sqrt{2}\sigma}+i\frac{\Delta_{cd}\sigma}{\sqrt{2}}\right)\right] \;,
\end{split}
\label{Eq:ResRes-eta_c(t) Sol Gaussian}
\end{align}
where $\text{erf}(x)\equiv (2/\sqrt{\pi})\int _{0}^{x} dz \exp (-z^2)$ is the error function. The residual photon population can be defined, based on solution~(\ref{Eq:ResRes-eta_c(t) Sol Gaussian}), as the steady-state response $n_{c}(\infty)\equiv \lim\limits_{t\to\infty} |\eta_c(t)|^2$. We then define a normalized leakage measure as the ratio of residual photon population to the intended maximum photon population as 
\begin{align}
L(\theta,\theta_{\text{G}})\equiv \frac{n_c(\infty)}{(\frac{\Omega_c}{2\Delta_{cd}})^2}=\frac{\pi}{2}\theta^2e^{-\theta^2}\left|1+\text{erf}\left(\frac{i\theta^2+\theta_{\text{G}}}{\sqrt{2}\theta}\right)\right|^2 \;,
\label{Eq:ResRes-Def of L(theta,theta_G)}
\end{align}
where $\theta\equiv \Delta_{cd}\sigma$ and $\theta_{\text{G}}\equiv \Delta_{cd}t_{\text{G}}$ are unitless angles denoting the width and center of the pulse, respectively, with the assumption $\theta_G \gg \theta>0$. According to Eq.~(\ref{Eq:ResRes-Def of L(theta,theta_G)}), for $\theta\ll 1$, the leakage grows as $\theta^2$, while an exponential suppression of the form $\exp (-\theta^2)$ is expected for $\theta \gg 1$ [see Fig.~\ref{Fig:ResRes-LeakageFunction}].  

%%%%%%%%%%%%%%%%%%%%%%%%%%%% Figure: Steady-State Photon Number %%%%%%%%%%%%%%%%%
\begin{figure}	
\centering
\includegraphics[scale=0.455]{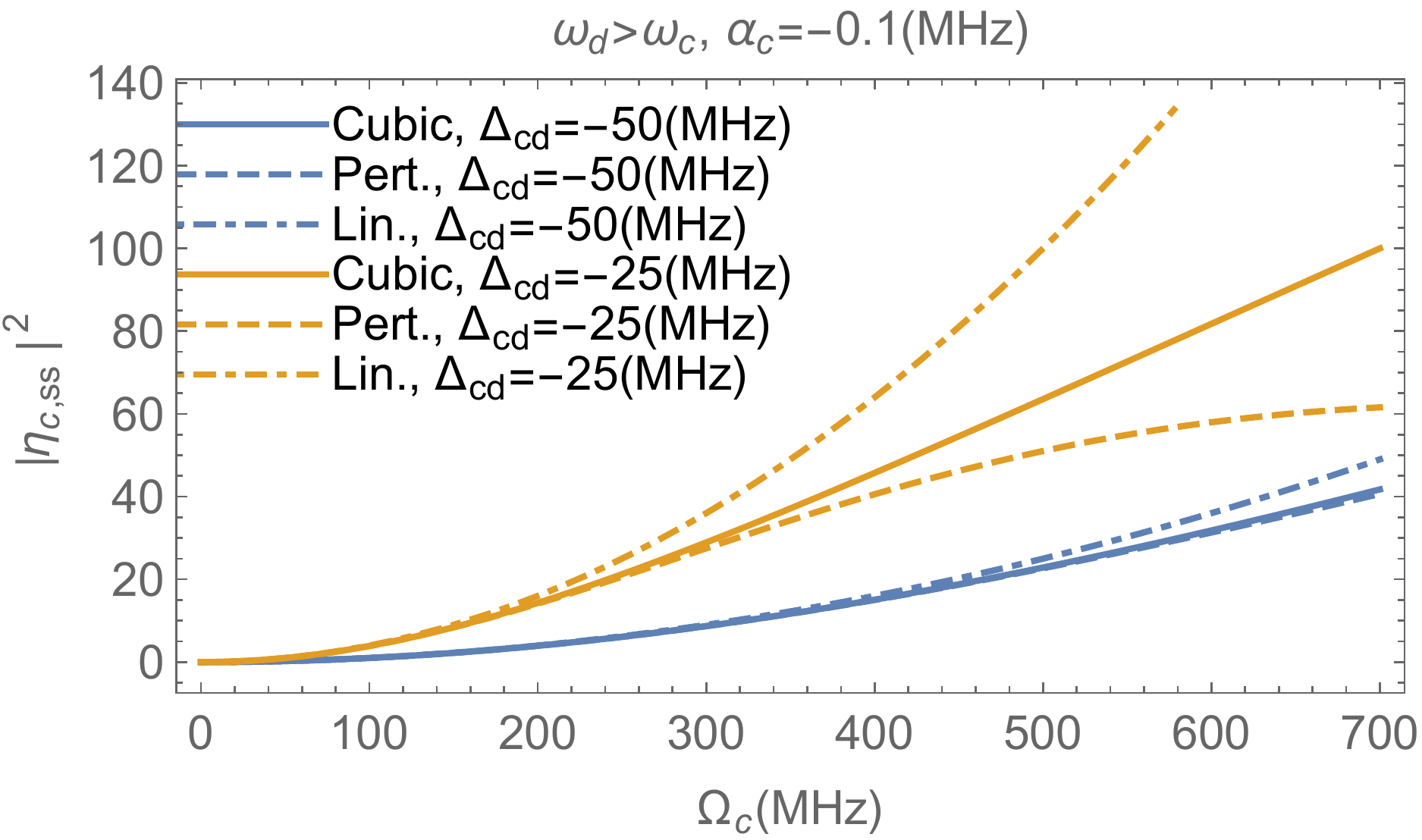}
\caption{Steady-state resonator photon number as a function of RIP drive for $\Delta_{cd}/2\pi=-50$ MHz (blue) and $\Delta_{cd}/2\pi=-25$ MHz (yellow). Resonator anharmonicity is set to $\alpha_c/2\pi=-100$ KHz. Solid lines are found via exact solution to Eq.~(\ref{Eq:ResRes-SS Eq. for etac}), dashed lines show the first order expansion in $|\alpha_c/\Delta_{cd}|$ based on Eq.~(\ref{Eq:ResRes-eta_c,ss SolWithAlpha}) and the dashed-dotted lines show the commonly used linear estimate $|\Omega_c/(2\Delta_{cd})|^2$. We find that for smaller resonator-drive detuning and stronger drive, it becomes more important to account for the resonator anharmonicity.}
\label{Fig:ResRes-SSPhNum}
\end{figure}
%%%%%%%%%%%%%%%%%%%%%%%%%%%%%%%%%%%%%%%%%%%%%%%%%%%%%%%%%%%%%%%%%%%%%%%%%%%%%%%%%

Secondly, based on Eq.~(\ref{Eq:DispTrans-Cond for eta_c(t)}) and up to lowest order in $|\alpha_c/\Delta_{cd}|$, the effect of resonator anharmonicity on leakage is a dynamic shift of the resonator-drive detuning as
\begin{align}
\Delta_{cd}(t)\equiv \Delta_{cd}+\alpha_c |\eta_c(t)|^2 \;.
\label{Eq:ResRes-Def of Del_cd(t)}
\end{align}
Following the \textit{linear} leakage measure~(\ref{Eq:ResRes-Def of L(theta,theta_G)}), if the anharmonicity increases (decreases) the effective $\Delta_{cd}(t)$, it leads to suppression (enhancement) of resonator leakage. Since $\alpha_c$ is intrinsically negative (softening Josephson nonlinearity), having $\Delta_{cd}<0$ (drive above resonator frequency) is beneficial. This heuristic argument is confirmed by numerical solution to Eq.~(\ref{Eq:DispTrans-Cond for eta_c(t)}), where we find this effect to become relevant at stronger photon number of the order $|\eta_c(t)|^2\sim |\Delta_{cd}/\alpha_c|$. A typical RIP design leads to $50 \ \text{KHz}<|\alpha_c /2\pi|<100 \ \text{KHz}$.

Furthermore, for strong RIP drive, the resonator anharmonicity must be accounted for in our estimate for the maximum photon number. Based on Eq.~(\ref{Eq:DispTrans-Cond for eta_c(t)}), the steady-state response satisfies
\begin{align}
\Delta_{cd}\eta_{c,\text{ss}}+\alpha_c |\eta_{c,\text{ss}}|^2 \eta_{c,\text{ss}}=-\frac{1}{2}\Omega_c \;,
\label{Eq:ResRes-SS Eq. for etac}
\end{align}
whose approximate solution can be obtained as   
\begin{align}
\eta_{c,\text{ss}} = -\frac{\Omega_c}{2\left[\Delta_{cd}+\alpha_c \left(\frac{\Omega_c}{2\Delta_{cd}}\right)^2\right]} +O(|\frac{\alpha_c}{\Delta_{cd}}|^2)\;.
\label{Eq:ResRes-eta_c,ss SolWithAlpha}
\end{align}
Figure~\ref{Fig:ResRes-SSPhNum} makes a comparison between the different estimates for steady-state resonator photon number. 

%%%%%%%%%%%%%%%%%%%%%%%%%%% Table: Dominant Contributions from Quartic Nonlinearity %%%%%%%%%%%%%%%%%%%%%%%%%%%%%
\begin{table*}[ht]
\begin{tabular}{|c|c|c|c|}
\hline	
Operator & Coefficient & Normal-mode expression & Estimate (MHz) \\
\hline\hline
$\hat{a}$ & $\lambda_{a}(t)$ & \makecell{$\frac{v_{ca}\Omega_c(t)}{2}-\frac{\epsilon_a\bar{\omega}_a}{4}u_{aa}u_{ac}\left[u_{aa}^2+u_{ab}^2+u_{ac}^2\left(1+|\eta_c(t)|^2\right)\right]\eta_c^*(t)$ \\ $-\frac{\epsilon_b\bar{\omega}_b}{4}u_{ba}u_{bc}\left[u_{ba}^2+u_{bb}^2+u_{bc}^2\left(1+|\eta_c(t)|^2\right)\right]\eta_c^*(t)$} & \makecell{$-0.052\Omega_c(t)-26.466\eta_c^*(t)$ \\ $-0.282|\eta_c(t)|^2\eta_c^*(t)$}\\
\hline
$\hat{b}$ & $\lambda_b(t)$ & \makecell{$\frac{v_{cb}\Omega_c(t)}{2}-\frac{\epsilon_a\bar{\omega}_a}{4}u_{ab}u_{ac}\left[u_{aa}^2+u_{ab}^2+u_{ac}^2\left(1+|\eta_c(t)|^2\right)\right]\eta_c^*(t)$ \\ $-\frac{\epsilon_b\bar{\omega}_b}{4}u_{bb}u_{bc}\left[u_{ba}^2+u_{bb}^2+u_{bc}^2\left(1+|\eta_c(t)|^2\right)\right]\eta_c^*(t)$} & \makecell{$-0.049\Omega_c(t)-26.654\eta_c^*(t)$\\ $-0.264|\eta_c(t)|^2\eta_c^*(t)$} \\
\hline\hline
$\hat{a}^{\dag}\hat{a}$ & $\delta_{Sa}$ & $-\frac{1}{4}\epsilon_a\bar{\omega}_a u_{aa}^2\left(u_{aa}^2+u_{ab}^2+u_{ac}^2\right)-\frac{1}{4}\epsilon_b\bar{\omega}_b u_{ba}^2\left(u_{ba}^2+u_{bb}^2+u_{bc}^2\right)$ & $-252.004$\\ 	
\hline
$\hat{b}^{\dag}\hat{b}$ & $\delta_{Sb}$ & $-\frac{1}{4}\epsilon_a\bar{\omega}_a u_{ab}^2\left(u_{aa}^2+u_{ab}^2+u_{ac}^2\right)-\frac{1}{4}\epsilon_b\bar{\omega}_b u_{bb}^2\left(u_{ba}^2+u_{bb}^2+u_{bc}^2\right)$ & $-272.309$	\\
\hline
$\hat{c}^{\dag}\hat{c}$ & $\delta_{Sc}$ & $-\frac{1}{4}\epsilon_a\bar{\omega}_a u_{ac}^2\left(u_{aa}^2+u_{ab}^2+u_{ac}^2\right)-\frac{1}{4}\epsilon_b\bar{\omega}_b u_{bc}^2\left(u_{ba}^2+u_{bb}^2+u_{bc}^2\right)$ & 	$-5.439$ \\
\hline\hline
$\hat{a}^{\dag}\hat{a}$ & $\delta_{Da}(t)$ & $-\frac{1}{2}\left(\epsilon_a\bar{\omega}_au_{aa}^2 u_{ac}^2+\epsilon_b\bar{\omega}_b u_{ba}^2 u_{bc}^2\right)|\eta_c(t)|^2$  & $-5.371 |\eta_c(t)|^2$\\ 	
\hline
$\hat{b}^{\dag}\hat{b}$ & $\delta_{Db}(t)$ & $-\frac{1}{2}\left(\epsilon_a\bar{\omega}_au_{ab}^2 u_{ac}^2+\epsilon_b\bar{\omega}_b u_{bb}^2 u_{bc}^2\right)|\eta_c(t)|^2$  & $-5.395|\eta_c(t)|^2$ 	\\
\hline
$\hat{c}^{\dag}\hat{c}$ & $\delta_{Dc}(t)$ & $-\frac{1}{2}\left(\epsilon_a\bar{\omega}_au_{ac}^4 +\epsilon_b\bar{\omega}_b u_{bc}^4\right)|\eta_c(t)|^2 $  & $-0.112 |\eta_c(t)|^2$	\\
\hline\hline
$\hat{a}^{\dag}\hat{b}$ & $\lambda_{a^*b}(t)$ & \makecell{$-\frac{\epsilon_a\bar{\omega}_a}{4}u_{aa}u_{ab}\left[u_{aa}^2+u_{ab}^2+u_{ac}^2\left(1+2|\eta_c(t)|^2\right)\right]$ \\ $-\frac{\epsilon_b\bar{\omega}_b}{4}u_{ba}u_{bb}\left[u_{ba}^2+u_{bb}^2+u_{bc}^2\left(1+2|\eta_c(t)|^2\right)\right]$} & $2.436+0.057|\eta_c(t)|^2$\\
\hline
$\hat{a}^{\dag}\hat{c}$ & $\lambda_{a^*c}(t)$ & \makecell{$-\frac{\epsilon_a\bar{\omega}_a}{4}u_{aa}u_{ac}\left[u_{aa}^2+u_{ab}^2+u_{ac}^2\left(1+2|\eta_c(t)|^2\right)\right]$ \\ $-\frac{\epsilon_b\bar{\omega}_b}{4}u_{ba}u_{bc}\left[u_{ba}^2+u_{bb}^2+u_{bc}^2\left(1+2|\eta_c(t)|^2\right)\right]$} & $-26.466-0.563|\eta_c(t)|^2$\\
\hline
$\hat{b}^{\dag}\hat{c}$ & $\lambda_{b^*c}(t)$ & \makecell{$-\frac{\epsilon_a\bar{\omega}_a}{4}u_{ab}u_{ac}\left[u_{aa}^2+u_{ab}^2+u_{ac}^2\left(1+2|\eta_c(t)|^2\right)\right]$ \\ $-\frac{\epsilon_b\bar{\omega}_b}{4}u_{bb}u_{bc}\left[u_{ba}^2+u_{bb}^2+u_{bc}^2\left(1+2|\eta_c(t)|^2\right)\right]$} & $-26.654-0.527|\eta_c(t)|^2$\\
\hline\hline
$\hat{a}^{\dag}\hat{a}\hat{a}$ & $\lambda_{a^*aa}(t)$ & $-\frac{1}{4}(\epsilon_a\bar{\omega}_a u_{aa}^3 u_{ac}+\epsilon_b\bar{\omega}_b u_{ba}^3 u_{bc}) \eta_c^*(t)$ & $-25.867 \eta_c^*(t)$\\
\hline
$\hat{b}^{\dag}\hat{b}\hat{b}$ & $\lambda_{b^*bb}(t)$ & $-\frac{1}{4}(\epsilon_a\bar{\omega}_a u_{ab}^3 u_{ac}+\epsilon_b\bar{\omega}_b u_{bb}^3 u_{bc})\eta_c^*	(t)$ & $-26.954 \eta_c^*(t)$\\
\hline
$\hat{c}^{\dag}\hat{c}\hat{c}$ & $\lambda_{c^*cc}(t)$ & $-\frac{1}{4}(\epsilon_a\bar{\omega}_a u_{ac}^4+\epsilon_b\bar{\omega}_b u_{bc}^4 )\eta_c^*(t)$ & $-0.056 \eta_c^*(t)$ \\
\hline\hline
$\hat{a}^{\dag}\hat{a}\hat{b}$ & $\lambda_{a^*ab}(t)$ & $-\frac{1}{2}(\epsilon_a\bar{\omega}_a u_{aa}^2 u_{ab} u_{ac}+\epsilon_b\bar{\omega}_b u_{ba}^2 u_{bb} u_{bc}) \eta_c^*(t)$ & $1.127\eta_c^*(t)$ \\
\hline
\boldmath $\hat{a}^{\dag}\hat{a}\hat{c}$ & \boldmath $\lambda_{a^*ac}(t)$ & \boldmath $-\frac{1}{2}(\epsilon_a\bar{\omega}_a u_{aa}^2 u_{ac}^2+\epsilon_b\bar{\omega}_b u_{ba}^2 u_{bc}^2) \eta_c^*(t)$ &\boldmath  $-5.371\eta_c^*(t)$\\
\hline
$\hat{b}^{\dag}\hat{b}\hat{a}$ & $\lambda_{b^*ba}(t)$ & $ -\frac{1}{2}(\epsilon_a\bar{\omega}_a u_{ab}^2 u_{aa} u_{ac}+\epsilon_b\bar{\omega}_b u_{bb}^2 u_{ba} u_{bc}) \eta_c^*(t)$ & $-0.636\eta_c^*(t)$ \\
\hline
\boldmath$\hat{b}^{\dag}\hat{b}\hat{c}$ & \boldmath$\lambda_{b^*bc}(t)$ & \boldmath$-\frac{1}{2}(\epsilon_a\bar{\omega}_a u_{ab}^2 u_{ac}^2+\epsilon_b\bar{\omega}_b u_{bb}^2 u_{bc}^2) \eta_c^*(t)$ & \boldmath $-5.395\eta_c^*(t)$\\
\hline
$\hat{c}^{\dag}\hat{c}\hat{a}$ & $\lambda_{c^*ca}(t)$ & $ -\frac{1}{2}(\epsilon_a\bar{\omega}_a u_{ac}^3 u_{aa}+\epsilon_b\bar{\omega}_b u_{bc}^3 u_{ba}) \eta_c^*(t)$ & $-0.564\eta_c^*(t)$\\
\hline
$\hat{c}^{\dag}\hat{c}\hat{b}$ & $\lambda_{c^*cb}(t)$ & $ -\frac{1}{2}(\epsilon_a\bar{\omega}_a u_{ac}^3 u_{ab}+\epsilon_b\bar{\omega}_b u_{bc}^3 u_{bb}) \eta_c^*(t)$ & $-0.527\eta_c^*(t)$\\
\hline\hline
$\frac{1}{2}\hat{a}^{\dag}\hat{a}^{\dag}\hat{a}\hat{a}$ & $\alpha_a$ & $-\frac{1}{4}\left(\epsilon_a\bar{\omega}_a u_{aa}^4+\epsilon_b\bar{\omega}_b u_{ba}^4\right)$ & $-249.164$ \\
\hline
$\frac{1}{2}\hat{b}^{\dag}\hat{b}^{\dag}\hat{b}\hat{b}$ & $\alpha_b$ & $-\frac{1}{4}\left(\epsilon_a\bar{\omega}_a u_{ab}^4+\epsilon_b\bar{\omega}_b u_{bb}^4\right)$ & $-269.458$\\
\hline
$\frac{1}{2}\hat{c}^{\dag}\hat{c}^{\dag}\hat{c}\hat{c}$ & $\alpha_c$  & $-\frac{1}{4}\left(\epsilon_a\bar{\omega}_a u_{ac}^4+\epsilon_b\bar{\omega}_b u_{bc}^4\right)$ & $-0.056$ \\
\hline\hline
$\hat{a}^{\dag}\hat{a}\hat{b}^{\dag}\hat{b}$ & $2\chi_{ab}$ & $-\frac{1}{2}\left(\epsilon_a\bar{\omega}_a u_{aa}^2u_{ab}^2+\epsilon_b\bar{\omega}_b u_{ba}^2u_{bb}^2\right)$ & $-0.309$ \\
\hline
$\hat{a}^{\dag}\hat{a}\hat{c}^{\dag}\hat{c}$ & $2\chi_{ac}$ & $-\frac{1}{2}\left(\epsilon_a\bar{\omega}_a u_{aa}^2u_{ac}^2+\epsilon_b\bar{\omega}_b u_{ba}^2u_{bc}^2\right)$ & $-5.371$\\
\hline
$\hat{b}^{\dag}\hat{b}\hat{c}^{\dag}\hat{c}$ & $2\chi_{bc}$ & $-\frac{1}{2}\left(\epsilon_a\bar{\omega}_a u_{ab}^2u_{ac}^2+\epsilon_b\bar{\omega}_b u_{bb}^2u_{bc}^2\right)$ & $-5.395$\\
\hline
\end{tabular}
\caption{Dominant multimode interaction contributions (causing at most single-excitation transitions) via normal ordering of the \textit{quartic} nonlinearity in Eq.~(\ref{Eq:DispTrans-Hs(t) Trans}). For brevity, the Hermitian conjugates of the unbalanced operators are \textit{not} shown. From top to bottom, the contributions include direct drive on qubits coming from both linear charge and nonlinear flux hybridization, static and dynamic frequency shifts, exchange interactions, self and cross number-quadrature interactions, and self-Kerr (anharmonicity) and cross-Kerr interactions. Terms equivalent to $\hat{\sigma}_{a,b}^{z}\hat{c}+\text{H.c.}$ of the dispersive JC model, mainly responsible for the effective RIP interaction, are highlighted in boldface. Importantly, we find $\lambda_{a^*ac}=2\chi_{ac}\eta_c^*(t)$, $\lambda_{b^*bc}=2\chi_{bc}\eta_c^*(t)$, which is consistent with adopting phenomenological models from the outset. For simplicity, the estimates are given after doing the RWA on the displacement. For instance, terms that are proportional to $|\eta_c(t)|^2\eta_c^*(t)$ come from $[\eta_c(t)e^{-i\omega_d t}+\eta_c^*(t)e^{+i\omega_d t}]^3$. Circuit parameters are the same as Fig~\ref{fig:Model-NormNormQuantities}. According to Eqs.~(\ref{Eq:NormModeHam-U Sol})--(\ref{Eq:NormModeHam-V Sol}), the flux and charge hybridization coefficients are found as $\mathbf{U}=\big(\begin{smallmatrix}
0.994229 & -0.0217913 & 0.103214 \\
 0.0112824 & 0.994923 & 0.0995246 \\
 -0.0805908 & -0.0853793 & 0.992933 \\
\end{smallmatrix}\big)$ and $\mathbf{V}=\big(\begin{smallmatrix}
 0.997153 & -0.0192382 & 0.0792791 \\
 0.0128348 & 0.996284 & 0.0867091 \\
 -0.104939 & -0.0978607 & 0.990185 \\
\end{smallmatrix}\big)$. Furthermore, based on Eq.~(\ref{Eq:NormModeHam-Def of tilde(omega)_j}), the normal mode frequencies are found as $\tilde{\omega}_a /2\pi \approx 5375.850$ MHz, $\tilde{\omega}_b /2\pi \approx 6107.200$ MHz and $\tilde{\omega}_c /2\pi \approx 7019.430$ MHz.}
\label{Tab:MLM-Terms in H4_displaced}
\end{table*}
%%%%%%%%%%%%%%%%%%%%%%%%%%%%%%%%%%%%%%%%%%%%%%%%%%%%%%%%%%%%%%%%%%%%%%%%%%%%%%%%%

Lastly, we discuss the benefit of DRAG \cite{Motzoi_Simple_2009, Gambetta_Analytic_2011, Malekakhlagh_Mitigating_2021} in mitigating the resonator leakage. Consider a complex-valued control pulse as $\Omega_c (t) = \Omega_{cy} - i \Omega_{cx}(t)$, and, for simplicity, assume that the contribution due to $\alpha_c$ in Eq.~(\ref{Eq:DispTrans-Cond for eta_c(t)}) is negligible. In the frequency domain, the solution reads $\tilde{\eta}_c(\omega)= -\tilde{\Omega}_c(\omega)/[2(\omega+\Delta_{cd})]$. To mitigate leakage, the pulse should have minimal frequency content at $\omega=-\Delta_{cd}$. Consider an Ansatz for a DRAG pulse as
\begin{subequations}
\begin{align}
&\Omega_{cy}(t)=\Omega_c P_{\text{nc}}(t;0,\tau) \;,
\label{Eq:ResRes-YDRAGAnsatz x}\\
&\Omega_{cx}(t)=\frac{1}{\Delta_D} \Omega_c \dot{P}_{\text{nc}}(t;0,\tau) 
\label{Eq:ResRes-YDRAGAnsatz y}\;,
\end{align}
\end{subequations}
with $\Delta_D$ as the DRAG parameter. Given that $\partial_t \leftrightarrow i\omega$, the corresponding transfer function for DRAG is $\tilde{\Omega}_c(\omega)=\Omega_c T_D(\omega)\tilde{P}_{\text{nc}}(\omega)$ where $T_D(\omega)=1+\omega/\Delta_D$. Therefore, DRAG acts as an effective notch filter and setting $\Delta_D=\Delta_{cd}$ results in a zero in the pulse spectrum. In Fig.~\ref{fig:Leak-2DSweepWithDRAG}, using the exact ab-initio simulation, it is shown that such a DRAG pulse can improve the background resonator leakage by at least one order of magnitude at small $\Delta_{cd}$. 

In summary, we have characterized the resonator response in terms of a driven classical Kerr oscillator. Comparing two pulse shapes, Gaussian and nested cosine, we confirmed that the nested cosine pulse results in slightly improved resonator leakage due to smoother rise and fall. We introduced a classical leakage measure in Eq.~(\ref{Eq:ResRes-Def of L(theta,theta_G)}) and Fig.~\ref{Fig:ResRes-LeakageFunction}, which characterizes the trade-off between pulse time and resonator-drive detuning. Furthremore, employing a DRAG pulse proves to be very helpful in suppressing background leakage (see Sec.~\ref{Sec:Leak} and Fig.~\ref{fig:Leak-2DSweepWithDRAG}).  
%%%%%%%%%%%%%%%%%%%%%%%%%%%%%%%%%%%%%%%%%%%%%%%%%%%%%%%%%%%%%%%%%%%%%%%%%%%%%%%%

%%%%%%%%%%%%%%%%%%%% App:Starting Hamiltonin for RIP gate %%%%%%%%%%%%%%%%%%%%%%%
\section{Approximate ab-initio model}
\label{App:MLM}

In this appendix, starting from the displaced Hamiltonian~(\ref{Eq:DispTrans-Hs(t) Trans}), we construct an \textit{approximate} ab-initio model Hamiltonian. There are numerous processes that originate from the Taylor expansion of the cosine nonlinearity. Here, up to the quartic expansion, we read off dominant interaction terms by first performing normal ordering of the nonlinearity using SNEG \cite{Zitko_Sneg_2011} and then regrouping the terms into zeroth order and interaction parts. The advantage of normal ordering is that it leads to a lower number of interaction forms while applying SWPT (discussed in Appendix~\ref{App:EffRIPHam}).

We group the terms in Eq.~(\ref{Eq:DispTrans-Hs(t) Trans}) as 
\begin{align}
\HO_s(t)=\HO_0+\HO_{\text{int}}(t) \;,
\end{align}
where $\HO_0$ contains \textit{time-independent diagonal} contributions and up to the quartic expansion is the same as the Kerr model:
\begin{align} 
\begin{split}
\HO_0&=\sum\limits_{j=a,b,c}\left[\omega_j\hat{n}_j+\frac{\alpha_j}{2}\hat{n}_j(\hat{n}_j-1)\right]\\
&+\sum\limits_{j,k=a,b,c \atop j > k} 2\chi_{jk}\hat{n}_j\hat{n}_k \;.
\end{split}
\label{Eq:MLM-Def of H0}
\end{align} 
In Eq.~(\ref{Eq:MLM-Def of H0}), $\omega_j\equiv \tilde{\omega}_j+\delta_{Sj}$ is the renormalized frequency for mode $j=a,b,c$ accounting for the static frequency shift. Effective anharmonicity of the normal modes and the pairwise cross-Kerr interactions are denoted by $\alpha_j$ and $2\chi_{jk}$ for $j,k=a,b,c$ (see Table~\ref{Tab:MLM-Terms in H4_displaced} for the definitions). Moreover, $\HO_{\text{int}}(t)$ is the perturbation consisting of the rest of the \textit{time-dependent} nonlinear contributions of the generic form
\begin{align}
(\hat{a}^{\dag})^{i}\hat{a}^{j}(\hat{b}^{\dag})^{k}\hat{b}^{l}(\hat{c}^{\dag})^{m}\hat{c}^{n}[\eta^*(t)]^{p}[\eta_c(t)]^q e^{i(p-q)\omega_d t}\;,
\label{Eq:MLM-interaction terms}
\end{align}
where the sum of exponents should match the order of expansion in the cosine nonlinearity. For our approximate model, we keep contributions that are either balanced (equal number of creation and annihilation) or have a comparably small energy difference (single-excitation) as
\begin{align} 
\begin{split}
\HO_{\text{int}}(t)&=\sum\limits_{j=a,b,c}\delta_{Dj}(t)\hat{n}_j\\
&+\sum\limits_{j=a,b}[\lambda_j(t)e^{i\omega_d t}\hat{j}+\text{H.c.}]\\
&+\sum\limits_{j,k=a,b,c \atop j>k}[\lambda_{j^*k}(t)\hat{j}^{\dag}\hat{k}+\text{H.c.}]\\
&+\sum\limits_{j,k=a,b,c}[\lambda_{j^*jk}(t)e^{i\omega_d t}\hat{n}_j\hat{k}+\text{H.c.}] \;. \\
\end{split}
\label{Eq:MLM-Hint(t) compact}
\end{align}
Table~\ref{Tab:MLM-Terms in H4_displaced} summarizes such interaction terms and provides an estimate based on quartic expansion of Eq.~(\ref{Eq:DispTrans-Hs(t) Trans}).

In summary, compared to the phenomenological Kerr model of Appendix~\ref{App:KM}, we accounted for a few additional interaction forms. Namely, the second line of Eq.~(\ref{Eq:MLM-Hint(t) compact}) is an unwanted drive term acting on the qubit modes. The third line contains exchange interactions between the normal modes. Finally, the last line contains all possible number-quadrature interactions. In Appendix~\ref{App:EffRIPHam}, we find that the dominant source of RIP interaction is an interplay between the interaction forms $\lambda_{a^*ac}(t)\hat{a}^{\dag}\hat{a}\hat{c}+\text{H.c.}$ and $\lambda_{b^*bc}(t)\hat{b}^{\dag}\hat{b}\hat{c}+\text{H.c.}$ . Moreover, from normal ordering of the nonlinearity, we obtain $\lambda_{a^*ac}(t)=2\chi_{ac} \hat{\eta}_c^*(t)$ and $\lambda_{b^*bc}(t)=2\chi_{bc} \hat{\eta}_c^*(t)$ in agreement with the phenomenological Kerr model (see Table~\ref{Tab:MLM-Terms in H4_displaced}).    

%%%%%%%%%%%%%%%%%%%%%%%%%%%%%%%%%%%%%%%%% App:L Effective multilevel RIP Ham%%%%%%%%%%%%%%%%%%%%%%%%%%%%%%%%%%%%%%%%%%%%
\section{Effective Hamiltonian based on the approximate ab-initio model}
\label{App:EffRIPHam}

Here, using time-dependent SWPT, we derive an effective RIP Hamiltonian based on the ab-initio model in Eqs.~(\ref{Eq:MLM-Def of H0}) and~(\ref{Eq:MLM-Hint(t) compact}). Compared to the phenomenological Kerr model in Appendix~\ref{App:KM}, we find corrections to both the \textit{intended} RIP interaction as well as a series of new effective interaction forms.

To simplify perturbation, we first obtain the interaction-frame Hamiltonian based on Eqs.~(\ref{Eq:MLM-Def of H0}) and~(\ref{Eq:MLM-Hint(t) compact}). Employing identities~(\ref{Eq:KM-f(nc)c=cf(nc-1)})--(\ref{Eq:KM-f(nc)c^d=c^df(nc+1)}), the annihilation operator for each mode transforms as
\begin{subequations}
\begin{align}
e^{i\HO_0 t}\hat{j} e^{-i\HO_0 t}=e^{-i\hat{\omega}_j t}\hat{j}, \quad j=a,b,c \;,
\label{Eq:EffRIPHam-IntPic of j}
\end{align}
where we have defined \textit{operator-valued} normal mode frequencies as 
\begin{align}
&\hat{\omega}_a \equiv \omega_a +\alpha_a \hat{n}_a+2\chi_{ab}\hat{n}_b+2\chi_{ac}\hat{n}_c \;,
\label{Eq:EffRIPHam-Def of hat(wa)}\\
&\hat{\omega}_b \equiv \omega_b +2\chi_{ab}\hat{n}_a+\alpha_b \hat{n}_b+2\chi_{bc}\hat{n}_c \;,
\label{Eq:EffRIPHam-Def of hat(wb)}\\
&\hat{\omega}_c \equiv \omega_c +2\chi_{ac}\hat{n}_a+2\chi_{bc}\hat{n}_b +\alpha_c \hat{n}_c \;.	
\label{Eq:EffRIPHam-Def of hat(wc)}
\end{align}
\end{subequations}
Using Eqs.~(\ref{Eq:EffRIPHam-IntPic of j})--(\ref{Eq:EffRIPHam-Def of hat(wc)}), we find the interaction-frame Hamiltonian as
\begin{align}
\begin{split}
\HO_{I}(t) &\equiv e^{i\HO_0 t}\HO_{\text{int}}(t)e^{-i\HO_0 t}\\
&=\sum\limits_{j=a,b,c}\delta_{Dj}(t)\hat{n}_j\\
&+\sum\limits_{j=a,b}[\lambda_j(t)e^{-i\hat{\Delta}_{jd} t}\hat{j}+\text{H.c.}]\\
&+\sum\limits_{j,k=a,b,c \atop j>k}[\lambda_{j^*k}(t)\hat{j}^{\dag}e^{i\hat{\Delta}_{jk} t}\hat{k}+\text{H.c.}]\\
&+\sum\limits_{j,k=a,b,c}[\lambda_{j^*jk}(t)e^{-i\hat{\Delta}_{kd} t}\hat{n}_j\hat{k}+\text{H.c.}] \;,
\end{split}
\label{Eq:EffRIPHam-H_I(t)}
\end{align}
with operator-valued detunings defined as $\hat{\Delta}_{jd}\equiv \hat{\omega}_j-\omega_d$ and $\hat{\Delta}_{jk}\equiv \hat{\omega}_j-\hat{\omega}_k$ for $j,k=a,b,c$.

Up to the first order, the effective Hamiltonian contains dynamic frequency shifts, the only diagonal contribution in Eq.~(\ref{Eq:EffRIPHam-H_I(t)}), as
\begin{align}
\HO_{I,\text{eff}}^{(1)}(t)=\sum\limits_{j=a,b,c}\delta_{Dj}(t)\hat{n}_j \;.
\label{Eq:EffRIPHam-H_I,eff^(1) Sol}
\end{align}
The SWPT generator $\hat{G}_1(t)$ is then the time-integral of the rest of off-diagonal terms as
\begin{align}
\begin{split}
\hat{G}_1(t)&=\sum\limits_{j=a,b}\int^{t} dt'[\lambda_j(t')e^{-i\hat{\Delta}_{jd} t'}\hat{j}+\text{H.c.}]\\
&+\sum\limits_{j,k=a,b,c \atop j>k}\int^{t} dt'[\lambda_{jk}(t')\hat{j}^{\dag}e^{i\hat{\Delta}_{jk}t'}\hat{k}+\text{H.c.}]\\
&+\sum\limits_{j,k=a,b,c}\int^{t} dt'[\lambda_{j^*jk}(t')e^{-i\hat{\Delta}_{kd} t'}\hat{n}_j\hat{k}+\text{H.c.}] \;.
\end{split}
\label{Eq:EffRIPHam-G1(t) Sol}
\end{align}

Replacing Eq.~(\ref{Eq:EffRIPHam-G1(t) Sol}) into the second order effective Hamiltonian~(\ref{Eq:TLM-H_I,eff^(2) Cond}), we find that the off-diagonal terms in $\hat{G}_1(t)$ can generate a variety of effective \textit{diagonal} interactions of the form
\begin{align} 
\begin{split}
&\HO_{I,\text{eff}}^{(2)}(t)=\sum\limits_{j=a,b,c}\hat{\delta}_{j}^{(2)}(t)\hat{n}_j+\sum\limits_{j=a,b,c}\frac{1}{2}\hat{\alpha}_j^{(2)}(t)\hat{n}_j^2\\
&+\sum\limits_{j=a,b,c}\hat{\beta}_j^{(2)}(t)\hat{n}_j^3+\sum\limits_{j,k=a,b,c\atop j>k}2\hat{\chi}_{jk}^{(2)}(t)\hat{n}_j\hat{n}_k\\
&+\sum\limits_{j,k=a,b,c\atop j>k}\hat{\chi}_{jjk}^{(2)}(t)\hat{n}_j^2\hat{n}_k +\chi_{abc}^{(2)}(t)\hat{n}_a\hat{n}_b\hat{n}_c\;.
\end{split}
\label{Eq:EffRIPHam-H_I,eff^(2) Sol}
\end{align}
The contributions in Eq.~(\ref{Eq:EffRIPHam-H_I,eff^(2) Sol}) can be summarized as second order dynamic frequency shifts, dynamic anharmonic shifts, RIP-like (number-number) interactions, higher order two-body interaction (number-squared-number) and three-body number interactions, respectively. Note that the normal resonator mode $\hat{c}$ denotes excitations on top of the coherent displacement $\eta_c(t)$, hence corrections involving $\hat{c}$ are expected to be orders of magnitude smaller than the ones dependent on $\eta_c(t)$.

In the following, we first provide the derivation for the intended RIP interaction, i.e. $\hat{\chi}_{ab}^{(2)}(t)$ and then quote the final result for the rest of the effective interaction rates. There are multiple processes that contribute to an effective $\chi_{ab}^{(2)}(t)$. From the dispersive JC model, we learned that the interplay between $\hat{\sigma}_a^{z}\hat{c}+\text{H.c.}$ and $\hat{\sigma}_b^{z}\hat{c}+\text{H.c.}$ is the dominant source. \textit{Multilevel} equivalent terms are $\hat{a}^{\dag}\hat{a}\hat{c}+\text{H.c.}$ and $\hat{b}^{\dag}\hat{b}\hat{c}+\text{H.c.}$, highlighted in boldface in Table~\ref{Tab:MLM-Terms in H4_displaced}. In addition, up to the 2nd order in SWPT, we find five more possibilities for mixing: (i) $\bm{\hat{a}^{\dag}\hat{a}\hat{c}+\textbf{H.c.}}$ and $\bm{\hat{b}^{\dag}\hat{b}\hat{c}+\textbf{H.c.}}$, (ii) $\hat{a}^{\dag}\hat{a}\hat{a}+\text{H.c.}$ and $\hat{b}^{\dag}\hat{b}\hat{a}+\text{H.c.}$, (iii) $\hat{b}^{\dag}\hat{b}\hat{b}+\text{H.c.}$ and $\hat{a}^{\dag}\hat{a}\hat{b}+\text{H.c.}$, (iv) $\hat{a}^{\dag}\hat{b}+\text{H.c.}$ and $\hat{a}^{\dag}\hat{b}+\text{H.c.}$, (v) $\hat{a}+\text{H.c.}$ and $\hat{b}^{\dag}\hat{b}\hat{a}+\text{H.c.}$, (vi) $\hat{b}+\text{H.c.}$ and $\hat{a}^{\dag}\hat{a}\hat{b}+\text{H.c.}$.

We next analyze case (i) in more detail. The non-zero contributions coming from the interplay of $\hat{a}^{\dag}\hat{a}\hat{c}+\text{H.c.}$ and $\hat{b}^{\dag}\hat{b}\hat{c}+\text{H.c.}$ is found from Eq.~(\ref{Eq:TLM-H_I,eff^(2) Cond}) as
\begin{widetext}
\begin{align}
\begin{split}
\frac{i}{2}\left[\int^{t} dt' \lambda_{a^*ac}(t')e^{-i\hat{\Delta}_{cd}t'}\hat{n}_a\hat{c},\lambda_{b^*bc}^*(t)\hat{n}_b\hat{c}^{\dag}e^{i\hat{\Delta}_{cd}t}\right] +\frac{i}{2}\left[\int^{t} dt' \lambda^*_{a^*ac}(t')\hat{n}_a\hat{c}^{\dag}e^{i\hat{\Delta}_{cd}t'},\lambda_{b^*bc}(t)e^{-i\hat{\Delta}_{cd}t}\hat{n}_b\hat{c}\right] \\
+\frac{i}{2}\left[\int^{t} dt' \lambda_{b^*bc}(t')e^{-i\hat{\Delta}_{cd}t'}\hat{n}_b\hat{c},\lambda_{a^*ac}^*(t)\hat{n}_a\hat{c}^{\dag}e^{i\hat{\Delta}_{cd}t}\right]+\frac{i}{2}\left[\int^{t} dt' \lambda^*_{b^*bc}(t')\hat{n}_b\hat{c}^{\dag}e^{i\hat{\Delta}_{cd}t'},\lambda_{a^*ac}(t)e^{-i\hat{\Delta}_{cd}t}\hat{n}_a\hat{c}\right] \;.
\end{split}
\label{Eq:EffRIPHam-aac&bbcInterplay}
\end{align}
The first and the third terms in Eq.~(\ref{Eq:EffRIPHam-aac&bbcInterplay}) are equal and the rest are Hermitian conjugate of the former. The first term can be simplified as
\begin{align}
\begin{split}
\frac{i}{2}\left[\int^{t} dt' \lambda_{a^*ac}(t')e^{-i\hat{\Delta}_{cd}t'}\hat{n}_a\hat{c},\lambda_{b^*bc}^*(t)\hat{n}_b\hat{c}^{\dag}e^{i\hat{\Delta}_{cd}t}\right]=-\frac{1}{2i} \int^{t} dt' \lambda_{a^*ac}(t')\lambda_{b^*bc}^*(t)\left[e^{-i\hat{\Delta}_{cd}t'}\hat{c},\hat{c}^{\dag}e^{i\hat{\Delta}_{cd}t}\right]\hat{n}_a\hat{n}_b \;.
\end{split}
\label{Eq:EffRIPHam-aac&bbcInterplay 1stTerm}
\end{align}
The commutator in Eq.~(\ref{Eq:EffRIPHam-aac&bbcInterplay 1stTerm}) can be evaluated using identities~(\ref{Eq:KM-f(nc)c=cf(nc-1)})--(\ref{Eq:KM-f(nc)c^d=c^df(nc+1)}) as
\begin{align}
\begin{split}
&\left[e^{-i\hat{\Delta}_{cd}t'}\hat{c},\hat{c}^{\dag}e^{i\hat{\Delta}_{cd}t}\right]=e^{i\hat{\Delta}_{cd}(t-t')}(\hat{n}_c+1)-e^{i(\hat{\Delta}_{cd}-\alpha_c)(t-t')}\hat{n}_c = e^{i\hat{\Delta}_{cd}(t-t')}+\left[1-e^{-i\alpha_c(t-t')}\right]e^{i\hat{\Delta}_{cd}(t-t')}\hat{n}_c \;.
\end{split}
\label{Eq:EffRIPHam-aac&bbc 1stTermCommutator}
\end{align}
Based on Eqs.~(\ref{Eq:EffRIPHam-aac&bbcInterplay 1stTerm})--(\ref{Eq:EffRIPHam-aac&bbc 1stTermCommutator}), on top of the desired $\hat{n}_a\hat{n}_b$ interaction, there also exists an effective $\hat{n}_a\hat{n}_b\hat{n}_c$ term. Substituting the $\hat{n}_a\hat{n}_b$ contribution into Eq.~(\ref{Eq:EffRIPHam-aac&bbcInterplay}) and further simplification yields
\begin{align}
-\Im\left\{2\int^{t}dt'\lambda_{a^*ac}(t')\lambda_{b^*bc}^*(t)e^{i\hat{\Delta}_{cd}(t-t')}\right\}\hat{n}_a\hat{n}_b \;.
\label{Eq:EffRIPHam-aac&bbcInterplay Sol}
\end{align}
Upon replacing $\lambda_{a^*ac}=2\chi_{ac}\eta_c^*(t)$ and $\lambda_{b^*bc}=2\chi_{bc}\eta_c^*(t)$, according to the normal ordering in Table~\ref{Tab:MLM-Terms in H4_displaced}, we find Eq.~(\ref{Eq:EffRIPHam-aac&bbcInterplay Sol}) to be in agreement with Eqs.~(\ref{Eq:KM-H_I,eff^(2) Sol})--(\ref{Eq:KM-Def of A_eta(na,nb)}) of the multilevel Kerr model.

All in all, from the six aforementioned possibilities, we obtain the following expression for $\hat{\chi}_{ab}^{(2)}(t)$:
\begin{align} 
\begin{split}
&\hat{\chi}_{ab}^{(2)}(t)=-\bm{\textbf{Im}\left\{\int^{t}dt'\lambda_{a^*ac}(t')\lambda_{b^*bc}^*(t)e^{i\hat{\Delta}_{cd}(t-t')}\right\}}\\
&-\Im\left\{\int^{t}dt'\lambda_{a^*aa}(t')\lambda_{b^*ba}^*(t)e^{i\hat{\Delta}_{ad}(t-t')}\left[1+e^{-i\alpha_a(t-t')}\right]\right\} -\Im\left\{\int^{t}dt'\lambda_{b^*bb}(t')\lambda_{a^*ab}^*(t)e^{i\hat{\Delta}_{bd}(t-t')}\left[1+e^{-i\alpha_b(t-t')}\right]\right\}\\
&-\Im\left\{\int^{t}dt'\lambda_{a}(t')\lambda_{b^*ba}^*(t) e^{i\hat{\Delta}_{ad}(t-t')}\left[1-e^{-i\alpha_a(t-t')}\right]\right\}-\Im\left\{\int^{t}dt'\lambda_{b}(t')\lambda_{a^*ab}^*(t) e^{i\hat{\Delta}_{bd}(t-t')}\left[1-e^{-i\alpha_b(t-t')}\right]\right\}\\
&-\frac{1}{2}\Im\left\{\int^{t}dt'\lambda_{a^*b}(t')\lambda_{a^*b}^*(t)\left[e^{-i\left(\hat{\Delta}_{ab}-\alpha_a+2\chi_{ab}\right)(t-t')}-e^{-i\left(\hat{\Delta}_{ab}+\alpha_b-2\chi_{ab}\right)(t-t')}\right]\right\}\;.
\end{split}
\label{Eq:EffRIPHam-Chi_ab^(2) Sol}
\end{align}
Under adiabatic approximation, Eq.~(\ref{Eq:EffRIPHam-Chi_ab^(2) Sol}) simplifies to
\begin{align} 
\begin{split}
\hat{\chi}_{ab}^{(2)}(t) \approx & \bm{-\frac{\lambda_{a^*ac}(t)\lambda_{b^*bc}^*(t)}{\hat{\Delta}_{cd}}} -\lambda_{a^*aa}(t)\lambda_{b^*ba}^*(t)\left(\frac{1}{\hat{\Delta}_{ad}}+\frac{1}{\hat{\Delta}_{ad}-\alpha_a}\right)-\lambda_{b^*bb}(t)\lambda_{a^*ab}^*(t)\left(\frac{1}{\hat{\Delta}_{bd}}+\frac{1}{\hat{\Delta}_{bd}-\alpha_b}\right)\\
&-\lambda_a(t)\lambda^*_{b^*ba}(t)\left(\frac{1}{\hat{\Delta}_{ad}}-\frac{1}{\hat{\Delta}_{ad}-\alpha_a}\right)-	\lambda_b(t)\lambda^*_{a^*ab}(t) \left(\frac{1}{\hat{\Delta}_{bd}}-\frac{1}{\hat{\Delta}_{bd}-\alpha_b}\right)\\
&-\frac{1}{2}\left|\lambda_{a^*b}(t)\right|^2\left(\frac{1}{\hat{\Delta}_{ab}-\alpha_a+2\chi_{ab}}-\frac{1}{\hat{\Delta}_{ab}+\alpha_b-2\chi_{ab}}\right)\;.
\end{split}
\label{Eq:EffRIPHam-Chi_ab^(2) AdiabaticSol}
\end{align}
Firstly, note the operator structure of the effective RIP interaction in Eq.~(\ref{Eq:EffRIPHam-Chi_ab^(2) AdiabaticSol}). The advantage of such a compact representation is that the resulting \textit{state-dependent rates} can be immediately read off of Eq.~(\ref{Eq:EffRIPHam-Chi_ab^(2) AdiabaticSol}) based on Eqs.~(\ref{Eq:EffRIPHam-Def of hat(wa)})--(\ref{Eq:EffRIPHam-Def of hat(wc)}). For instance, in the computational basis, the only non-zero matrix element is found as  
\begin{align}
\begin{split}
&\bra{1_a1_b0_c}2\hat{\chi}_{ab}^{(2)}(t)\hat{n}_a\hat{n}_b\ket{1_a1_b0_c}= \bm{-\frac{2\lambda_{a^*ac}(t)\lambda_{b^*bc}^*(t)}{\Delta_{cd}+2\chi_{ac}+2\chi_{bc}}} -2\lambda_{a^*aa}(t)\lambda_{b^*ba}^*(t)\left(\frac{1}{\Delta_{ad}+\alpha_a+2\chi_{ab}}+\frac{1}{\Delta_{ad}+2\chi_{ab}}\right)\\
&-2\lambda_{b^*bb}(t)\lambda_{a^*ab}^*(t)\left(\frac{1}{\Delta_{bd}+\alpha_b+2\chi_{ab}}+\frac{1}{\Delta_{bd}+2\chi_{ab}}\right)-2\lambda_{a}(t)\lambda_{b^*ba}^*(t)\left(\frac{1}{\Delta_{ad}+\alpha_a+2\chi_{ab}}-\frac{1}{\Delta_{ad}+2\chi_{ab}}\right)\\
&-2 \lambda_{b}(t)\lambda_{a^*ab}^*(t) \left(\frac{1}{\Delta_{bd}+\alpha_b+2\chi_{ab}}-\frac{1}{\Delta_{bd}+2\chi_{ab}}\right)-|\lambda_{a^*b}(t)|^2 \left(\frac{1}{\Delta_{ab}-\alpha_b+2\chi_{ab}}-\frac{1}{\Delta_{ab}+\alpha_a-2\chi_{ab}}\right)
\end{split}
\label{Eq:EffRIPHam-Chi_ab^(2) AdiabaticSol ExpVal}
\end{align}
Secondly, since $\lambda_{a^*ac}=2\chi_{ac}\eta_c^*(t)$ and $\lambda_{b^*bc}=2\chi_{bc}\eta_c^*(t)$, the first term in Eq.~(\ref{Eq:EffRIPHam-Chi_ab^(2) AdiabaticSol}) matches the first term of Eq.~(\ref{Eq:KM-H_I,eff^(2) AdiabSol}) from the multilevel Kerr model. 

Following a similar derivation as the one provided in Eqs.~(\ref{Eq:EffRIPHam-aac&bbcInterplay})--(\ref{Eq:EffRIPHam-Chi_ab^(2) Sol}), we find expressions for other effective interaction terms in Hamiltonian~(\ref{Eq:EffRIPHam-H_I,eff^(2) Sol}). For instance, the effective RIP-like interaction between modes a and c is found as
\begin{subequations}
\begin{align} 
\begin{split}
&\hat{\chi}_{ac}^{(2)}(t)=-\Im\left\{\int^{t}dt'\lambda_{a^*ab}(t')\lambda_{c^*cb}^*(t)e^{i\hat{\Delta}_{bd}(t-t')}\right\}-\Im\left\{\int^{t}dt'\lambda_{a^*aa}(t')\lambda_{c^*ca}^*(t)e^{i\hat{\Delta}_{ad}(t-t')}\left[1+e^{-i\alpha_a(t-t')}\right]\right\}\\
&-\Im\left\{\int^{t}dt'\lambda_{c^*cc}(t')\lambda_{a^*ac}^*(t)e^{i\hat{\Delta}_{cd}(t-t')}\left[1+e^{-i\alpha_c(t-t')}\right]\right\}-\Im\left\{\int^{t}dt'\lambda_{a}(t')\lambda_{c^*ca}^*(t) e^{i\hat{\Delta}_{ad}(t-t')}\left[1-e^{-i\alpha_a(t-t')}\right]\right\} \\
&-\frac{1}{2}\Im\left\{\int^{t}dt'\lambda_{a^*c}(t')\lambda_{a^*c}^*(t)\left[e^{-i\left(\hat{\Delta}_{ac}-\alpha_a+2\chi_{ac}\right)(t-t')}-e^{-i\left(\hat{\Delta}_{ac}+\alpha_c-2\chi_{ac}\right)(t-t')}\right]\right\} \;.
\end{split}
\label{Eq:EffRIPHam-Chi_ac^(2) Sol}
\end{align}
Similarly, $\hat{\chi}_{bc}^{(2)}(t)$ reads
\begin{align} 
\begin{split}
&\hat{\chi}_{bc}^{(2)}(t)=-\Im\left\{\int^{t}dt'\lambda_{b^*ba}(t')\lambda_{c^*ca}^*(t)e^{i\hat{\Delta}_{ad}(t-t')}\right\}-\Im\left\{\int^{t}dt'\lambda_{b^*bb}(t')\lambda_{c^*cb}^*(t)e^{i\hat{\Delta}_{bd}(t-t')}\left[1+e^{-i\alpha_b(t-t')}\right]\right\}\\
&-\Im\left\{\int^{t}dt'\lambda_{c^*cc}(t')\lambda_{b^*bc}^*(t)e^{i\hat{\Delta}_{cd}(t-t')}\left[1+e^{-i\alpha_c(t-t')}\right]\right\}-\Im\left\{\int^{t}dt'\lambda_{b}(t')\lambda_{c^*cb}^*(t) e^{i\hat{\Delta}_{bd}(t-t')}\left[1-e^{-i\alpha_b(t-t')}\right]\right\} \\
&-\frac{1}{2}\Im\left\{\int^{t}dt'\lambda_{b^*c}(t')\lambda_{b^*c}^*(t)\left[e^{-i\left(\hat{\Delta}_{bc}-\alpha_b+2\chi_{bc}\right)(t-t')}-e^{-i\left(\hat{\Delta}_{bc}+\alpha_c-2\chi_{bc}\right)(t-t')}\right]\right\}\;.
\end{split}
\label{Eq:EffRIPHam-Chi_bc^(2) Sol}
\end{align}
\end{subequations}

Next, we consider possible interplays that lead to an effective second order dynamic frequency shift for each normal mode. For instance, for qubit a, we find that the interplay between (i) $\hat{a}+\text{H.c.}$ and $\hat{a}+\text{H.c.}$, (ii) $\hat{a}+\text{H.c.}$ and $\hat{a}^{\dag}\hat{a}\hat{a}+\text{H.c.}$, (iii) $\hat{b}+\text{H.c.}$ and $\hat{a}^{\dag}\hat{a}\hat{b}+\text{H.c.}$, (iv) $\hat{a}^{\dag}\hat{b}+\text{H.c.}$ and $\hat{a}^{\dag}\hat{b}+\text{H.c.}$, (v) $\hat{a}^{\dag}\hat{c}+\text{H.c.}$ and $\hat{a}^{\dag}\hat{c}+\text{H.c.}$ results in the second order dynamic frequency shift
\begin{subequations}
\begin{align} 
\begin{split}
&\hat{\delta}_{a}^{(2)}(t)=-\Im\left\{\int^{t} dt' \lambda_a(t')\lambda_{a}^*(t)e^{i\hat{\Delta}_{ad}(t-t')}\left[1-e^{-i\alpha_a (t-t')}\right]\right\}\\
&-\Im\left\{\int^{t} dt' 2\lambda_a(t')\lambda_{a^*aa}^*(t)e^{i\hat{\Delta}_{ad}(t-t')}\left[1+e^{-i\alpha_a (t-t')}\right]\right\}-\Im\left\{\int^{t} dt' 2\lambda_b(t')\lambda_{a^*ab}^*(t)e^{i\hat{\Delta}_{bd}(t-t')}\right\}\\
&-\Im\left\{\int^{t}dt' \lambda_{a^*b}(t')\lambda_{a^*b}^*(t) e^{-i(\hat{\Delta}_{ab}-\alpha_a+2\chi_{ab})(t-t')} \right\}-\Im\left\{\int^{t}dt' \lambda_{a^*c}(t')\lambda_{a^*c}^*(t) e^{-i(\hat{\Delta}_{ac}-\alpha_a+2\chi_{ac})(t-t')} \right\} \;.
\end{split}
\label{Eq:EffRIPHam-del_a^(2) Sol}
\end{align}
In a similar manner, the dynamic frequency shift for qubit b reads
\begin{align} 
\begin{split}
&\hat{\delta}_{b}^{(2)}(t)=-\Im\left\{\int^{t} dt' \lambda_b(t')\lambda_{b}^*(t)e^{i\hat{\Delta}_{bd}(t-t')}\left[1-e^{-i\alpha_b (t-t')}\right]\right\}\\
&-\Im\left\{\int^{t} dt' 2\lambda_b(t')\lambda_{b^*bb}^*(t)e^{i\hat{\Delta}_{bd}(t-t')}\left[1+e^{-i\alpha_b (t-t')}\right]\right\}-\Im\left\{\int^{t} dt' 2\lambda_a(t')\lambda_{b^*ba}^*(t)e^{i\hat{\Delta}_{ad}(t-t')}\right\}\\
&-\Im\left\{\int^{t}dt' \lambda_{a^*b}^*(t')\lambda_{a^*b}(t) e^{-i(-\hat{\Delta}_{ab}-\alpha_b+2\chi_{ab})(t-t')} \right\}-\Im\left\{\int^{t}dt' \lambda_{b^*c}(t')\lambda_{b^*c}^*(t) e^{-i(\hat{\Delta}_{bc}-\alpha_b+2\chi_{bc})(t-t')} \right\}.
\end{split}
\label{Eq:EffRIPHam-del_b^(2) Sol}
\end{align}
Moreover, for the normal resonator mode we find	
\begin{align} 
\begin{split}
&\hat{\delta}_{c}^{(2)}(t)=-\Im\left\{\int^{t} dt' 2\lambda_a(t')\lambda_{c^*ca}^*(t)e^{i\hat{\Delta}_{ad}(t-t')}\right\}-\Im\left\{\int^{t} dt' 2\lambda_b(t')\lambda_{c^*cb}^*(t)e^{i\hat{\Delta}_{bd}(t-t')}\right\}\\
&-\Im\left\{\int^{t}dt' \lambda_{a^*c}^*(t')\lambda_{a^*c}(t) e^{-i(-\hat{\Delta}_{ac}-\alpha_c+2\chi_{ac})(t-t')} \right\}-\Im\left\{\int^{t}dt' \lambda_{b^*c}^*(t')\lambda_{b^*c}(t) e^{-i(-\hat{\Delta}_{bc}-\alpha_c+2\chi_{bc})(t-t')} \right\} \;.
\end{split}
\label{Eq:EffRIPHam-del_c^(2) Sol}
\end{align}
\end{subequations}
A second order anharmonic shift can come from any of the mixings (i) $\hat{j}+\text{H.c.}$ and $\hat{j}^{\dag}\hat{j}\hat{j}+\text{H.c.}$, (ii) $\hat{j}^{\dag}\hat{j}\hat{j}+\text{H.c.}$  and $\hat{j}^{\dag}\hat{j}\hat{j}+\text{H.c.}$, (iii) $\hat{j}^{\dag}\hat{j}\hat{k}+\text{H.c.}$  and $\hat{j}^{\dag}\hat{j}\hat{k}+\text{H.c.}$, $k\neq j$, 
leading to the effective rate
\begin{align}
\begin{split}
&\hat{\alpha}_j^{(2)}(t) = -\Im\left\{\int^{t} dt' 4\lambda_{j}(t')\lambda_{j^*jj}^*(t)e^{i\hat{\Delta}_{jd}(t-t')}\left[1-e^{-i\alpha_j (t-t')}\right]\right\} \\
&-\Im\left\{\int^{t} dt' 2\lambda_{j^*jj}(t')\lambda_{j^*jj}^*(t)e^{i\hat{\Delta}_{jd}(t-t')}\right\}-\Im\left\{\sum\limits_{k\neq j}\int^{t} dt' 2\lambda_{j^*jk}(t')\lambda_{j^*jk}^*(t)e^{i\hat{\Delta}_{kd}(t-t')}\right\} \;, 
\end{split}
\end{align}
for each mode $j=a,b,c$. A similar correction of the form $\hat{\beta}_j^{(2)}(t)\hat{n}_j^3$ is also found as
\begin{align}
\hat{\beta}_j^{(2)}(t)= -\Im\left\{\int^{t} dt' \lambda_{j^*jj}(t')\lambda_{j^*jj}^*(t)e^{i\hat{\Delta}_{jd}(t-t')}\left[1-e^{-i\alpha_j (t-t')}\right]\right\} \;, \quad  j=a,b,c \;,
\end{align}
as a result of mixing between $\hat{j}^{\dag}\hat{j}\hat{j}+\text{H.c.}$  and $\hat{j}^{\dag}\hat{j}\hat{j}+\text{H.c.}$ . Higher order two-body interactions of the form $\hat{\chi}_{jjk}\hat{n}_j^2\hat{n}_k$ are obtained as
\begin{align}
\begin{split}
\hat{\chi}_{jjk}^{(2)}(t) = &-\Im\left\{\int^{t} dt' 2\lambda_{j^*jj}(t')\lambda_{k^*kj}^*(t)e^{i\hat{\Delta}_{jd}(t-t')}\left[1-e^{-i\alpha_j (t-t')}\right]\right\}\\
&-\Im\left\{\int^{t} dt' \lambda_{j^*jk}(t')\lambda_{j^*jk}^*(t)e^{i\hat{\Delta}_{kd}(t-t')}\left[1-e^{-i\alpha_k (t-t')}\right]\right\}  \;, \quad j\neq k \in \{a,b,c\}.
\end{split}
\end{align}
Lastly, there will also be an effective three body interaction of the form $\hat{\chi}_{abc}^{(2)}(t)$ as
\begin{align} 
\begin{split}
\hat{\chi}_{abc}^{(2)}(t)=&-\Im\left\{\int^{t}dt' 2\lambda_{b^*ba}(t)\lambda_{c^*ca}^*(t) e^{i\hat{\Delta}_{ad}(t-t')}\left[1-e^{-i\alpha_a(t-t')}\right] \right\} \\
&-\Im\left\{\int^{t}dt' 2\lambda_{a^*ab}(t)\lambda_{c^*cb}^*(t) e^{i\hat{\Delta}_{bd}(t-t')}\left[1-e^{-i\alpha_b(t-t')}\right] \right\} \\
&-\Im\left\{\int^{t}dt' 2\lambda_{a^*ac}(t)\lambda_{b^*bc}^*(t) e^{i\hat{\Delta}_{cd}(t-t')}\left[1-e^{-i\alpha_c(t-t')}\right] \right\} \;. 
\end{split}
\label{Eq:EffRIPHam-del_abc^(2) Sol}
\end{align}
\end{widetext}

To summarize, starting from the approximate ab-initio Hamiltonian in Eqs.~(\ref{Eq:MLM-Def of H0}) and~(\ref{Eq:MLM-Hint(t) compact}), which account for additional \textit{single-excitation} contributions compared to the Kerr model, and using SWPT, we have characterized effective interaction forms in Eqs.~(\ref{Eq:EffRIPHam-H_I,eff^(1) Sol}) and~(\ref{Eq:EffRIPHam-H_I,eff^(2) Sol}). The result of this appendix is used for comparison with the Kerr and the JC effective gate parameters in Fig.~\ref{fig:EffHam-GateParamsComparison} of the main text.  
%%%%%%%%%%%%%%%%%%%%%%%%%%%%%%%%%%%%%%%%%%%%%%%%%%%%%%%%%%%%%%%%%%%%%%%%%%%%%%%%

%%%%%%%%%%%%%%%%%%%%%%%%%%%%%%%%%%%%%% App: Mechanisms for Leakage %%%%%%%%%%%%%%
\section{Quantum processes for leakage}
\label{App:LeakMech}
We consider a three-level toy model and study quantum processes for leakage. The discussion here is based on the Magnus expansion \cite{Magnus_Exponential_1954, Blanes_Magnus_2009, Blanes_Pedagogical_2010} up to the 2nd order. 

\begin{figure}[t!]
\centering
\includegraphics[scale=0.30]{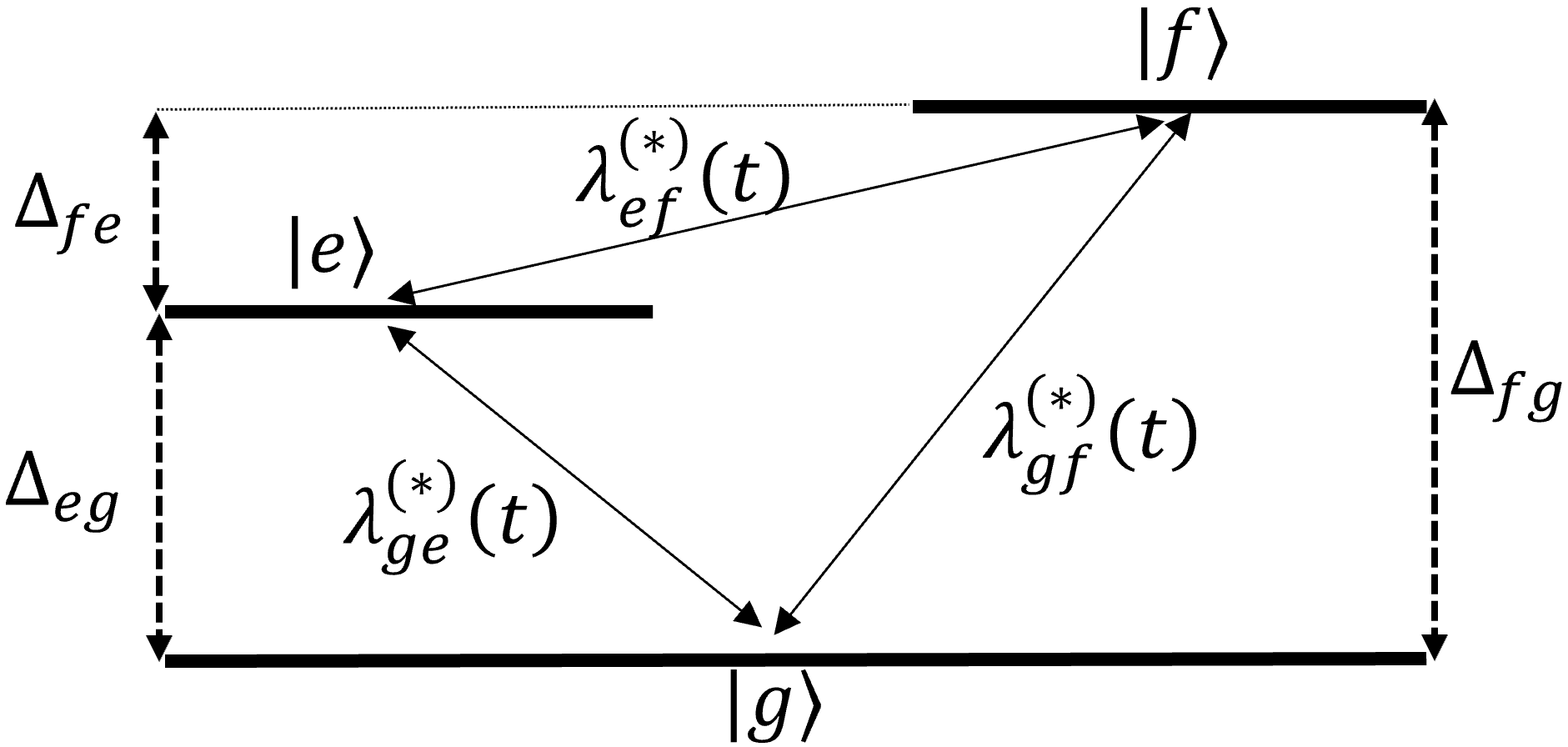}
\caption{A three-level toy model with  pairwise time-dependent interaction. Leakage may occur both directly (first order) but also via in-between transitions (higher order).}
\label{Fig:ToyLeak-ThreeLevelLeakage}
\end{figure}

The system Hamiltonian for the three-level system shown in Fig.~\ref{Fig:ToyLeak-ThreeLevelLeakage} reads
\begin{subequations}
\begin{align}
\HO_{0}=\begin{bmatrix}
0 & 0 & 0 \\
0 & \Delta_\text{eg} & 0 \\
0 & 0 & \Delta_\text{fg} 
\end{bmatrix} \;,
\label{Eq:ToyLeak-Def of H0}
\end{align}
which is represented over the system eigenbasis $\{\ket{g}, \ket{e} , \ket{f} \}$. Quantum states $\ket{g}$, $\ket{e}$ and $\ket{f}$ can in principle be any arbitrary states in a multilevel quantum system. Moreover, we consider a pairwise interaction Hamiltonian as
\begin{align}
\HO_{\text{int}}(t)=\begin{bmatrix}
0 & \lambda_{\text{ge}}^*(t) & \lambda_{\text{gf}}^*(t) \\
\lambda_{\text{ge}}(t)  & 0 & \lambda_{\text{ef}}^*(t) \\
\lambda_{\text{gf}}(t)  & \lambda_{\text{ef}}(t) & 0 
\end{bmatrix} \;,
\label{Eq:ToyLeak-Def of Hint}
\end{align}
\end{subequations}
where $\lambda_{jk}$(t) $(j\neq k)$ is the \textit{generic} time-dependent coupling between states $\ket{j}$ and $\ket{k}$. Here, we define leakage as the probability of ending up in state $\ket{f}$ at the end of the pulse, i.e. $t=\tau$, as
\begin{align}
p_{L,j} \equiv \left|\bra{f}\hat{U}_{I}(\tau,0)\ket{j}\right|^2 \;, 
\label{Eq:ToyLeak-Def of Pleak}
\end{align}
for $j=g, \ e$.
 
%%%%%%%%%%%%%%%%%%%%%%%%% Fig: ThreeLevelLeakage %%%%%%%%%%%%%%%%%%%%%%%%%%%%%%%
%%%%%%%%%%%%%%%%%%%%%%%%%%%%%%%%%%%%%%%%%%%%%%%%%%%%%%%%%%%%%%%%%%%%%%%%%%%%%%%%%

The interaction-frame Hamiltonian $\HO_{\text{I}}(t)\equiv e^{i \HO_0 t}\HO_{\text{int}}(t)e^{-i \HO_0 t}$ is found as
\begin{align}
\HO_{\text{I}}(t)=\begin{bmatrix}
0 & \lambda_{\text{ge}}^*(t)e^{-i\Delta_\text{eg}t} & \lambda_{\text{gf}}^*(t)e^{-i\Delta_{\text{fg}}t} \\
\lambda_{\text{ge}}(t)e^{i\Delta_{\text{eg}}t} & 0 & \lambda_{\text{ef}}^*(t)e^{-i\Delta_{\text{fe}}t} \\
\lambda_{\text{gf}}(t)e^{i\Delta_{\text{fg}}t} & \lambda_{\text{ef}}(t)e^{i\Delta_{\text{fe}}t} & 0 
\end{bmatrix} \;,
\label{Eq:ToyLeak-Def of HI}
\end{align}
where for simplicity we have defined $\Delta_{\text{fe}}\equiv \Delta_{\text{fg}}-\Delta_{\text{eg}}$. The time-evolution operator in the interaction frame reads
\begin{align}
\hat{U}_I(t,0)=\mathbb{T}\exp \left[-i \int_{0}^{t}dt' \HO_I(t')\right] \;,
\label{Eq:ToyLeak-Def of UI(t,0)}
\end{align}
where $\mathbb{T}$ is the time-ordering operator. In the Magnus method, we solve for the generator of $\hat{U}_I(t,0)$ perturbatively as 
\begin{subequations}
\begin{align}
\begin{split}
\hat{U}_{I}(t,0) & \equiv e^{-i\hat{K}(t,0)} =\hat{I}-i\hat{K}_1(t,0)\\
&-i\hat{K}_2(t,0)-\frac{1}{2}\hat{K}_1^2(t,0)+O\left(\HO_I^3(t)\right) \;,
\end{split}
\label{Eq:ToyLeak-MagExp}
\end{align}
where $\hat{K}_n(t,0)$ is the $n$th order solution for the generator. Up to the 2nd order we find \cite{Blanes_Pedagogical_2010}
\begin{align}
& \hat{K}_1(t,0) \equiv \int_{0}^{t} dt' \HO_I(t') \;,
\label{Eq:ToyLeak-Def of K1(t,0)}\\
& \hat{K}_2(t,0) \equiv -\frac{i}{2}\int_{0}^{t} dt' \int_{0}^{t'} dt'' \left[\HO_I(t'),\HO_I(t'')\right] \;.
\label{Eq:ToyLeak-Def of K2(t,0)}
\end{align}
\end{subequations}
  
Using Eqs.~(\ref{Eq:ToyLeak-MagExp})--(\ref{Eq:ToyLeak-Def of K2(t,0)}), we calculate the transition probabilities. For instance, the probability amplitude for $\ket{g}\rightarrow\ket{f}$ at $t=\tau$ reads
\begin{align}
\begin{split}
&\bra{\text{f}}\hat{U}_{I}(\tau,0)\ket{g}=-i\int_{0}^{\tau}dt' \lambda_{\text{gf}}(t')e^{i\Delta_\text{fg}t'}\\
&-\frac{1}{2}\left(\int_{0}^{\tau}dt'\lambda_{\text{ef}}(t')e^{i\Delta_{\text{fe}}t'}\right)\left(\int_{0}^{\tau} dt'\lambda_{\text{ge}}(t')e^{i\Delta_{\text{eg}}t'}\right)\\
&+\frac{1}{2} \int_{0}^{\tau} dt'  \int_{0}^{t'} dt'' \lambda_{\text{ge}}(t') \lambda_{\text{ef}}(t'') e^{i\Delta_{\text{eg}}t'} e^{i\Delta_{\text{fe}}t''}\\
&-\frac{1}{2} \int_{0}^{\tau} dt' \int_{0}^{t'} dt'' \lambda_{\text{ef}}(t')\lambda_{\text{ge}}(t'')e^{i\Delta_{\text{fe}}t'}e^{i\Delta_{\text{eg}}t''}\;.
\end{split}
\label{Eq:ToyLeak-<g|UI(t,0)|f> Sol}
\end{align}
%%%%%%%%%%%%%%%%%%%%%%%%%%%%%%%%%%%%%%%%%%%%%%%%%%%%%%%%%%%%%%%%%%%%%%%%%%%%%%%%%
which is a sum over all possible paths connecting states $\ket{g}$ and $\ket{f}$ either as a result of a direct interaction (first order) or through intermediate states (second order and higher). Other transitions have the same generic form. 

According to Eq.~(\ref{Eq:ToyLeak-<g|UI(t,0)|f> Sol}), leakage occurs when the pulse spectrum has a non-negligible overlap with system transition frequencies. For instance, the first order term in Eq.~(\ref{Eq:ToyLeak-<g|UI(t,0)|f> Sol}) can be expressed in frequency domain as
\begin{align}
\begin{split}
&-i \int_{0}^{\tau} dt' \lambda_{\text{gf}}(t') e^{i\Delta_\text{fg}t'}= \\
&-\int_{-\infty}^{\infty} \frac{d\omega'}{2\pi}\frac{\tilde{\lambda}_{\text{gf}}(\omega')}{\omega'+\Delta_{\text{fg}}}\left[e^{i \Delta_{\text{fg}} \tau}-1\right] \;.
\end{split}
\label{Eq:ToyLeak-gTof FreqRep}
\end{align}
In Sec.~\ref{Sec:Leak} of the main text, we employ this fact to explain leakage to specific high-excitation qubit states in the parameter space of the RIP gate.  	
%%%%%%%%%%%%%%%%%%%%%%%%%%%%%%%%%%%%%%%%%%%%%%%%%%%%%%%%%%%%%%%%%%%%%%%%%%%%%%%%

%%%%%%%%%%%%%%%%%%%%%%%%%%%%%% Appendix: Leakage Error %%%%%%%%%%%%%%%%%%%%%%%%%
\section{Leakage error}
\label{App:FidITOLeak}
Here we show that the average error of an arbitrary quantum operation has an approximate lower bounded set by the corresponding average leakage probability \cite{Wood_Quantification_2018}. Our assumption is that the \text{physical} evolution occurs in an extended D-dimensional Hilbert space, while the intended operation is defined over a d-dimensional subspace. 

We employ projection operators $\hat{P}$ and $\hat{Q}$, onto the computational and leakage subspaces respectively, that satisfy
\begin{align}
\begin{split}
&\hat{P}^2=\hat{P} \;,  \quad \quad \ \hat{Q^2}=\hat{Q} \;,\\
& \hat{P}+ \hat{Q}=\hat{I} \;, \quad \hat{P}\hat{Q}=\hat{Q}\hat{P}=0 \;.
\end{split}
\label{eqn:FidITOLeak-P,Q Identity}
\end{align}
We then rewrite the overall time-evolution as
\begin{align}
\begin{split}
\hat{U} & =(\hat{P}+\hat{Q})\hat{U}(\hat{P}+\hat{Q})\\
& = \hat{P}\hat{U}\hat{P} + \hat{P}\hat{U}\hat{Q} + \hat{Q}\hat{U}\hat{P} + \hat{Q}\hat{U}\hat{Q} \;,
\end{split}
\label{eqn:FidITOLeak-Def of reduction}
\end{align}
where the first term provides the intended reduced operation (not necessarily unitary) in the computational subspace as
\begin{align}
\hat{U}_r\equiv \hat{P} \hat{U} \hat{P} \;.
\label{eqn:FidITOLeak-Def of Ur}
\end{align}

We assume the system starts in the computational subspace with initial state $\ket{\Psi_s}$ and the final state is obtained as $\ket{\Psi_f}\equiv \hat{U}\ket{\Psi_s}$. The corresponding leakage probability reads
\begin{align}
p_{L,\Psi_s} \equiv \bra{\Psi_f} \hat{Q}^{\dag}\hat{Q}\ket{\Psi_f} =\bra{\Psi_s} \hat{U}^{\dag}\hat{Q} \hat{U}\ket{\Psi_s} \;.
\label{eqn:FidITOLeak-Def of pL}
\end{align}
Accordingly, we can define an average leakage probability by tracing over all possible initial states in the computational subspace as
\begin{align}
\bar{p}_{L} \equiv \frac{1}{d}\Tr \left(\hat{P}\hat{U}^{\dag}\hat{Q} \hat{U}\hat{P}\right) \;.
\label{eqn:FidITOLeak-Def of avg pL}
\end{align}
In Eq.~(\ref{eqn:FidITOLeak-Def of avg pL}), the trace is defined over the extended Hilbert space, explaining the need for additional $\hat{P}$ projections. 

Average gate fidelity between the reduced operation $\hat{U}_r$ and the ideal (intended) gate operation $\hat{U}_i$ can be obtained as \cite{Pedersen_Fidelity_2007}
\begin{align}
\bar{F}(\hat{U}_{r},\hat{U}_i)\equiv \frac{\Tr\left(\hat{U}_{r}\hat{U}_{r}^{\dag}\right)}{d(d+1)}+\frac{\left|\Tr\left(\hat{U}_{r}^{\dag}\hat{U}_{i}\right)\right|^2}{d(d+1)} \;.
\label{eqn:FidITOLeak-Def of Fid}
\end{align}
The first term in Eq.~(\ref{eqn:FidITOLeak-Def of Fid}) captures the infidelity due to $\hat{U}_r$ not being unitary, which e.g. can be caused by leakage, while the second term captures the infidelity as a result of deviation between the reduced physical and ideal operations. In the following, we derive a rather simple relation between the average leakage probability~(\ref{eqn:FidITOLeak-Def of avg pL}) and average fidelity~(\ref{eqn:FidITOLeak-Def of Fid}). 

Employing Eqs.~(\ref{eqn:FidITOLeak-P,Q Identity})--(\ref{eqn:FidITOLeak-Def of Ur}) and the cyclic property of the trace operation we can write
\begin{align}
\begin{split}
\Tr \left(\hat{U}_r \hat{U}_r^{\dag}\right) &= \Tr \left(\hat{P}\hat{U} \hat{P}\hat{P}^{\dag}\hat{U}^{\dag} \hat{P}^{\dag}\right) \\
&=\Tr \left(\hat{P}\hat{U}^{\dag}\hat{P}\hat{U} \hat{P}\right)\\
&=\Tr \left(\hat{P}\hat{U}^{\dag} (\hat{I}-\hat{Q})\hat{U}\hat{P}\right)\\
&=\Tr \left(\hat{P}\hat{U}^{\dag} \hat{U}\hat{P}\right)-\Tr \left(\hat{P}\hat{U}\hat{Q}\hat{U}^{\dag} \hat{P}\right)\\
&=\Tr\left(\hat{P}\right)-\Tr \left(\hat{P}\hat{U}\hat{Q}\hat{U}^{\dag} \hat{P}\right) \;.
\end{split}
\label{eqn:FidITOLeak-Simple Tr(Ur^d Ur)} 
\end{align}
Comparing Eq.~(\ref{eqn:FidITOLeak-Simple Tr(Ur^d Ur)}) with the average leakage probability~(\ref{eqn:FidITOLeak-Def of avg pL}) and using $\Tr (\hat{P})=d$ we find
\begin{align}
\Tr \left(\hat{U}_r \hat{U}_r^{\dag} \right) = d(1-\bar{p}_L) \;.
\label{eqn:FidITOLeak-Tr(Ur^d Ur) final}
\end{align}
Equation~(\ref{eqn:FidITOLeak-Tr(Ur^d Ur) final}) shows that the amount by which the reduced operation is non-unitary is proportional to the average leakage probability. 

Knowledge of the average leakage probability is \textit{not} sufficient for finding an estimate for the second term in Eq.~(\ref{eqn:FidITOLeak-Def of Fid}). This is dependent on the underlying noise model that sets the relation between $\hat{U}_r$ and $\hat{U}_i$ (see Ref.~\cite{Wood_Quantification_2018} for a detailed analysis). If we assume that $\hat{U}_r \propto \hat{U}_i$, based on Eq.~(\ref{eqn:FidITOLeak-Tr(Ur^d Ur) final}) and the fact that $\hat{U}_i$ is unitary, we can deduce 
\begin{align}
\hat{U}_r \approx \sqrt{1-\bar{p}_L}\hat{U}_i \;.
\label{eqn:FidITOLeak-Ur ITP Ui}
\end{align}
Replacing Eq.~(\ref{eqn:FidITOLeak-Ur ITP Ui}) in the second term of Eq.~(\ref{eqn:FidITOLeak-Def of Fid}) we find	
\begin{align}
\left|\Tr \left(\hat{U}_r^{\dag}\hat{U}_i\right)\right|^2 \approx d^2 (1-\bar{p}_L) \;.
\label{eqn:FidITOLeak-|Ur^d Ui|}
\end{align}
Substituting Eqs.~(\ref{eqn:FidITOLeak-Tr(Ur^d Ur) final}) and~(\ref{eqn:FidITOLeak-|Ur^d Ui|}) into Eq.~(\ref{eqn:FidITOLeak-Def of Fid}) yields
\begin{align}
\begin{split}
\bar{F}(\hat{U}_r,\hat{U}_i) & \approx \frac{1}{d(d+1)}\left[d(1-\bar{p}_L)+d^2(1-\bar{p}_L)\right]\\
& =1-\bar{p}_L\;.
\end{split}
\label{eqn:FidITOLeak-Fid ITO pL}
\end{align}

In summary, the average leakage probability sets an approximate lower bound for the corresponding average error $\bar{E} (\hat{U}_r,\hat{U}_i )\equiv 1-\bar{F} (\hat{U}_r,\hat{U}_i )$ as
\begin{align}
\bar{E}(\hat{U}_r,\hat{U}_i) \gtrsim \bar{p}_L \;.
\label{eqn:FidITOLeak-Infid ITO avg pL}
\end{align}
In the case of a two-qubit gate operation with $d=4$ and target error given by $\bar{E}(\hat{U}_r,\hat{U}_i) \leq 10^{-4}$, the average leakage probability must be kept below $\bar{p}_L \lesssim 10^{-4}$ as well.

%%%%%%%%%%%%%%%%%%% Appendix: Numerical methods %%%%%%%%%%%%%%%%%%%%%%%%%%%%%%%%%
\section{Numerical methods}
\label{App:NumMet}

Numerical simulations for leakage in Sec.~\ref{Sec:Leak} were performed using the Schr\"odinger equation solver of the \texttt{QuTiP} package \cite{Johansson_Qutip_2012}, called \texttt{sesolve}, plus standard Python scientific computing dependencies such as \texttt{numpy} \cite{Harris_Array_2020}, \texttt{scipy} \cite{Virtanen_Scipy_2020} and \texttt{pandas} \cite{Mckinney_Pandas_2011}. Simulations were run in parallel over multiple servers: 3 servers with 48 cores and 196 GB of RAM, 1 server with 56 cores and 248 GB of RAM, plus a controlling server with 4 cores and 8 GB of RAM. These servers were running Red Hat Enterprise Linux 7 and IBM Spectrum LSF for job queuing. Simulations were managed using \texttt{dask} and \texttt{dask\textunderscore jobqueue} \cite{dask_2016} to submit jobs to \texttt{LSF}. \texttt{QuTiP}, \texttt{dask\textunderscore jobqueue}, and their dependencies were managed using \texttt{conda} \cite{conda_2017} and the packages provided by \texttt{conda-forge} \cite{conda_forge_community_2015_4774216}. The servers were provisioned on the IBM Cloud by the IBM Research Hybrid Cloud service.
%%%%%%%%%%%%%%%%%%%%%%%%%%%%%%%%%%%%%%%%%%%%%%%%%%%%%%%%%%%%%%%%%%%%%%%%%%%%%%%%

%%%%%%%%%%%%%%%%%%%%%%%%%%%%%% Bibliography %%%%%%%%%%%%%%%%%%%%%%%%%%%%%%%%%%%%
\bibliographystyle{unsrt}
\bibliography{RIPGateBibliography}
%%%%%%%%%%%%%%%%%%%%%%%%%%%%%%%%%%%%%%%%%%%%%%%%%%%%%%%%%%%%%%%%%%%%%%%%%%%%%%%%
\end{document}